\documentclass[a4paper,10pt]{article}

\usepackage{amsmath}
\usepackage{amsthm}
\usepackage{amsfonts}
\usepackage{hyperref}
\usepackage{color}
\usepackage{multirow}
\usepackage[dvips]{graphicx}
\usepackage{subfigure}
\usepackage{amssymb}
\usepackage{epsfig}
\usepackage{natbib}

\newcommand{\comment}[1]{}
\numberwithin{equation}{section}

\begin{document}

\bibliographystyle{plainnat}

\title{\textbf{Numerical Study of Drop Motion on a Surface with Wettability Gradient
and Contact Angle Hysteresis}}

\author{$\textbf{Jun-Jie Huang}^{1, 2,  3\footnote{Corresponding author. E-mail: jjhuang1980@gmail.com; jjhuang@cqu.edu.cn.}}, 
\textbf{Haibo Huang}^{4},
\textbf{Xinzhu Wang}^{1, 2}$\\
\\
$ ^1$  Department of Engineering Mechanics, \\
Chongqing University, Chongqing 400044, China \\
$ ^2$ Chongqing Key Laboratory of Heterogeneous Material Mechanics \\
(Chongqing University), Chongqing 400044, China\\
$ ^3$ State Key Laboratory of Mechanical Transmission, \\
Chongqing University, Chongqing 400044, China \\
$ ^4$ Department of Modern Mechanics, \\
University of Science and Technology of China, Hefei, Anhui 230026, China}

\maketitle

\textbf{Abstract}

In this work, the motion of a two-dimensional drop on a surface with given wettability gradient
is studied numerically by a hybrid lattice-Boltzmann finite-difference method
using the multiple-relaxation-time collision model. 
We incorporate the geometric wetting boundary condition 
that allows accurate implementation of a contact angle hysteresis model.
The method is first validated through three benchmark tests, 
including the layered Poiseuille flow with a viscosity contrast,
the motion of a liquid column in a channel with specified wettability gradient
and the force balance for a static drop attached to a surface with hysteresis 
subject to a body force.
Then, simulations of a drop on a wall with given wettability gradient
are performed under different conditions.
The effects of the Reynolds number, 
the viscosity ratio, the wettability gradient,
as well as the contact angle hysteresis
on the drop motion are investigated in detail.
It is found that the capillary number of the drop in steady state
is significantly affected by the viscosity ratio, the magnitudes of the wettability gradient
and the contact angle hysteresis,
whereas it only shows very weak dependence on the Reynolds number.

\textit{Wettability Gradient}, 
\textit{Contact Angle Hysteresis},
\textit{Wetting Boundary Condition},
\textit{Hybrid Lattice Boltzmann Method}.

\newpage

\section{Introduction}\label{sec:intro}

The motion of a drop is encountered in nature, in our daily life
and in many industries as well.
It may be caused by body forces like gravity, by a difference in pressure,
or by a difference in surface forces (for example,
the Marangoni effect due to surface tension gradient
and a migrating drop on a surface with wettability gradient (WG)).
As the size of the drop decreases, the surface forces
become more important in determining the motion of the drop.
To drive and control the motion of discrete drops
through modifications of surface wettability
possesses many advantages at small scales.
Such problems have received more and more attention in recent years
because of their significance in digital microfluidics
and the development of lab-on-a-chip as well as some other technologies
~\citep{Darhuber2005}.
For instance, recently,~\cite{loc10-wg-drop-mixing} employed
a surface with WG to accelerate a droplet before its collision with
another one to enhance the mixing between them,
and~\cite{jmm13-wgd-pump} developed a micropump by using
axisymmetric WG to drive droplets, which has potential applications in
some microelectromechanical systems.
Besides, gradient surfaces or directional surfaces, 
including those having WG, 
have been found to be used for droplet transport in many natural phenomena
~\citep{afm12-directional-surf-review}.
Therefore, the study of drop motion caused by WG
has important implications in many areas.
Due to the presence of either geometrical or chemical heterogeneities (or both),
the motion of an interface on solid substrates usually shows certain hysteresis,
i.e., the contact angle when the interface is moving forward 
(called the \textit{advancing contact angle}, denoted as $\theta_{A}$)
is larger than that when it is moving backward 
(called the \textit{receding contact angle}, denoted as $\theta_{R}$).
This phenomenon is known as contact angle hysteresis (CAH).
It is characterized by the difference between the two angles, $\theta_{A}-\theta_{R}$,
and it may strongly affect a drop driven by WG
especially at small scales.

The study of drops under WG began more than two decades ago.
~\cite{l89-drop-gradient} analyzed the motion of two-dimensional (2-D) and three-dimensional (3-D) droplets
on substrates with small gradient in wettability or temperature
at different scales and obtained the formula for the droplet velocity
through the balance of driving and resistance forces 
under the quasi-steady assumption
as well as some other simplifying assumptions.
~\cite{jphysii91-wg-drop} did experimental studies about
the dynamics of a 2-D drop (liquid ridge) sitting
across a wettability discontinuity.
The displacements and constant angles of the two contact lines
of the drop were measured
and several stages of motion were identified,
including a steady stage with constant velocity
and constant advancing and receding angles.
~\cite{sci92-wg-drop} demonstrated 
the (continuous) WG-driven
(water) drop in experiment even on a substrate tilted 
to the horizontal by $15^{\circ}$,
showing that the driving force caused by a strong chemical gradient 
can become large enough to 
overcome both the gravity and the hydrodynamic resistance.
They also pointed out that the effect
of CAH must be small;
otherwise, the drop might not move.
~\cite{prl95-frun-drop} demonstrated spontaneous droplet motion
on a surface that is modified by some agents inside the droplet through some reaction.
A quite broad range of droplet velocities were reported,
and the relations between the droplet velocity and its size as well as the receding contact angle
were obtained and compared favorably with the theoretical ones.
To guide the droplet's motion, a track between two hydrophobic regions was employed
by~\cite{prl95-frun-drop}.
~\cite{l02-vib-wg-drop} showed that the resistance caused by CAH
may be greatly reduced by applying a periodic force
to the drop generated from a in-plane vibration 
that resulted from an audio speaker.
Later,~\cite{l04-drop-grad-surf-ratchet} extended this study
to more fluids with different surface tensions and viscosities
and explored the vibration parameters
(including the wave form, amplitude and frequency). 
They found an interesting ratcheting motion of a drop on gradient surfaces
that results from shape fluctuation and that the drop velocity 
increases linearly with the amplitude but nonlinearly with the frequency.
By employing the wedge and lubrication approximations respectively, 
~\cite{l05-wg-drop} derived two sets of approximate results
for a 3-D drop driven by WG
on the resistance acting on the drop and 
also on the quasi-steady drop velocity.
Subsequently,~\cite{l06-wg-drop-exp-theor} reported experiments 
for a 3-D drop on a surface with spatially varying WG,
and the respective results were compared with the theoretical predictions
presented in~\citep{l05-wg-drop} assuming quasi-steady state.
The varitions of the velocity of the drop with its position
were obtained in the experiments. 
The theoretical results were shown to agree reasonably well
with the experimental ones when the hysteresis effect was included,
but larger discrepancy was seen if the hysteresis effect
was not considered in the theoretical ones.
~\cite{l05-rw-driven-drop-lbm} employed reactive-wetting,
which modifies the wetting property once a drop covers the surface,
to drive a 3-D drop along a (tilted) surface, and they reported 
the measured drop velocities at different angle of inclination.
Besides,~\cite{l05-rw-driven-drop-lbm} also performed computer simulations
using the lattice-Boltzmann method (LBM) to visualize the drop motion, 
the flow and pressure fields.
Hysteresis effect was reported to be negligible
because of the specifically selected and prepared gold substrates.
Hysteresis was not considered in the simulations
by~\cite{l05-rw-driven-drop-lbm}, either.
~\cite{l05-wg-drop-ew} demonstrated reversible droplet transport 
through experiments that used dynamic bias voltage
to adjust electrochemical reactions and ultimately to generate WG.
~\cite{pof06-wg-drop-theor} developed an asymptotic theory
for a WG-driven 2-D droplet based the lubrication assumption.
The droplet's shape and velocity were obtained as functions
of the WG and the volume of the droplet.
The CAH effect was not included in their theoretical analysis.
~\cite{pof08-wg-drop} did experimental studies 
on emulsion separation induced by an abrupt change in wettability
and also did LBM simulations
of a 3-D droplet on a surface with a step WG.
Enhanced separation was reported in confined geometry
and it was also highlighted that smaller droplets
are more easily guided by the step WG.
In their simulations, the substrates were assumed to be ideally flat
and the hysteresis effects did not come into play.
~\cite{jchemphys08-md-wg-drop} carried out molecular dynamics (MD) study
of the dynamics of a nanodroplet on a surface with different types
of WG. Systems of nanometer scale,
including a Lennard-Jones system and water on a self-assembled monolayer,
were investigated and the observables reported by~\cite{jchemphys08-md-wg-drop}
included the shape, the center-of-mass position, the velocity,
the base length, and the advancing and receding angles of the droplet
during the motion. Reasonable agreement were obtained between
the MD simulation and theoretical prediction on the droplet velocity.
Considerable focus was given by~\cite{jchemphys08-md-wg-drop}
to the CAH, the inclusion of which improved the agreement.
It is noted that the work by~\cite{jchemphys08-md-wg-drop}
may be regarded as a kind of (virtual) experiment and there were no explicit CAH model
because of the extremely small size.
~\cite{jcis08-lbm-wg-drop} performed 3-D LBM simulations of a droplet
on substrates with different wettability distribution
and temporal control (of the wettability),
and identified suitable spatiotemporal wettability control parameters 
for unidirectional droplet transport.
The effects of CAH were not included by~\cite{jcis08-lbm-wg-drop},
which may lead to large discrepancies for comparisons with experiments.
Recently,~\cite{aps2010-cn-wg-drop} employed LBM to simulate a 2-D droplet
on a substrate with a step WG, which was set to follow the droplet's motion
to ensure a continually acting driving force (to some extent,
mimicking the situation of reactive wetting used in~\citep{prl95-frun-drop}).
Variations of the droplet velocity and the dynamic contact angles were extracted
from the simulations and were shown to agree reasonable well with theoretical predictions
when the WG was small.
~\cite{l10-wg-drop} employed the smoothed particle hydrodynamics technique
based on the diffuse interface method to study the dynamics of a 3-D drop
on an inclined surface with WG.
The effects of the drop size, the angle of surface inclination
and the strength of WG on the drop motion were investigated,
and several possible outcomes were reported, depending on these conditions.
~\cite{l10-wg-drop} presented some results about CAH for the drop even though
they did not introduce any surface heterogeneities or use any explicit CAH model
(thus we suspect that the reported CAH is actually due to some dynamic effects
or simply reflects the interaction between the gravity and WG,
and is not like the CAH in other studies).
~\cite{pre12-wg-drop} presented systematically a phase-field-based 
thermohydrodynamic model for one-component two-phase fluids
with certain boundary conditions derived from various balance equations,
and employed it to numerically study the 2-D droplet motion on substrates
having given WG
with/without phase transition and substrate temperature change.
They investigated the droplet's shape, migration velocity $V_{\textrm{mig}}$, 
the velocity profile at selected sections,
and the distribution of slip velocity.
They obtained the relations between $V_{\textrm{mig}}$, 
the magnitude of WG (denoted by
$S = \frac{d}{dx} \cos \theta$ following the notation by~\cite{l05-wg-drop}
where $\theta$ is the contact angle of the wall and $x$ is the coordinate
along the direction of WG)
and the slip length $l_{s}$,
which agree with previous theoretical predictions by~\cite{l89-drop-gradient}.
The effects of CAH were not considered by~\cite{pre12-wg-drop},
which makes the $V_{\textrm{mig}}-S$ line pass the origin
(in the presence of CAH, $V_{\textrm{mig}}$ may remain to be zero
before $S$ reaches certain minimum value capable of driving the droplet).
~\cite{jsm12-wg-drop-lbm-sim} carried out LBM simulations of a 2-D droplet
inside a microchannel  with a stepwise change in wettability,
which differs from most of previous studies
on droplets under open geometry.
The simulated evolutions of the droplet velocity were found to agree well
with analytical predictions developed by~\cite{jsm12-wg-drop-lbm-sim}.
They focused on the effects of the channel height, the ratios of fluid viscosity 
and density, the channel geometry (for grooved channels),
as well as the appearance of an obstacle inside the channel.
CAH was not considered in this work.

Despite abundant research on WG-driven drops
there are still certain open questions and characteristics 
of such drop motions that remain to be explored.
As noted above, most existing simulations of WG-driven drops
have not included (explicitly) the effects of CAH through a \textit{continuum} model
even though CAH has been long identified as an important factor
in the motion of WG-driven drop
~\citep{sci92-wg-drop, l06-wg-drop-exp-theor}.
How CAH affects the migration velocity and the flow
need to be investigated more thoroughly or 
to be further confirmed by simulations.
Besides, most previous studies did not consider the effects of the surrounding fluid.
This is reasonable for air-liquid systems with large density and viscosity contrasts,
but may be questionable for liquid-liquid (e.g., water-oil) systems.
Just for curiosity, it is also of interest to know how the surrounding fluid affects the drop motion caused by WG.
In this work, we study a 2-D drop on a surface with specified WG
numerically by a phase-field-based hybrid lattice-Boltzmann method
that incorporates a CAH model.
Our aim is mainly to explore the effects of the Reynolds number,
the wettability gradient of different strengths,
the viscosity ratio, 
and the CAH 
on the motion of the drop. 
Even though phenomenological CAH models have been included
in simulations of various multiphase flows,
including a drop subject to a shear flow, a pressure gradient 
or under the action of gravity (see, e.g.,
~\citep{jcp05-cah-gfm, Spelt06, ijmf8-cah-vof, jfm08-drop-shear, 
jcp10-cah-vof, pre13cah-lbm}),
it appears to us that they have not been used in the study of WG-driven drops.
Thus, this work is a further and essential step 
towards more accurate and realistic simulations of WG-driven drops.

The paper is organized as follows.
Section~\ref{sec:method}  introduces
the phase-field model for binary fluids and the wetting boundary condition (WBC)
on a wall together with the CAH model.
The numerical method (simplified from~\citep{hybrid-mrt-lb-fd-axisym}) and
the implementation of the WBC and CAH model are also described briefly in this section.
In Section~\ref{sec:res-dis},
several validation tests are presented first and then
investigations on a drop driven by a stepwise WG are carried out 
under different conditions, and the results are discussed
and compared with some theoretical predictions. 
Section~\ref{sec:conclusion} summaries the findings and concludes this paper.

\section{Theoretical and Numerical Methodology}\label{sec:method} 

The present simulations are based on the phase-field modeling 
of two-phase flows.
The physical governing equations are solved by a hybrid 
lattice-Boltzmann finite-difference method.
For flows of binary fluids,
there are two fundamental dynamics: 
the hydrodynamics for fluid flow and
the interfacial dynamics.
We introduce the phase-field model for interfacial dynamics first.

\subsection{Phase-Field Model}\label{sec:pfm-if}

In the phase-field model, two immiscible fluids are distinguished 
by an order parameter field $\phi$.
For a system of binary fluids, 
a free energy functional $\mathcal{F}$ may be defined based on $\phi$ as,
\begin{equation}
  \label{eq:fe-functional-def}
  \mathcal{F} (\phi, \boldsymbol{\nabla} \phi)
  = \int_{V} \bigg( \Psi (\phi) 
  + \frac{1}{2} \kappa \vert \boldsymbol{\nabla} \phi \vert ^2 
  \bigg) dV 
  + \int_{S} \varphi (\phi_{S} ) dS ,
\end{equation}
where $\Psi (\phi)$ is the \textit{bulk free energy} density. 
The popular form of $\Psi (\phi)$ is the double-well form,
\begin{equation}
  \label{eq:bulk-fe}
  \Psi (\phi) = a (\phi^{2} - 1)^{2} ,
\end{equation}
with $a$ being a constant.
With this form of $\Psi (\phi)$,
$\phi$ varies between $1$ in one of the fluids 
(named \textit{fluid A} for convenience)
and $-1$ in the other (named \textit{fluid B}).
The second term in the bracket on the right-hand-side (RHS) of Eq. (\ref{eq:fe-functional-def})
is the \textit{interfacial energy} density
with $\kappa$ being another constant,
and the last term on the RHS of Eq. (\ref{eq:fe-functional-def})
in the surface integral, $\varphi (\phi_{S} )$, is the 
\textit{surface energy} density 
with $\phi_{S}$ being the order parameter on the surface (i.e., solid wall).

The chemical potential $\mu$ is obtained 
by taking the variation of the free energy functional $\mathcal{F}$
with respect to the order parameter $\phi$,
\begin{equation}
  \label{eq:chem-potential}
\mu = \frac{\delta \mathcal{F}}{\delta \phi} 
= \frac{d \Psi (\phi)}{d \phi} -  \kappa \nabla ^2 \phi
= 4 a \phi (\phi ^2 -1) -  \kappa \nabla ^2 \phi.
\end{equation}
The coefficients $a$ (in the bulk free energy) and 
$\kappa$ (in the interfacial energy)
are related to the interfacial tension $\sigma$ and 
interface width $W$ as~\citep{ijnmf09-pflbm-mobility},
\begin{equation}
  \label{eq:a-kappa-sigma-W}
  a = \frac{3 \sigma}{4 W}, \quad
  \kappa = \frac{3 \sigma W}{8}.
\end{equation}
Equivalently, the interfacial tension $\sigma$ and interface width $W$ 
can be expressed in terms of $a$ and $\kappa$ as,
\begin{equation}
  \label{eq:sigma-W-a-kappa}
\sigma  = \frac{4}{3} \sqrt{ 2 \kappa a}, \quad 
W = \sqrt{\frac{2 \kappa}{a}}.
\end{equation}
Usually, it is assumed that the diffusion of the order parameter 
is driven by the gradient of the chemical potential.
By also including the contribution due to convection,
one obtains the following evolution equation of the order parameter~\citep{jacqmin99jcp},
\begin{equation}
  \label{eq:che}
  \frac{\partial \phi}{\partial t}
  + (\boldsymbol{u} \cdot \boldsymbol{\nabla}) \phi
  = \boldsymbol{\nabla} \cdot (M  \boldsymbol{\nabla} \mu)   ,
\end{equation}
where $M$ is the diffusion coefficient 
called \textit{mobility} (taken as constant in this work),
and $\boldsymbol{u}$ is the local 
fluid velocity.

Suitable boundary conditions are needed for
Eqs. (\ref{eq:chem-potential}) and (\ref{eq:che}).
Here we mainly focus on the conditions near a (rigid) wall,
which are closely related to the wetting phenomenon and 
the motion of contact line.
For the fluid velocity $\boldsymbol{u}$ that appears in Eq. (\ref{eq:che}),
we assume that the no-slip condition applies on a wall.
In what follows, we concentrate on the conditions 
for the phase-field variables, $\phi$ and $\mu$.
It is noted that in phase-field simulations interface slip on a wall
is allowed due to the diffusion in Eq. (\ref{eq:che})~\citep{jacqmin00jfm}.

\subsection{Wetting Boundary Condition}\label{ssec:wbc}

On a wall, the boundary condition for the chemical potential $\mu$
is simply the no-flux condition,
\begin{equation}
  \label{eq:bc-chem-potential}
  \boldsymbol{n}_{w} \cdot \boldsymbol{\nabla} \mu \vert _{S}
  = \frac{\partial \mu}{\partial n_{w}} \bigg\vert _{S} 
 = 0  ,
\end{equation}
where $\boldsymbol{n}_{w}$ denotes the unit normal vector on the wall
pointing \textit{into} the fluid.
For the order parameter $\phi$, 
there are different kinds of boundary conditions
in the literature with varying degree of complexity
~\citep{jacqmin00jfm, pre03-gnbc, FELBM-CL04b, 
pre07-geom-wbc, ijmpc09wall-fe-bc, 
pof09-dyn-wetting, cpc11wbc-lbm, pof11-wall-energy-relax}. 
\cite{wbc-dewetting} compared several types of boundary conditions
for the study of drop dewetting.
Here the wetting boundary condition (WBC) in geometric formulation 
proposed by~\cite{pre07-geom-wbc}
is adopted because of its certain advantages.
The geometrical WBC abandons the surface energy integral 
in Eq. (\ref{eq:fe-functional-def}) and starts from some
geometric considerations. It assumes that
the contours of the order parameter in the diffuse interface are
parallel to each other, including in the region near
the surface. Then, the unit vector normal to the interface, 
denoted by $\boldsymbol{n}_{s}$, 
may be written in terms of the gradient of the order parameter 
as~\citep{pre07-geom-wbc},
\begin{equation}
  \boldsymbol{n}_{s} = \frac{\boldsymbol{\nabla} \phi}{\vert 
    \boldsymbol{\nabla} \phi\vert} .
\end{equation}
By noting that the vector $\boldsymbol{\nabla} \phi$ may be decomposed as,
\begin{equation}
  \boldsymbol{\nabla} \phi = (\boldsymbol{n}_{w} \cdot \boldsymbol{\nabla}
  \phi) \boldsymbol{n}_{w} 
+ (\boldsymbol{t}_{w} \cdot \boldsymbol{\nabla} \phi) \boldsymbol{t}_{w}  ,
\end{equation}
where $\boldsymbol{t}_{w}$ is the unit tangential vector along the wall,
one finds that the contact angle $\theta$ 
at the contact line may be expressed by,
\begin{equation}
  \label{eq:geom-ca}
  \tan \bigg( \frac{\pi}{2}-\theta \bigg) 
  = \frac{- \boldsymbol{n}_{w} \cdot \boldsymbol{\nabla} \phi}{\vert 
    \boldsymbol{\nabla} \phi 
    - (\boldsymbol{n}_{w} \cdot \boldsymbol{\nabla} \phi) \boldsymbol{n}_{w}
    \vert} 
  = \frac{- \boldsymbol{n}_{w} \cdot \boldsymbol{\nabla}
    \phi}{\vert (\boldsymbol{t}_{w} \cdot \boldsymbol{\nabla} \phi)
    \boldsymbol{t}_{w} \vert}  .
\end{equation}
Thus, one has,
\begin{equation}
  \label{eq:bc-op-geom}
  \frac{\partial \phi}{\partial n_{w}} \bigg\vert _{S} 
  = - \tan \bigg( \frac{\pi}{2}-\theta \bigg)  
  \vert  \boldsymbol{t}_{w} \cdot \boldsymbol{\nabla} \phi \vert  .
\end{equation}
In the design of this geometric WBC, 
the following fact has been taken into account: 
the tangential component of $\phi$'s graident cannot be modified
during simulation and the local (microscopic) contact angle can only
be enforced through the change of the normal
component~\citep{pre07-geom-wbc}.
Owning to this, the geometric WBC performs better than other 
surface-energy-based boundary conditions
in assuring that the local contact angle matches the specified one.

\subsection{Contact Angle Hysteresis Model}\label{ssec:cah}

The above boundary conditions are applicable for ideally smooth surfaces
with given contact angle.
In reality, however, perfectly smooth surfaces are rarely encountered
and the CAH can play an important role.
There exist some investigations on the relation between the CAH
and the underlying surface heterogeneities at small scales
(e.g., see the work by~\cite{l07-lbm-cah}).
Here we do not intend to consider the CAH by directly including 
the surface heterogeneities;
instead, we assume the surface is surficiently smooth at a relatively large scale
and employ a phenomenological CAH model for contact lines. 
Specifically, the method presented 
by~\cite{jfm08-drop-shear} (in phase-field simulation)
and also used by~\cite{pre13cah-lbm} (in LBM simulation) is employed. 
In this method, the effects of CAH are considered as follows,
\begin{equation}
  \label{eq:effective-cah}
  \left.
  \begin{matrix} 
    \theta = \theta_{A}  \quad 
    &&\textrm{if} \quad U_{\textrm{cl}} > 0 \\
    \theta_{R} < \theta < \theta_{A}   \quad 
    &&\textrm{if} \quad U_{\textrm{cl}} = 0\\
    \theta = \theta_{R}  \quad 
    &&\textrm{if} \quad U_{\textrm{cl}} < 0 
  \end{matrix} \right\} ,
\end{equation}
where $U_{\textrm{cl}}$ is the contact line velocity.
The implementation details will be described next.

\subsection{Governing Equations for Hydrodynamics and 
Numerical Method}\label{ssec:gvn-eqn-num-method}

In the above, the basics of phase-field model for binary fluids, 
the wetting boundary condition as well as the contact angle hysteresis
model have been presented.
In this section, the governing equations of the fluid flow
and the methods for the numerical solutions of all equations
are briefly introduced. 

When the interfacial tension effects are modeled by the phase-field model,
the governing equations of the incompressible flow of binary fluids
with \textit{uniform density} and 
\textit{variable viscosity} may be written as,
\begin{equation}
  \label{eq:INSCHE_continuity-general}
\boldsymbol{\nabla} \cdot \boldsymbol{u} = 0   ,
\end{equation}
\begin{equation}
  \label{eq:INSCHE_momentum-general}
\partial_{t} \boldsymbol{u}
+ (\boldsymbol{u} \cdot \boldsymbol{\nabla}) \boldsymbol{u}
= - \boldsymbol{\nabla} S_{p} 
+ \boldsymbol{\nabla} \cdot [\nu (\phi) (\boldsymbol{\nabla} \boldsymbol{u} 
+ (\boldsymbol{\nabla} \boldsymbol{u})^{T}) ]
- \phi \boldsymbol{\nabla} \mu + \boldsymbol{G}  ,
\end{equation}
where $S_{p}$ is a term similar to the hydrodynamic pressure
in single-phase incompressible flow~\citep{jacqmin99jcp},
$\boldsymbol{G}$ is a body force (which may be zero or a function of space and time),
and $\nu (\phi)$ is the kinematic viscosity
which is a function of the order parameter.
In this work, the following function is adopted to interpolate the viscosity
from the order parameter,
\begin{equation}
  \label{eq:nu-phi-jcp10}
  \nu (\phi) 
= \bigg[ \frac{1 + \phi}{2} \frac{1}{\nu_{A}} 
+ \frac{1 - \phi}{2} \frac{1}{\nu_{B}} \bigg]^{-1} ,
\end{equation}
where $\nu_{A}$ and $\nu_{B}$ are the kinematic viscosities of fluid A
(represented by $\phi = 1$) and fluid B (represented by $\phi = -1$), 
respectively.
As pointed out by~\cite{jcp10lbm-drop-impact} and~\cite{pre13-pf-lbm-vardv},
who employed Eq. (\ref{eq:nu-phi-jcp10}) or its equivalent form
in LBM simulations of binary fluids,
this form of function for $\nu (\phi)$ performs better than
the commonly used linear function  
in phase-field simulations, which reads
~\citep{jcp03pfm, jcp06pf-fem-adaptive},
\begin{equation}
  \label{eq:nu-phi-lin}
  \nu (\phi)
= \frac{1 + \phi}{2} \nu_{A} + \frac{1 - \phi}{2} \nu_{B} .
\end{equation}
We note that~\cite{jcp97-layered-couette-vis} analyzed
the issue of viscosity interpolation 
and proposed the use of Eq. (\ref{eq:nu-phi-jcp10}) 
much earlier 
in the volume-of-fluid simulation of two-phase flows.

The complete set of governing equations of binary fluids considered
in this work consists of Eqs. (\ref{eq:INSCHE_continuity-general}), 
(\ref{eq:INSCHE_momentum-general}) and (\ref{eq:che}).
The first two are solved by the lattice-Boltzmann method
and the third is solved by the finite-difference method 
for spatial discretization 
and the $4^{th}-$order Runge-Kutta method for time marching.
The whole method is called the hybrid lattice-Boltzmann finite-difference method.
The present formulation is simplified from 
another axisymmetric version presented by~\cite{hybrid-mrt-lb-fd-axisym},
but with some extension for binary fluids with \textit{variable viscosity}.
Most of the details of this hybrid method can be found
in Ref.~\citep{hybrid-mrt-lb-fd-axisym};
for conciseness, they will not be fully repeated; 
here we mainly describe the extension for variable viscosity
and the implementation of the geometric WBC with CAH model.
It is noted that there are different choices for some of the components
of the hybrid method in~\citep{hybrid-mrt-lb-fd-axisym}.
The present work uses the multiple-relaxation-time (MRT) collision model for LBM,
the centered formulation (instead of the GZS formulation) 
for the forcing term, 
and the isotropic discretization based on the D2Q9 velocity model 
(i.e., the \texttt{iso} scheme in~\citep{hybrid-mrt-lb-fd-axisym}) 
to evaluate the spatial gradients of the phase-field variables.
The effects of variable viscosity are taken into account
through the modification of one of the relaxation parameters
in the MRT collision model, specifically the parameter $\tau_{f}$,
\begin{equation}
  \label{eq:tau-phi-jcp10}
  \frac{1}{\tau_{f} (\phi) - 0.5} 
= \frac{1 + \phi}{2 (\tau_{f, A} - 0.5)} 
+ \frac{1 - \phi}{2 (\tau_{f, B} - 0.5)}  ,
\end{equation}
where $\tau_{f, A}$ and $\tau_{f, B}$ are two relaxation parameters 
related to the kinematic viscosities of fluids A and B
(i.e., $\nu_{A}$ and $\nu_{B}$) as,
\begin{equation}
  \label{eq:nu-tau-A-B}
  \nu_{A} = c_{s}^{2} (\tau_{f, A} - 0.5) \delta_{t}   , \quad
  \nu_{B} = c_{s}^{2} (\tau_{f, B} - 0.5) \delta_{t}   ,
\end{equation}
where $c_{s} = c / \sqrt{3}$ is the \textit{lattice sound speed} in LBM
(for the adopted D2Q9 velocity model), $\delta_{t}$ is the time step,
and $c = \delta_{x} / \delta_{t} $ is the lattice velocity
($\delta_{x}$, the grid size).

The spatial domain of simulation is a rectangle specified by
$0 \leq x \leq L_{x}, \  0 \leq y \leq L_{y}$,
and this domain is discretized into 
$N_{x} \times N_{y}$ uniform squares of side length $h$($=\delta_{x}$),
giving $L_{x} = N_{x} h$ and $L_{y} = N_{y} h$.
The distribution functions in LBM
and the discrete phase-field variables, 
$\phi_{i,j}$ and $\mu_{i,j}$,
are both located at the centers of the squares
(like the cell center in the finite-volume method).
The indices $(i,j)$ for the bulk region 
(i.e., within the computational domain)
are $1\leq i \leq N_{x}, \ 1\leq j \leq N_{y}$.
To faciliate the implementation of boundary conditions,
a ghost layer is added on each side of the domain.

The WBC involves the enforcement
of the normal gradient of the order parameter $\phi$ on the wall.
Consider the case with the lower side being a wall with 
a given contact angle $\theta$.
The enforcement of $\phi$'s normal gradient is realized 
through a ghost layer of squares,
the centers of which are $h/2$ below the wall with the index $j=0$.
When the geometric WBC is used, upon discretization of 
Eq. (\ref{eq:bc-op-geom}), one has,
\begin{equation}
\label{eq:phi0-geom}
\phi_{i,0} = \phi_{i,1} + \tan \bigg( \frac{\pi}{2}-\theta \bigg)
\vert  \boldsymbol{t}_{w} \cdot \boldsymbol{\nabla} \phi \vert  h.
\end{equation}
Eq. (\ref{eq:phi0-geom}) contains the tangential component of 
$\phi$'s gradient on the wall
$\boldsymbol{t}_{w} \cdot \boldsymbol{\nabla} \phi \vert_{S}$, 
and it is evaluated by the following extrapolation scheme,
\begin{equation}
\boldsymbol{t}_{w} \cdot \boldsymbol{\nabla} \phi \vert_{S}
= 1.5 \boldsymbol{t}_{w} \cdot \boldsymbol{\nabla} \phi \vert_{i,1} 
- 0.5 \boldsymbol{t}_{w} \cdot \boldsymbol{\nabla} \phi \vert_{i,2}  ,
\end{equation}
where the tangential gradients on the right-hand-side
are calculated by the central difference scheme, e.g.,
\begin{equation}
\boldsymbol{t}_{w} \cdot \boldsymbol{\nabla} \phi \vert_{i,1} 
= \frac{\partial \phi}{\partial t_{w}} \bigg\vert_{i,1} 
= \frac{\phi_{i+1,1} - \phi_{i-1,1}}{2 h}  .
\end{equation}
Once the order parameter in the ghost layer below the wall is specified
according to Eq (\ref{eq:phi0-geom}),
the normal gradient condition for $\phi$ is enforced. 
For a wall along some other directions, the formulas are similar
(only some changes to the indices are required).
It is noted that the above schemes for finite differencing
and extrapolation are $2^{nd}$-order accurate.

The actual implementation of the CAH model given in Eq. (\ref{eq:effective-cah})
is as follows~\citep{jfm08-drop-shear}.
First, an initial approximation of the local contact angle on the wall,
$\theta^{i}$, is obtained by using Eq. (\ref{eq:geom-ca}).
Based on the range $\theta^{i}$ belongs to (i.e.,
one of the three ranges divided by the advancing
and receding angles, $\theta_{A}$ and $\theta_{R}$),
$\phi_{i,0}$ is specified in one of the following three manners
(also take the lower side as an example):
\begin{itemize}
\item {(1) if $\theta^{i} \geq \theta_{A}$,  set $\theta = \theta_{A}$
and then update $\phi_{i,0}$ by using Eq. (\ref{eq:phi0-geom});}
\item {(2) if $\theta_{R} < \theta^{i} < \theta_{A}$,  
keep $\phi_{i,0}$ unchanged;}
\item {(3) if $\theta^{i} \leq \theta_{R}$,  set $\theta = \theta_{R}$
and then update $\phi_{i,0}$ by using Eq. (\ref{eq:phi0-geom}).}
\end{itemize}

We note that the present work only considers 2-D problems
and the implementation of the WBC with this CAH model is relatively easy
as compared with the situation for 3-D problems
(which may be delt with in future).

\section{Results and Discussions}\label{sec:res-dis}

\subsection{Characteristic Quantities,
Dimensionless Numbers and Numerical Parameters}\label{ssec:char-quant-diml-num-setup}

Before showing the results,
we introduce several important characteristic quantities and dimensionless numbers.
In each of the problems below (including the validation cases and the WG-driven drop), 
a relevant length scale 
(for instance, the drop radius $R$ or the channel height $H$ (or some fraction of it, e.g., $0.25 H$))
is chosen to be the characteristic length $L_{c}$.
Note that if the drop is only part of a circle, we take $R$ as the radius of the \textit{full} circle.
The constant density is selected as the characteristic density $\rho_{c}$.
The interfacial tension is $\sigma$.
As given in Section \ref{ssec:gvn-eqn-num-method},
the kinematic viscosity of fluid A (making up the drop) is $\nu_{A}$ 
(its dynamic viscosity is $\eta_{A} = \rho_{c} \nu_{A}$)
whereas that of fluid B (the ambient fluid) is $\nu_{B}$
(its dynamic viscosity is $\eta_{B} = \rho_{c} \nu_{B}$).
The viscosity ratio is thus $r_{\nu} = \nu_{A} / \nu_{B} = r_{\eta}$.
Based on the fluid properties, one can derive a characteristic velocity $U_{c}$
~\citep{DIMSpreading07},
\begin{equation}
	\label{eq:Uc}
  U_{c} = \frac{\sigma}{\rho_{c} \nu_{A}}   .
\end{equation}
Then, the characteristic time $T_{c}$ is,
\begin{equation}
	\label{eq:Tc}
  T_{c} = \frac{L_{c}}{U_{c}} = \frac{L_{c} \rho_{c} \nu_{A}}{\sigma}  .
\end{equation}
The quantities of length, time and velocity may be scaled by 
$L_{c}$, $T_{c}$ and $U_{c}$, respectively.
In two-phase flows, the capillary number $Ca$ and the Reynolds number $Re$ are commonly used
to characterize a problem.
The capillary number $Ca$ reflects the relative importance of the viscous force 
as compared with the interfacial tension force,
and the Reynolds number $Re$ reflects the ratio of the inertial force over the viscous force.
With the above characteristic quantities, 
the capillary number is found to be,
\begin{equation}
	\label{eq:ca-def1}
  Ca = \frac{\rho_{c} \nu_{A} U_{c}}{\sigma} = 1  ,
\end{equation}
and the Reynolds number is,  
\begin{equation}
	\label{eq:re-def1}
  Re = \frac{U_{c} L_{c}}{\nu_{A}} = \frac{\sigma}{\rho_{c} \nu_{A}} \frac{L_{c}}{\nu_{A}} 
  = \frac{\sigma L_{c}}{\rho_{c} \nu_{A}^{2}}  .
\end{equation}
It is noted that the capillary number and 
the Reynolds number given in Eqs. (\ref{eq:ca-def1}) and (\ref{eq:re-def1})
do not reflect the \emph{actual} physics of the problem
because the velocity scale $U_{c}$ is purely derived from the physical properties of the fluid
rather than taken as the characteristic dynamic velocities during the fluid motion.
Nevertheless, they are helpful in setting up the simulation.

For drop problems with $L_{c} = R$, 
another set of characteristic quantities may be derived
~\citep{jfm03-drop-coal},
\begin{equation}
	\label{eq:Uc-inv}
  U_{c, \textrm{inv}} = \sqrt{\frac{\sigma}{\rho_{c} R}}   ,
  \end{equation}
  \begin{equation}
	\label{eq:Tc-inv}
  T_{c, \textrm{inv}} = \frac{L_{c}}{U_{c, \textrm{inv}}} 
  = \frac{R}{U_{c, \textrm{inv}}}  = \sqrt{\frac{\rho_{c} R^{3}}{\sigma}}  .
\end{equation}
They contain no viscosity and are typically used in inviscid dynamics.
It is easy to find that,
\begin{equation}
	\label{eq:UcTc-inv-UcTc-rel}
  U_{c, \textrm{inv}} = \frac{U_{c}}{\sqrt{Re}}   , \quad
  T_{c, \textrm{inv}} = \sqrt{Re} T_{c} .
  \end{equation}
Following~\cite{jfm05drop-coal}, one may define another 
capillary number and Reynolds number based on $U_{c, \textrm{inv}}$,
\begin{equation}
	\label{eq:ca-inv-def}
  Ca_{\sigma} = \frac{\rho_{c} \nu_{A} U_{c, \textrm{inv}}}{\sigma} , 
    \end{equation}
  \begin{equation}
  \label{eq:re-inv-def}
  Re_{\sigma} = \frac{U_{c, \textrm{inv}} R}{\nu_{A}} . 
\end{equation}
From Eqs. (\ref{eq:ca-def1}), (\ref{eq:re-def1}) and (\ref{eq:UcTc-inv-UcTc-rel}),
it is easy to find that,
\begin{equation}
  Re_{\sigma} =   \sqrt{Re} , \quad Ca_{\sigma}  = \frac{1}{\sqrt{Re}}  = \frac{1}{Re_{\sigma} }.
\end{equation}
In addition, the Ohnesorge number $Oh$
is also often used for drop dynamics~\citep{arfm06-drop-impact}.
It is defined as 
(note here the drop radius, instead of the diameter in~\citep{arfm06-drop-impact}, is used),
\begin{equation}
\label{eq:oh-def}
  Oh = \frac{\rho_{c} \nu_{A}}{\sqrt{\rho_{c} \sigma R}} ,
\end{equation}
and it is related to the other dimensionless numbers as 
$Oh = 1 / \sqrt{Re} = 1 / Re_{\sigma}  = Ca_{\sigma} $.
No body force is included in the study of WG-driven drop,
but in some validation cases a constant body force (per unit mass) $g$ may be applied on the drop.
For a drop of radius $R$ under the action of a body force $g$,
the Bond number may be defined as, 
\begin{equation}
\label{eq:bo-def}
Bo = \frac{\rho_{c} g R^{2}}{\sigma}  ,
\end{equation}
which reflects the ratio of the body force over the interfacial tension force.

In phase-field-based simulations of two-phase flows, 
two additional parameters are
introduced: (1) the Cahn number 
(the ratio of interface width over the characteristic length),
\begin{equation}
\label{eq:cn-def}
Cn = \frac{W}{L_{c}}  ,
\end{equation}
and (2) the Peclet number
(the ratio of convection over diffusion in the CHE),
\begin{equation}
\label{eq:pe-def}
Pe = \frac{U_{c} L_{c}^{2}}{M \sigma}  .
\end{equation}
There exist a few previous studies that investigated or discussed the issue
on how to select $Cn$ and $Pe$ to get reliable results
for various problems~\citep{jacqmin99jcp, jfm10-sil-che-cl}. 
In this work, some investigations about the Cahn number will also be carried out
while the effects of the Peclet number will not be a major focus 
(only some suitable value will be used).

The simulations are performed in the range
$0 \leq t \leq t_{e}$,
where $t_{e}$ 
denotes the time at the end of the simulation.
Suppose the characteristic length $L_{c}$ 
is discretized by $N_{L}$ uniform segments
and the characteristic time $T_{c}$ 
is discretized by $N_{t}$ uniform segments,
then one has,
\begin{equation}
  \delta_{x} = \frac{L_{c}}{N_{L}} 
\bigg( = \frac{L_{x}}{N_{x}} = \frac{L_{y}}{N_{y}} = h\bigg), 
\quad \delta_{t} = \frac{T_{c}}{N_{t}}  .
\end{equation}

\subsection{Validation}

As mentioned in Section \ref{sec:intro},
the present numerical method is a simplified version (from axisymmetric to 2-D geometry) 
of that given in~\citep{hybrid-mrt-lb-fd-axisym}.
The hybrid method has been validated through the study of several drop problems in that work.
Here three more validation tests are performed to check the major extensions
in the present work, including 
(1) the extension to handle binary fluids with different viscosities;
(2) the capability to simulate drops on substrates with WG;
(3) the CAH model.

\subsubsection{Layered Poiseuille Flow}

\begin{figure}[htp]
  \centering
 \includegraphics[trim = 10mm 10mm 70mm 10mm, clip, width=6cm, scale = 1.0]{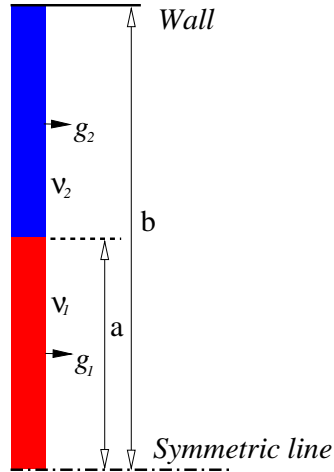}
  \caption{Problem setup for the layered Poiseuille flow inside a channel.}
  \label{fig:psle-2d}
\end{figure}

In this test, the layered two-phase flow inside an infinitely long horizontal channel is considered.
The channel height is $H= 2 b$ and the $x-$axis is located at the center of the channel.
The middle part ($-a \leq y < a$ where $0 < a < b$) 
is filled with one of the fluids (denoted as fluid 1)
and the remaining regions ($-b \leq y < -a$ and $a \leq y \leq b$)
are filled with the other fluid (denoted as fluid 2).
Due to the symmetry about the $x-$axis, only the upper half ($0 \leq y \leq b$) 
is considered.
The flow is driven by constant body forces in the horizontal direction
with different magnitudes $g_{1}$ and $g_{2}$ acting on the inner and outer fluids, respectively.
The two fluids have the same density and their kinematic viscosities are $\nu_{1}$ and $\nu_{2}$.
This problem is essentially 1-D with variations only in the vertical direction.
In simulation, along the horizontal direction (with no variations) only four grid points were used 
and periodic boundary conditions were applied.
The upper side is a stationary solid wall and 
the lower side is a symmetric line. 
Figure \ref{fig:psle-2d} illustrates the setup of the problem.
Initially, the velocities were zero everywhere.
Under the action of the body forces, a steady velocity profile is gradually developed.
Upon reaching steady state, the velocity profile $u (y)$ may be found by analytical means
~\citep{pof09-lbm-porousmedia},
\begin{equation}
	\label{eq:uy-psle-2d}
	u(y) = \left \{  
		\begin{matrix}
			A_{1}  y^{2} + C_{1}  &&  0 \leq y < a\\ 
			A_{2} y^{2} + B_{2} y + C_{2}  && a \leq y \leq b 
		\end{matrix}
	\right .  ,
\end{equation}
where the coefficients are given by,
\begin{equation}
	\begin{split}
	& A_{1} = - \frac{g_{1}}{2 \nu_{1}}, \quad A_{2} = - \frac{g_{2}}{2 \nu_{2}},  \quad
	 B_{2} = \bigg(- 2 A_{2} + 2 \frac{\nu_{1}}{\nu_{2}} A_{1} \bigg) a, \\
& C_{1} = (A_{2} - A_{1}) a^{2} - B_{2} (b - a) - A_{2} b^{2},
	\quad C_{2} = - A_{2} b^{2} - B_{2} b  .
	\end{split}
\end{equation}

\begin{figure}[htp]
  \centering
(a)  \includegraphics[scale = 0.6]{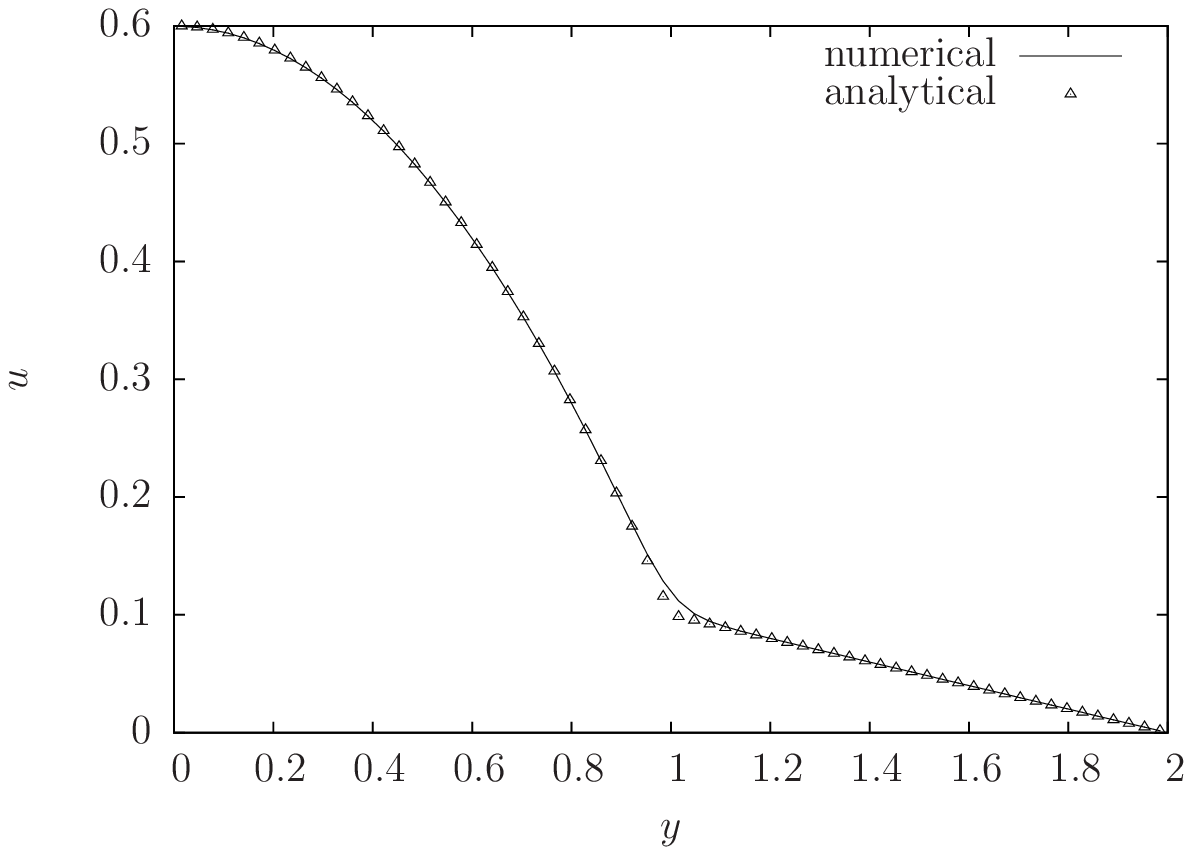}
(b)  \includegraphics[scale = 0.6]{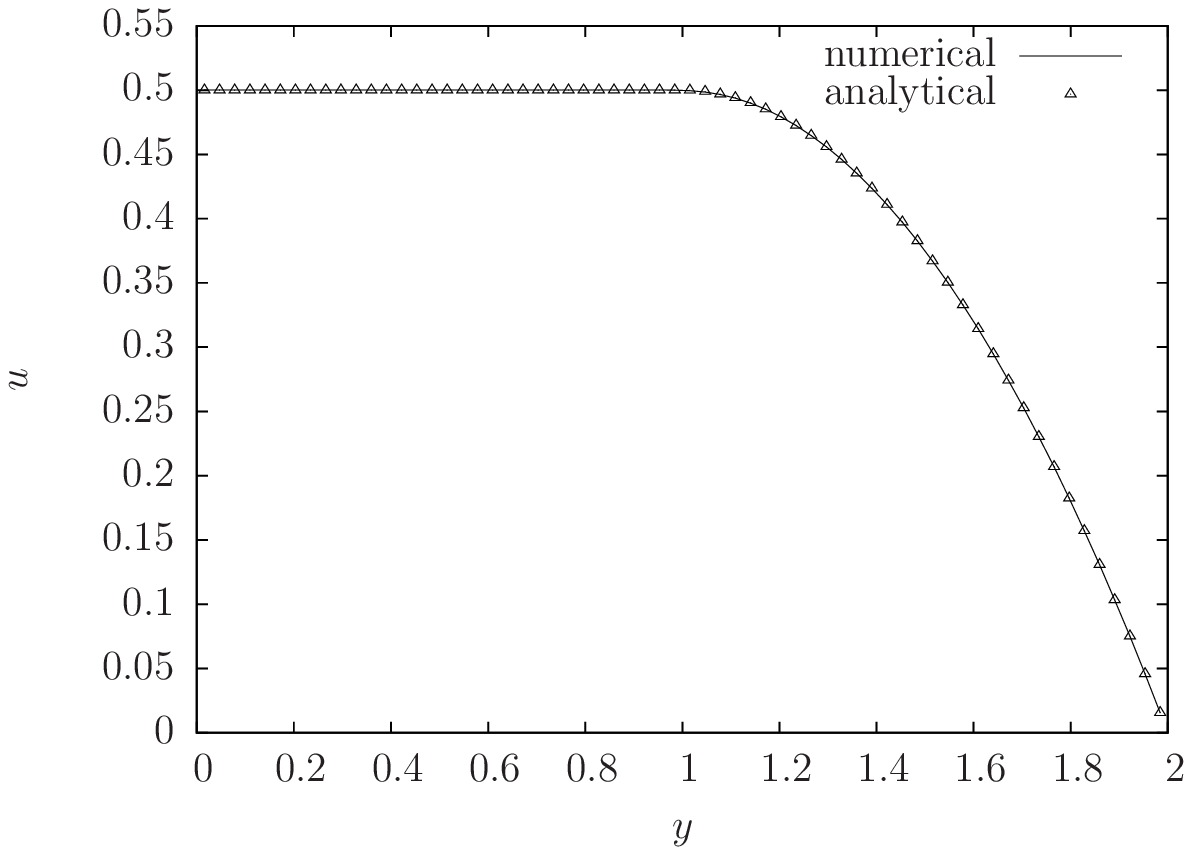}
(c)  \includegraphics[scale = 0.6]{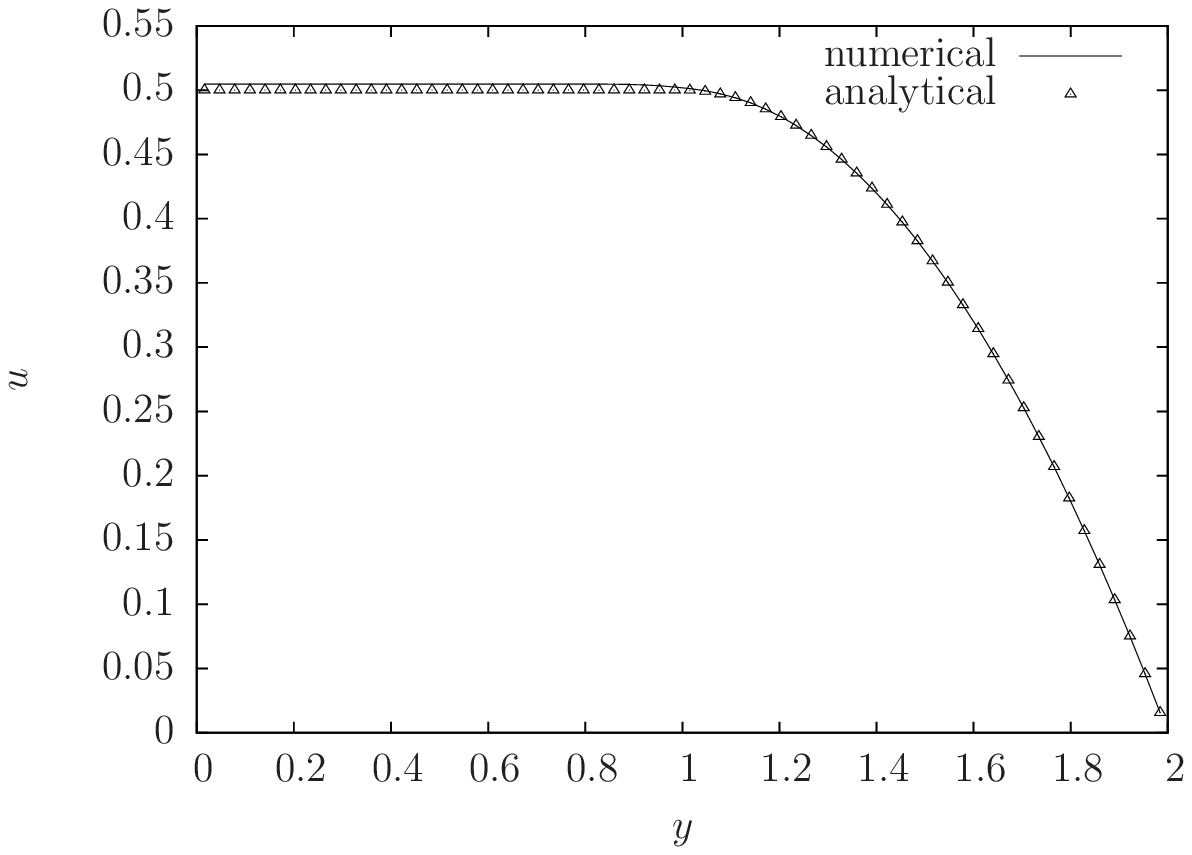}
(d)  \includegraphics[scale = 0.6]{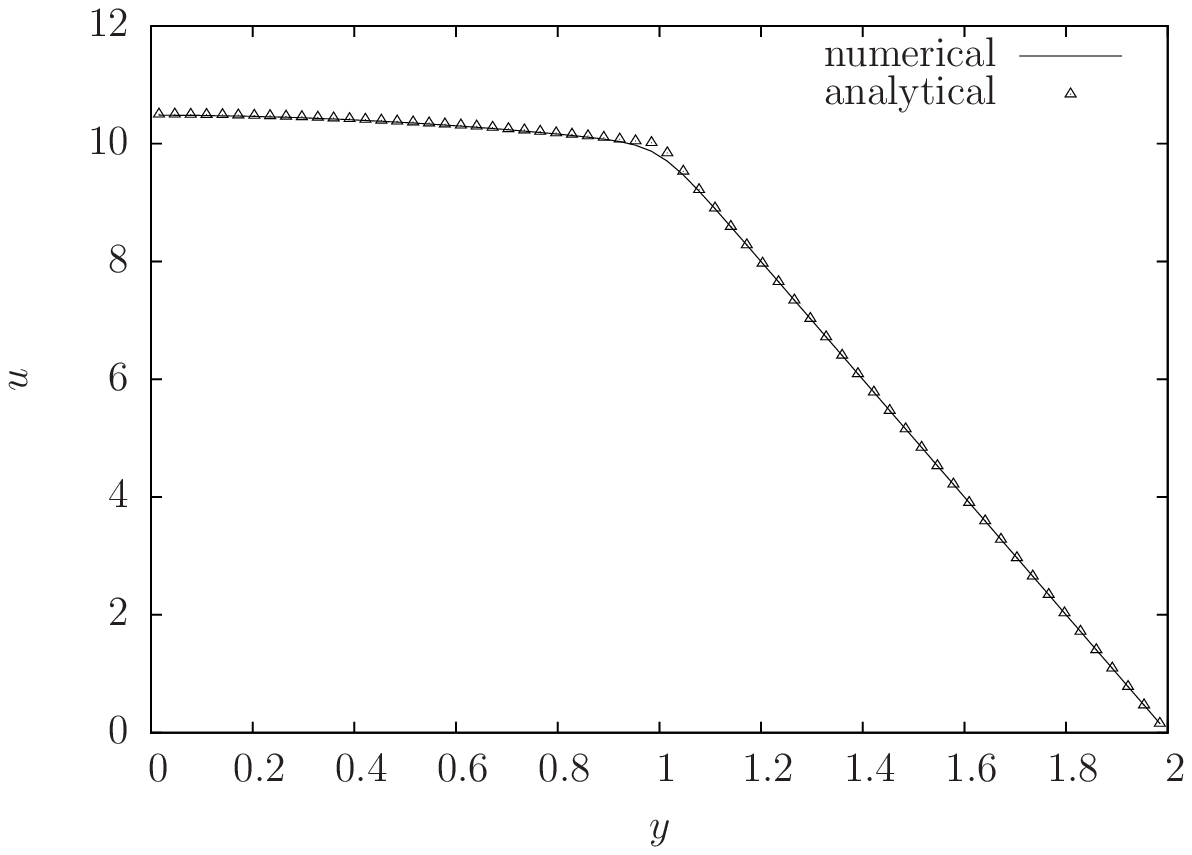}
  \caption{Comparison of the velocity profiles of layered Poiseuille flow
  with analytical solutions given in Eq. (\ref{eq:uy-psle-2d})
  under four different conditions:
  (a) $g_{1} = 1.46 \times 10^{-8}, \  g_{2} = 0, \   \nu_{1} / \nu_{2} = 0.1$; 
(b) $g_{1} = 0, \  g_{2} = 1.46 \times 10^{-8}, \   \nu_{1} / \nu_{2} = 10$; 
(c) $g_{1} = 0, \  g_{2} = 1.46 \times 10^{-6}, \   \nu_{1} / \nu_{2} = 0.1$; 
(d) $g_{1} = 1.46 \times 10^{-6}, \  g_{2} = 0, \   \nu_{1} / \nu_{2} = 10$. 
The parameters are $Cn = 0.125$, $Pe = 5 \times 10^{3}$,
$N_{L} = 32$, $N_{t} = 128$.}
  \label{fig:cmp-uy-psle-2d}
\end{figure}

The parameter $a$ is taken as $a = b / 2 = H / 4$ and is also chosen as the characteristic length, i.e., $L_{c} = a$.
Four cases with different force magnitudes and distributions at different viscosity ratios were studied:
(a) $g_{1} = 1.46 \times 10^{-8}, \  g_{2} = 0, \   \nu_{1} / \nu_{2} = 0.1$; 
(b) $g_{1} = 0, \  g_{2} = 1.46 \times 10^{-8}, \   \nu_{1} / \nu_{2} = 10$; 
(c) $g_{1} = 0, \  g_{2} = 1.46 \times 10^{-6}, \   \nu_{1} / \nu_{2} = 0.1$; 
(d) $g_{1} = 1.46 \times 10^{-6}, \  g_{2} = 0, \   \nu_{1} / \nu_{2} = 10$. 
The magnitudes of the body force are given in lattice units.
The Reynolds numbers as defined in Eq. (\ref{eq:re-def1})
are $1000$, $1000$, $100$ and $100$ for case (a), (b), (c) and (d), respectively
(note that we always applied the non-zero body force on fluid A, which may 
be located either on the inner side (fluid 1) or the outer sider (fluid 2)).
Figure \ref{fig:cmp-uy-psle-2d} compares the velocity profiles obtained from the present simulations
and those given by Eq. (\ref{eq:uy-psle-2d}) for the above four cases.
Note that the numerical solutions were obtained after the whole velocity field
became steady and the velocities in Fig. \ref{fig:cmp-uy-psle-2d} were scaled by the coefficient:
$\max (g_{1} / \nu_{1} , g_{2} / \nu_{2}) a^{2}$.
From Fig. \ref{fig:cmp-uy-psle-2d} it is observed that the numerical results agree quite well
with the theoretical solutions for all the cases.

\subsubsection{Liquid Column in a Channel with Given WG}\label{sssec:lc-wg}

\begin{figure}[htp]
  \centering
 \includegraphics[trim = 30mm 10mm 120mm 10mm, clip, width=6cm, scale = 1.0, angle = 90]{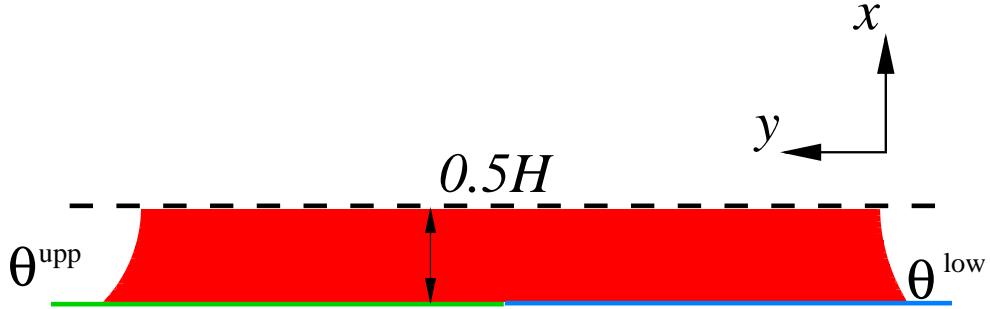}
  \caption{Problem setup for a liquid column inside a channel with a stepwise WG
  (the figure is rotated by $90^{\circ}$ in the anti-clockwise direction).}
  \label{fig:wg-rec-drop-setup}
\end{figure}

In the second test, a liquid column confined between two vertical flat plates 
located at $x=0$ and $x=H$ is considered.
The problem is symmetric about the middle vertical line $x = 0.5 H$, 
thus only the left half ($0 \leq x \leq 0.5 H$) is used in simulation.
The characteristic length is chosen to be $L_{c} = H$.
The problem setup is illustrated in Fig. \ref{fig:wg-rec-drop-setup}.
Initially, the liquid column has a (nominal) width of $W_{\textrm{lc}} = 4 H$ 
(the distance between the two three-phase points (TPPs) in the vertical direction)
and the $y-$coordinate of the middle point between the two TPPs is $y^{\textrm{mid}} = 3.5 H$,
giving the $y-$coordinates of the upper and lower TPPs: 
$y^{\textrm{low}} = 1.5 H$ and $y^{\textrm{upp}} = 5.5 H$.
In the region with $y> y^{\textrm{mid}}$ the wettability of the plate is 
specified by a contact angle $\theta^{\textrm{upp}}$,
and for $y \leq y^{\textrm{mid}}$ the CA is  $\theta^{\textrm{low}}$, 
which is kept to be larger than $\theta^{\textrm{upp}}$.
The initial upper and lower interface shapes were specified 
according to $\theta^{\textrm{upp}}$ and $\theta^{\textrm{low}}$.
Both the upper and lower parts of the plate are assumed to be smooth (i.e., having no CAH).
Because of the difference in the CAs, 
the liquid column is driven by the interfacial tension forces to move upwards
(i.e., towards the more hydrophilic part).

To make sure that the liquid column is always under the action of the WG,
$y^{\textrm{mid}} = (y^{\textrm{low}} + y^{\textrm{upp}}) / 2$ is updated at each step
and the wettability distribution is updated based on $y^{\textrm{mid}}$ to maintain the WG.
After some time, the liquid column gradually reaches a steady state,
which indicates a balance between the (driving) interfacial tension forces and the hydrodynamic resistances.
It is noted that a similar problem was investigated by~\cite{jsm12-wg-drop-lbm-sim},
who provided an approximate analytical solution
for the development of the centroid velocity of the liquid column $v_{\textrm{lc}}$ as,
\begin{equation}
\label{eq:vlc-theo}
v_{\textrm{lc}} = \frac{\sigma H [2 (\cos \theta^{\textrm{upp}} - \cos \theta^{\textrm{low}})] 
}{12 \rho_{c} [\nu_{A} W_{\textrm{lc}} + \nu_{B} (L_{y} - W_{\textrm{lc}}) ]}  
( 1 - e^{-t/t_{s}}),
\end{equation}
where $t_{s} = H^{2} L_{y} / [12 (\nu_{A} W_{\textrm{lc}} + \nu_{B} (L_{y} - W_{\textrm{lc}}) )]$.
Corresponding to the above settings, 
boundary conditions for a stationary wall are applied on the left ($x = 0$)
and on the right ($x = 0.5 H$) symmetric boundary conditions are used.
Periodic boundaries are assumed on the upper and lower sides of the simulation domain.

\begin{figure}[htp]
  \centering
 \includegraphics[scale = 1.0]{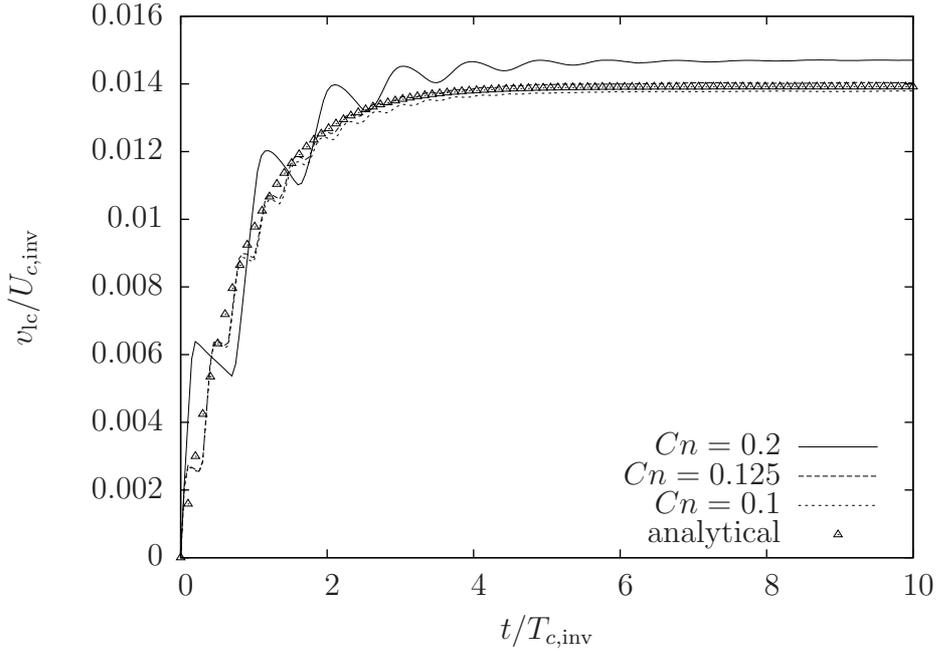}
  \caption{Comparison of the evolutions of the centroid velocity of the liquid column $v_{\textrm{lc}}$
  driven by a stepwise WG obtained at three Cahn numbers ($Cn=0.2$, $0.125$ and $0.1$)
  with that predicted by Eq. (\ref{eq:vlc-theo}).
The other common parameters are $L_{x} = 0.5$, $L_{y}= 20$, $Re = 100$, $r_{\nu} = 1$,
$\theta^{\textrm{upp}} = 47^{\circ}$, $\theta^{\textrm{low}} = 59^{\circ}$,
$Pe = 5 \times 10^{3}$.}
  \label{fig:cmp-vlc-wg-rec}
\end{figure}

For this problem, three Cahn numbers were tried, including $Cn=0.2$, $0.125$ and $0.1$.
The discretization parameter $N_{L}$ takes $20$, $32$ and $40$ for these Cahn numbers respectively,
so that the interface width (measured in the grid size $h$) is always $4.0$.
The common parameters are $L_{x} = 0.5$, $L_{y}= 20$, $Re = 100$, $r_{\nu} = 1$,
$\theta^{\textrm{upp}} = 47^{\circ}$, $\theta^{\textrm{low}} = 59^{\circ}$,
$Pe = 5 \times 10^{3}$.
Figure \ref{fig:cmp-vlc-wg-rec} shows 
the evolutions of the centroid velocity of the liquid column $v_{\textrm{lc}}$  
obtained at the above three Cahn numbers for $0 \leq t \leq 10 T_{c, \textrm{inv}}$.
Note that for this problem $U_{c, \textrm{inv}}$ and $T_{c, \textrm{inv}}$
were derived as in Eqs. (\ref{eq:Uc-inv}) and (\ref{eq:Tc-inv}) 
but with the drop radius $R$ replaced by the channel height $H$. 
In Fig. \ref{fig:cmp-vlc-wg-rec}
the velocity and time are scaled by $U_{c, \textrm{inv}}$ and $T_{c, \textrm{inv}}$ respectively.
It is seen from Fig. \ref{fig:cmp-vlc-wg-rec} that at $Cn=0.2$
the velocity shows relatively large fluctuations initially
and then gradually approaches a constant value,
which is slightly larger than the steady velocity
predicted by Eq. (\ref{eq:vlc-theo}).
When $Cn$ was reduced to $0.125$, the amplitude of the fluctuation became much reduced
and the velocity evolution obtained numerically became much closer to that by Eq. (\ref{eq:vlc-theo}).
To further reduce $Cn$ to $0.1$ only changed the results slightly.

\subsubsection{Drop subject to a body force}\label{sssec:drop-body-force}

In the third test, we consider a drop attached to a solid wall subject to a body force.
When there is CAH on the wall, 
the drop may stay attached to the wall even under the action of the body force.
This depends on the magnitude and direction of the body force,
as well as the magnitude of the hysteresis effect
(more specifically, on the advancing and receding angles, 
$\theta_{A}$ and $\theta_{R}$, on the wall).
In this problem, we assume that initially the drop is a semi-circle
on the left wall with the center (of the circle) being 
$(x_{c}, y_{c}) = (0, 1.5)$. 
This shape corresponds to an initial contact angle of $\theta^{i} = 90^{\circ}$.
The body force acts along the $y$-direction on the drop only and its density (per unit mass) is $g$.
The magnitude of the body force $g$ was varied by changing the Bond number.
The simulations were performed in a rectangular box with $L_{x} = 2$ and $L_{y} = 4$.
Stationary wall was assumed on all boundaries and the left wall has CAH
with $\theta_{A} = 105^{\circ}$ and $\theta_{R} = 75^{\circ}$. 
The common physical parameters are $Re = 16$, $r_{\nu} = 1$,
and the numerical parameters are $Cn=0.125$, $Pe = 5 \times 10^{3}$, 
$N_{L} = 32$, $N_{t} = 160$.
Seven $Bo$ numbers ($Bo = 2^{n+1} \times 10^{-3}$ with $n=1, 2, \cdots, 7$) were tried.
For this test we are mainly interested in the force balance 
when the drop is static.
For all the $Bo$ numbers considered, the drop finally reached a (nearly) static state.
Under the action of the body force the drop deformed slightly and its centroid
moved upwards a little bit, but the two three-phase-points (TPPs) were pinned 
due to the presence of CAH.

\begin{figure}[htp]
  \centering
 \includegraphics[scale = 0.5]{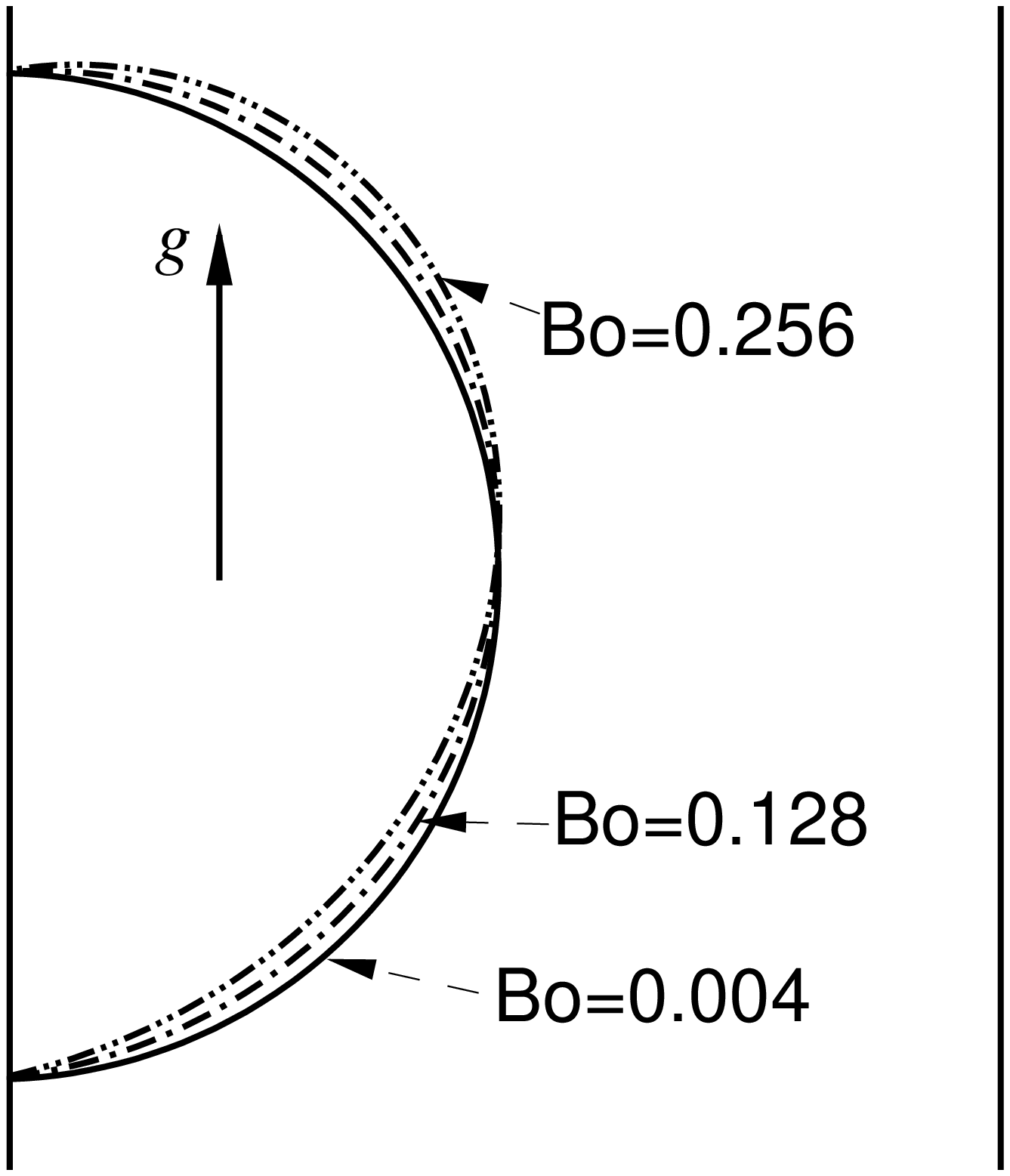}
  \caption{Drop shape (in equilibrium) on a wall with CAH 
  under the action of a body force at three different Bond numbers:
   $Bo = 0.004, \ 0.128, \ 0.256$.}
  \label{fig:cmp-drop-shape-bo}
\end{figure}

\begin{figure}[htp]
  \centering
(a) \includegraphics[scale = 0.6]{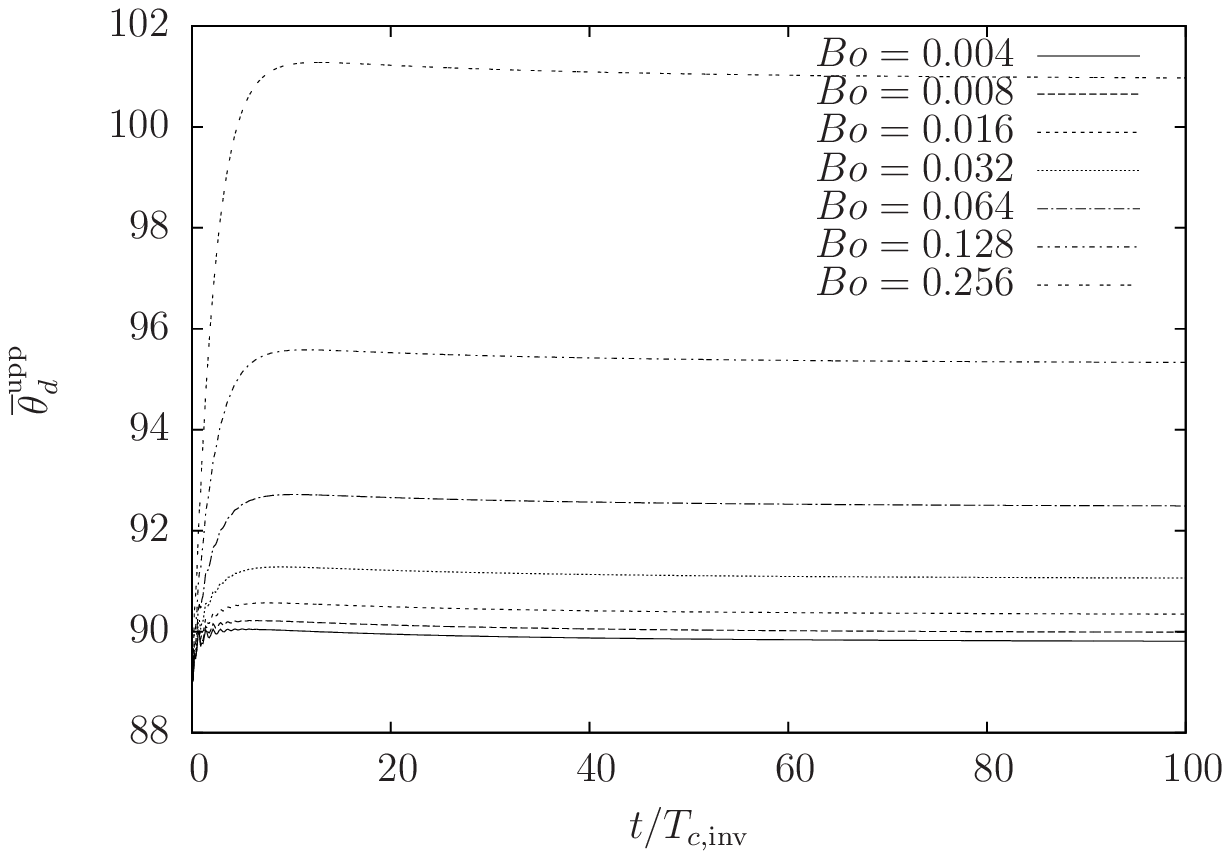}
(b) \includegraphics[scale = 0.6]{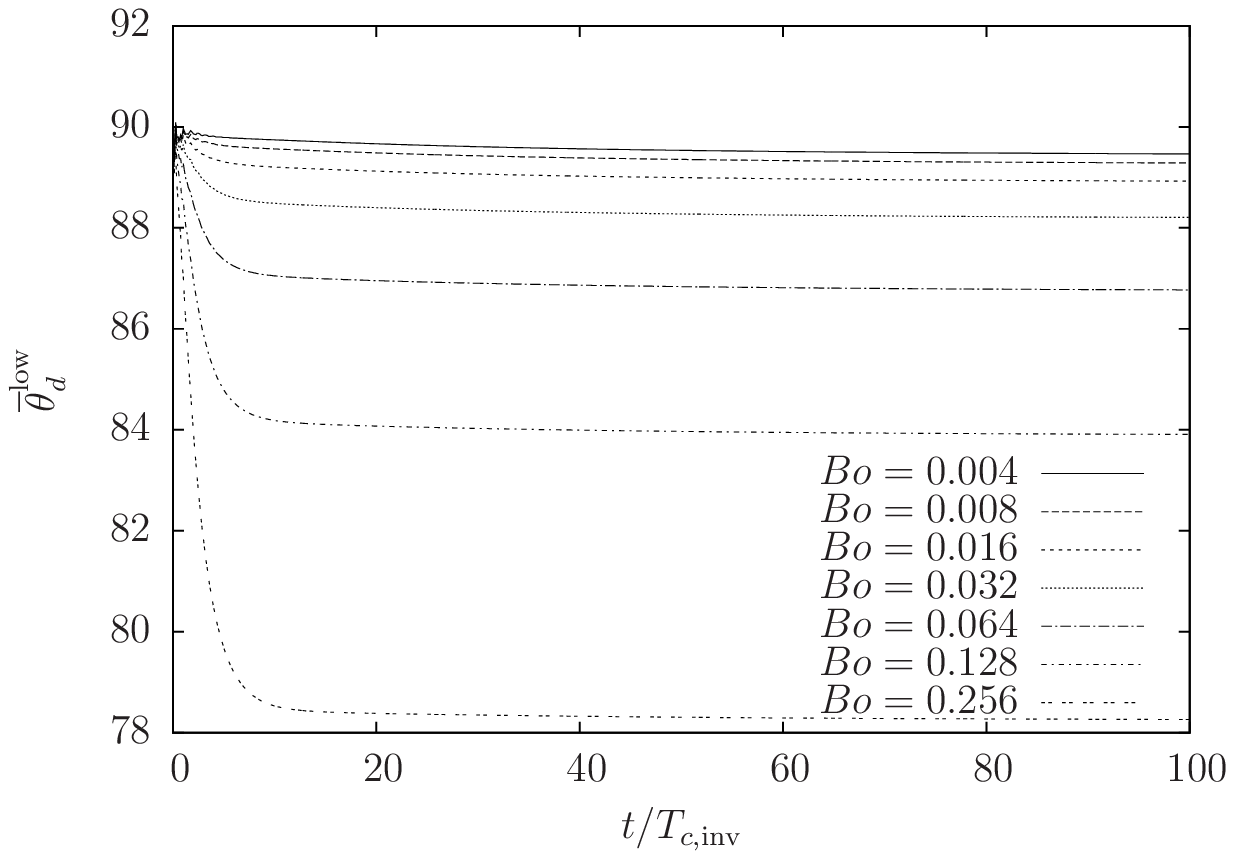}
  \caption{Evolutions of the (averaged) dynamic contact angles 
  at the upper and lower TPPs (on the wall) of the drop, 
  $\overline{\theta}^{\textrm{upp}}_{d}$ and $\overline{\theta}^{\textrm{low}}_{d}$,
  under the action of a body force at seven $Bo$ numbers:
  $Bo = 2^{n+1} \times 10^{-3}$ with $n=1, 2, \cdots, 7$.}
  \label{fig:cmp-ca-Re16-ca90-a105-r075-cn0d125-Pe5k}
\end{figure}

\begin{figure}[htp]
  \centering
 \includegraphics[scale = 1.0]{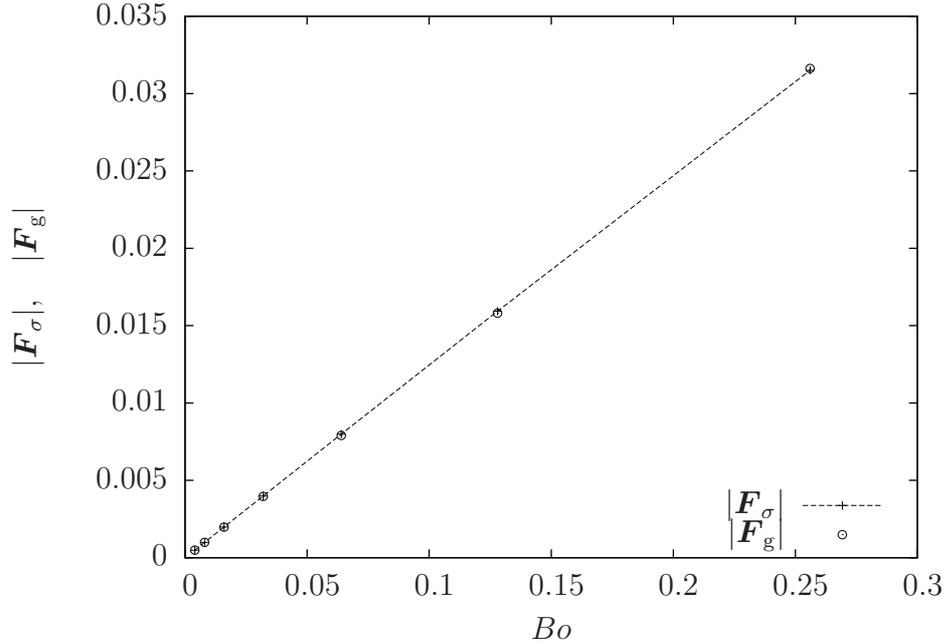}
  \caption{Variations of the magnitudes of 
  the net interfacial tension force $\vert \boldsymbol{F}_{\sigma} \vert$ (acting downwards)
  and the total body force $\vert \boldsymbol{F}_{g} \vert$ (acting upwards)
  on the drop (when in static equilibrium) with the Bond number.}
  \label{fig:bo-fst-fgrav-Re16-ca90-a105-r075-cn0d125-Pe5k}
\end{figure}

Figure \ref{fig:cmp-drop-shape-bo} shows the shapes of the drop
at three selected Bond numbers ($Bo = 0.004, \ 0.128, \ 0.256$)
after the interfacial tension force balanced the body force and the drop reached static equilibrium.
The increasing drop deformation with larger Bond number
is well captured, as seen in Fig. \ref{fig:cmp-drop-shape-bo}.
Figure \ref{fig:cmp-ca-Re16-ca90-a105-r075-cn0d125-Pe5k} 
compares the evolutions the local dynamic contact angles
at the upper and lower TPPs of the drop, 
$\overline{\theta}^{\textrm{upp}}_{d}$ and $\overline{\theta}^{\textrm{low}}_{d}$.
Note that the angles were averaged over the interfacial region spanning a few grid points.
Besides, the time is scaled by $T_{c, \textrm{inv}}$.
Although the contours of the order parameter in this region should ideally be parallel to each other,
we found that this could be slightly violated in the presence of CAH. 
Through such an average, the accuracy of the interfacial force calculation becomes improved.
It is observed from Fig. \ref{fig:cmp-ca-Re16-ca90-a105-r075-cn0d125-Pe5k}
that at the beginning $\overline{\theta}^{\textrm{upp}}_{d}$ increases with time
whereas $\overline{\theta}^{\textrm{low}}_{d}$ decreases as time evolves.
After the initial stage, the changes in both angles become quite small.
At the same time, it is seen that for all of the $Bo$ numbers 
$\overline{\theta}^{\textrm{upp}}_{d}$ remains to be smaller than the advancing angle
$\theta_{A} = 105^{\circ}$
and $\overline{\theta}^{\textrm{low}}_{d}$ is always larger than the receding angle
$\theta_{R} = 75^{\circ}$.
The magnitude of the net interfacial tension force (per unit length) 
acting on the drop may be calculated as
$\vert \boldsymbol{F}_{\sigma} \vert= 
\sigma (\cos \overline{\theta}^{\textrm{low}}_{d} - \cos \overline{\theta}^{\textrm{upp}}_{d})$
and this force pulls the drop downwards.
The total body force $\boldsymbol{F}_{g}$ on the drop may be calculated by a simple integration
over the area covered by the drop and it points upwards.
Figure \ref{fig:bo-fst-fgrav-Re16-ca90-a105-r075-cn0d125-Pe5k} 
shows the variations of the magnitudes of these two forces 
on the drop (when in static equilibrium) with the $Bo$ number.
From Fig. \ref{fig:bo-fst-fgrav-Re16-ca90-a105-r075-cn0d125-Pe5k}, 
it is easy to see that the two kinds of forces
has almost the same magnitudes for all the $Bo$ numbers tested.
This means that the balance condition for the drop was satisfied in the current simulations
under all the $Bo$ numbers.

\subsection{Drop Driven by WG}

\subsubsection{Problem setup}

\begin{figure}[htp]
  \centering
 \includegraphics[trim = 30mm 10mm 120mm 10mm, clip, width=6cm, scale = 1.0, angle = 90]{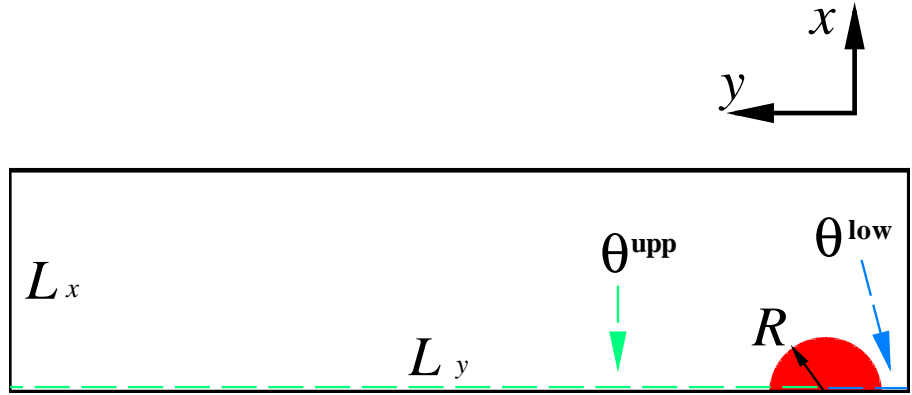}
  \caption{Problem setup for a (semi-circular) drop 
  inside a rectangular box with a stepwise WG.
  Note that the figure is rotated by $90^{\circ}$ in the anti-clockwise direction.
  The radius of the circle is $R$ and its initial origin is $(x_{c}, y_{c}) = (0, 1.5 R)$.
  The domain size is $L_{x} \times L_{y} = 4R \times 16R$.
  The contact angle of the wall in touch with the drop is
   $\theta^{\textrm{upp}}$ for $y > y_{c}$
  and $\theta^{\textrm{low}}$ for $y < y_{c}$.}
  \label{fig:wgdrop-setup}
\end{figure}

Now we study the main problem in this work, namely,
a drop on a wall 
subject to a stepwise WG.
Figure \ref{fig:wgdrop-setup} illustrates the overall setup.
There is no body force in this problem
(i.e., it is assumed that the Bond number is negligibly small).
This problem resembles that in Section \ref{sssec:lc-wg}
to some extent except that the domain is now a rectangular box
with solid walls on all boundaries
and the object under consideration is a drop in touch with one wall only.
In addition, the wall in touch with the drop may have hysteresis effects.
Therefore, in addition to the contact angles at the upper and lower parts,
$\theta^{\textrm{upp}}$ and $\theta^{\textrm{low}}$,
four additional parameters may come into play in this problem.
They are the advancing and receding angles of the upper and lower parts:
$\theta^{\textrm{upp}}_{A}$,  $\theta^{\textrm{upp}}_{R}$,
$\theta^{\textrm{low}}_{A}$,  $\theta^{\textrm{low}}_{R}$.
In fact, for walls with CAH, it should suffice to just give the advancing and receding angles, 
$\theta_{A}$ and $\theta_{R}$,
instead of the contact angle $\theta$, 
which may possibly take any value between $\theta_{A}$ and $\theta_{R}$.
However, here we still keep the usual contact angle
because it represents the limiting case when the CAH approaches zero
(i.e., $\theta_{A} \rightarrow \theta$ and $\theta_{R} \rightarrow \theta$).
In this way, the CAH effect may be better appreciated.
For convenience, when $\theta_{H} = \theta_{A} - \theta_{R}$ is not zero,
we assume that $\theta =  (\theta_{A} + \theta_{R}) / 2$.
Note that in Fig. \ref{fig:wgdrop-setup} the drop has a \textit{semi-circular} shape
(with the origin of the circle located on the $y-$axis, i.e., $x_{c} = 0$).
This initial shape corresponds to an initial contact angle $\theta^{\textrm{i}} = 90^{\circ}$,
and it is used below as the default setting. 

\subsubsection{Observables of interest}\label{sssec:obs-of-int}

In this problem, we are mainly interested in the following observables:
(1) the (instantaneous) average drop velocity (or the centroid velocity of the drop), 
$v_{\textrm{drop}} (t)$;
(2) the dynamic contact angles (DCAs) \textit{near} the upper and lower three-phase-points (TPPs),
$\theta^\textrm{upp}_{d, \textrm{nw}} (t)$ and $\theta^\textrm{low}_{d, \textrm{nw}} (t)$.
The centroid velocity of the drop $v_{\textrm{drop}} (t)$ was calculated by,
\begin{equation}
  \label{eq:dropVc}
  v_{\textrm{drop}} (t) = \frac{\int \int v (x, y, t) N(\phi) dx dy}{
    \int \int N(\phi) dx dy}   
  \approx \frac{\sum_{i,j} v_{i,j}(t) N(\phi_{i,j})}{
    \sum_{i,j} N(\phi_{i,j}) }  ,
\end{equation}
where $N(\phi)$ is defined by,
\begin{equation}
  \label{eq:Nphi}
   N(\phi) = \left \{ \begin{matrix} 1 && (\phi > 0)\\
   0 && (\phi \leq 0)
   \end{matrix} \right .  .
\end{equation}
The two DCAs,
$\theta^\textrm{upp}_{d, \textrm{nw}}$ and $\theta^\textrm{low}_{d, \textrm{nw}}$,
were measured at the interface (where $\phi = 0$)
along the layer next to the outermost one
(i.e., along the line which is $1.5 \delta_{x}$ away from the wall with WG).
In most cases studied here, the drop eventually reached a (nearly) steadily moving state.
We will focus especially on the steady state,
which in theory is reached only when $t_{e} \rightarrow \infty$ 
(if the drop is in an infinitely large domain).
In practice, it is determined by the following criterion for the drop velocity 
$v_{\textrm{drop}} (t)$ (with $t \geq T_{c,\textrm{inv}}$),
\begin{equation}
\label{eq:vdrop-steady}
\bigg \vert \frac{v_{\textrm{drop}} (t) - v_{\textrm{drop}} (t - T_{c,\textrm{inv}})}{v_{\textrm{drop}} (t)} \bigg \vert 
< 0.25\%   .
\end{equation}
That is, the relative change of $v_{\textrm{drop}}$ in one $T_{c,\textrm{inv}}$ is less than 
$0.25\%$.
The time to reach the steady state depends on various physical parameters:
in many cases below, $t_{e} = 50 T_{c, \textrm{inv}}$ guarantees that Eq. (\ref{eq:vdrop-steady})
can be satisfied; for some cases (e.g., at very low $Re$), even longer time 
(e.g., $t_{e} = 300 T_{c, \textrm{inv}}$) may be required.
It is noted that at large viscosity ratio the velocity showed some fluctuations
(which made it difficult to satisfy Eq. (\ref{eq:vdrop-steady}))
even though it seemed to have already entered a stable stage.
In those cases, the simulations were terminated at certain time (say, $t_{e} = 50 T_{c, \textrm{inv}}$)
when the fluctuations became reasonably small (e.g., the criterion is relaxed to $0.75\%$).
The steady drop velocity is denoted by $V_{\textrm{drop}}$
(equivalent to $V_{\textrm{mig}}$ defined by~\cite{pre12-wg-drop})
and it is no longer a function of time.

In Section \ref{ssec:char-quant-diml-num-setup}, 
we defined two Reynolds numbers, $Re$ and $Re_{\sigma}$,
based on two characteristic velocities $U_{c}$ and $U_{c, \textrm{inv}}$
(c.f. Eqs. (\ref{eq:re-def1}) and (\ref{eq:re-inv-def}))
rather than the \textit{actual} drop velocity.
To more realistically reflect the physics of the problem,
one may define yet another Reynolds number
based on $V_{\textrm{drop}}$ as,
\begin{equation}
\label{eq:re-drop-def}
Re_{\textrm{drop}} = \frac{V_{\textrm{drop}} R}{\nu_{A}} ,
\end{equation}
which is related to the other two as, 
\begin{equation}
Re_{\textrm{drop}} = \frac{V_{\textrm{drop}}}{U_{c}} Re = \frac{V_{\textrm{drop}}}{U_{c, \textrm{inv}}} Re_{\sigma}  .
\end{equation}
Similarly, another capillary number may be defined based on $V_{\textrm{drop}}$,
\begin{equation}
\label{eq:ca-drop-def}
Ca_{\textrm{drop}} = \frac{\rho_{c} \nu_{A} V_{\textrm{drop}}}{\sigma} ,
\end{equation}
which is related to the above two capillary numbers
and Reynolds numbers as,
\begin{equation}
\label{eq:ca-drop-relation}
Ca_{\textrm{drop}} = \frac{V_{\textrm{drop}}}{U_{c}} Ca = \frac{V_{\textrm{drop}}}{U_{c}}
= \frac{Re_{\textrm{drop}}}{Re}, \quad
Ca_{\textrm{drop}} = \frac{V_{\textrm{drop}}}{U_{c, \textrm{inv}}} Ca_{\sigma}  
= \frac{V_{\textrm{drop}}}{U_{c, \textrm{inv}}} \frac{1}{Re_{\sigma}}.
\end{equation}
For a real problem, the Reynolds number $Re$ (and $Re_{\sigma} = \sqrt{Re}$)
may be calculated once the drop dimension and the fluid properties are specified.
Since $Re_{\textrm{drop}} = Re \ Ca_{\textrm{drop}}$, 
we will mainly focus on $Ca_{\textrm{drop}}$.
In general, $Ca_{\textrm{drop}}$ may depend on the size of the domain to certain extent.
For simplicity, we concentrate on the situation in which the drop stays in a confined space
with the domain size being $L_{x} \times L_{y} =4R \times 16R$.
Then, one may write 
$Ca_{\textrm{drop}}$ 
as a function of all the remaining physical factors
that appear in this problem,
\begin{equation}
\label{eq:ca-drop-func}
Ca_{\textrm{drop}} = f (Re_{\sigma}, r_{\nu}, 
\theta^{\textrm{upp}}, \theta_{H}^{\textrm{upp}},
\theta^{\textrm{low}}, \theta_{H}^{\textrm{low}})  .
\end{equation}

\subsubsection{Parameter setting}

For numerical simulations, the results may depend on the spatial and temporal 
discretization parameters $N_{L}$ and $N_{t}$ as well
(i.e., convergence in space and time).
In addition, for phase-field-based simulations, the results depend to some extent on more factors
including the Cahn number $Cn$ and the Peclect number $Pe$ 
(i.e., convergence towards the sharp-interface limit)~\citep{jacqmin99jcp, jfm10-sil-che-cl}.
For conciseness, here we do not provide detailed investigations on all these factors;
instead, we just use some suitable values 
which we find in our tests provide reasonable results
without incurring too much computational costs.
Specifically, the spatial discretization parameter is fixed at $N_{L} = 32$
(i.e., the radius of the drop is discretized into $32$ uniform segments),
the Cahn number 
is fixed at $Cn=0.125$
which means that the interfacial width is about an eighth of the drop radius 
and spans about $N_{L} \times Cn= 4$ grid points,
and the Peclect number is fixed at $Pe = 5 \times 10^{3}$.
It is noted that the present definition of Cahn number differs from some others:
if one adopts the definition used by~\cite{jfm10-sil-che-cl} and also by~\cite{jcp07-dim-ldr}
where the interface width $\varepsilon$ is related to the present one as 
$\varepsilon = W / (2 \sqrt{2})$,
one would have an even smaller Cahn number of $0.044$.
In the computational domain, both the characteristic length $L_{c} = R$
and the characteristic time $T_{c}$ are fixed to be unity
(so is the characteristic velocity $U_{c}$).
When the spatial discretization parameter is fixed at $N_{L} = 32$,
in general it is not viable to use a fixed temporal discretization parameters $N_{t}$.
From Eqs. (\ref{eq:nu-tau-A-B}) and (\ref{eq:re-def1}), and also by noting that,
\begin{equation*}
c = \frac{\delta_{x}}{\delta_{t}} = \frac{L_{c}}{N_{L}} \frac{1}{T_{c} / N_{t}} = \frac{N_{t}}{N_{L}} U_{c}, 
\end{equation*}
one can derive the following relations
to determine the relaxation parameters $\tau_{f, A}$ and $\tau_{f, B}$,
\begin{equation}
\label{eq:tauf-re-vr}
\tau_{f, A} = 0.5 + \frac{3 N_{L}^{2}}{Re N_{t}}, \quad
\tau_{f, B} = 0.5 + \frac{3 N_{L}^{2}}{r_{\nu} Re N_{t}}.
\end{equation}
In LBM simulations, it is important to keep the relaxation parameter
in an appropriate range to guarantee the stability and accuracy.
Thus, we used different temporal discretization parameters $N_{t}$ for different cases
(depending on the Reynolds number and the viscosity ratio, as seen in Eq. (\ref{eq:tauf-re-vr}))
to make sure that $0.5 < \tau_{f, A} < 2.0$ and $0.5 < \tau_{f, B} < 2.0$.
The details about $N_{t}$ for different cases will be given later.

\subsubsection{Effects of the Reynolds number and viscosity ratio}

In this part, the effects of the Reynolds number $Re$ ($Re_{\sigma}$)
and the viscosity ratio $r_{\nu}$ are investigated
while the other factors in Eq. (\ref{eq:ca-drop-func}) are fixed at
$\theta^{\textrm{upp}} = 75^{\circ}$, $\theta^{\textrm{low}} = 105^{\circ}$,
$\theta_{H}^{\textrm{upp}} = \theta_{H}^{\textrm{low}} = 0$
(i.e., there is no CAH).

\begin{figure}[htp]
  \centering
 \includegraphics[scale = 1.0]{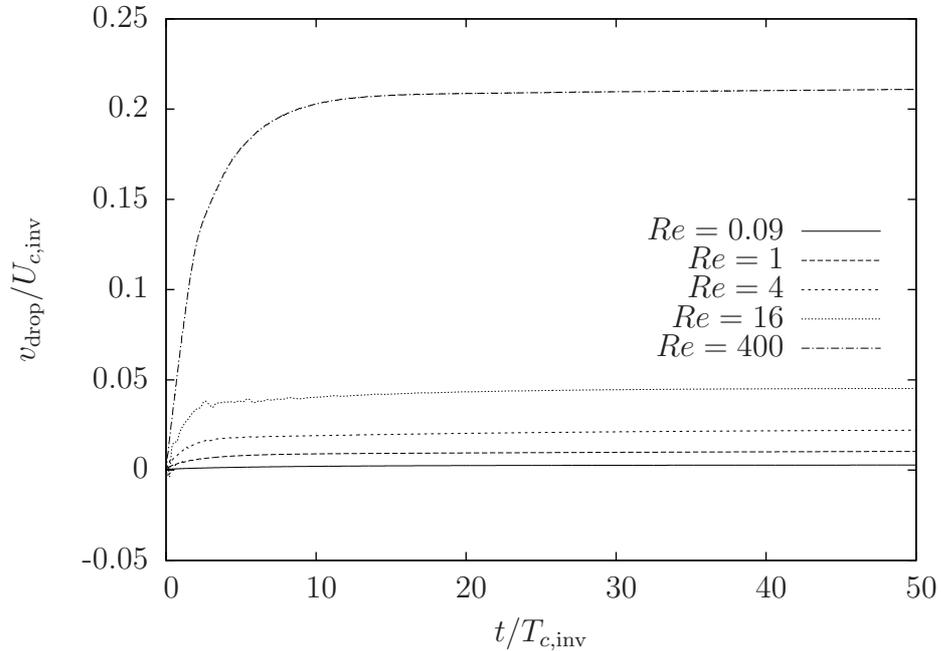}
  \caption{Evolutions of the centroid velocity of the drop $v_{\textrm{drop}}$
subject to a WG on a wall at different Reynolds numbers:
$Re = 0.09, \ 1, \ 4, \ 16$, and $400$.
The common parameters are $r_{\nu} = 1$, 
$\theta^{\textrm{upp}} = 75^{\circ}$, $\theta^{\textrm{low}} = 105^{\circ}$,
$\theta_{H}^{\textrm{upp}} = \theta_{H}^{\textrm{low}} = 0$,
$L_{x} = 4$, $L_{y} = 16$, 
$Cn=0.125$, $Pe = 5 \times 10^{3}$, 
$N_{L} = 32$.}
  \label{fig:cmp-vd-re-cainit90-cn0d125-Pe5k}
\end{figure}

\begin{figure}[htp]
  \centering
 \includegraphics[scale = 1.0]{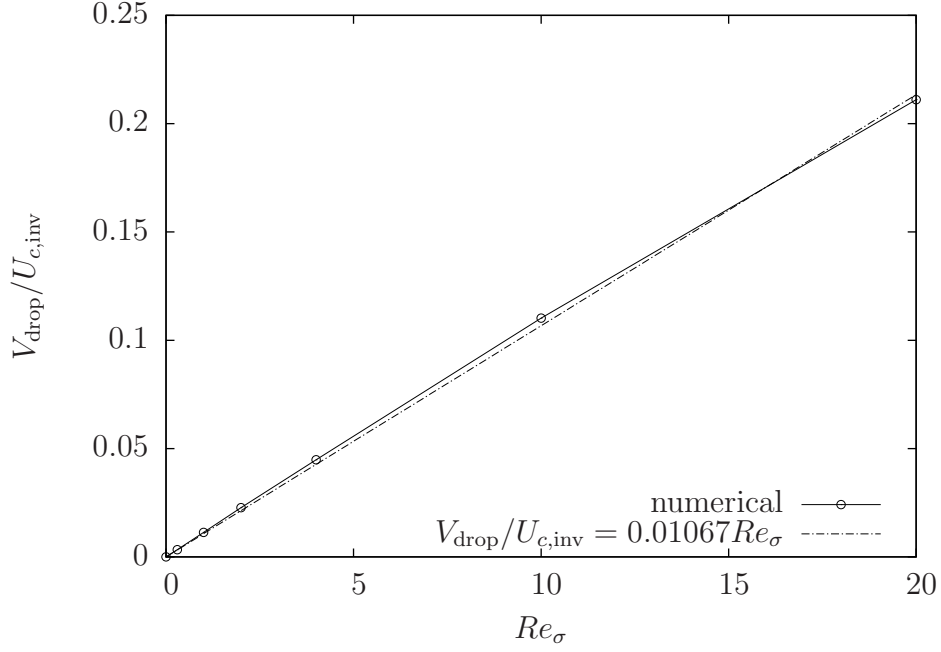}
  \caption{Variation of the centroid velocity of the drop in steady state $V_{\textrm{drop}}$
  subject to a stepwise WG with the Reynolds number $Re_{\sigma}$.
The common parameters are $r_{\nu} = 1$, 
$\theta^{\textrm{upp}} = 75^{\circ}$, $\theta^{\textrm{low}} = 105^{\circ}$,
$\theta_{H}^{\textrm{upp}} = \theta_{H}^{\textrm{low}} = 0$,
$L_{x} = 4$, $L_{y} = 16$, 
$Cn=0.125$, $Pe = 5 \times 10^{3}$, 
$N_{L} = 32$.}
  \label{fig:cmp-vdsteady-re-cainit90-cn0d125-Pe5k}
\end{figure}

First, we vary the Reynolds number while keeping the viscosity ratio at $r_{\nu} = 1$.
Six Reynolds numbers spanning a wide range were considered, including
$Re = 0.09, \ 1, \ 4, \ 16, \ 100$, and $400$.
The corresponding values of $Re_{\sigma} = \sqrt{Re}$ are
$0.3, \ 1, \ 2, \ 4, \ 10$, and $20$, respectively,
and the ratio of the maximum and minimum values of $Re_{\sigma}$ is about $67$.
The temporal discretization parameter $N_{t}$ was varied for different $Re$:
$N_{t} = 160$, $320$, $640$, $3200$ and $25600$
for $Re \geq 100$, $Re = 16$, $Re = 4$, $Re = 1$, and $Re = 0.09$, respectively.
The simulation time $t_{e}$ (measured in $T_{c, \textrm{inv}}$) was 
$50$, $100$, $150$ and $300$
for $Re \geq 100$, $Re = 16$,  $Re = 4$ and $Re \leq 1$, respectively.
Figure \ref{fig:cmp-vd-re-cainit90-cn0d125-Pe5k}
shows the evolutions the drop velocity $v_{\textrm{drop}}$
in $0 \leq t \leq 50 T_{c, \textrm{inv}}$
at five different Reynolds numbers:
$Re = 0.09, \ 1, \ 4, \ 16$, and $400$.
As found in Fig. \ref{fig:cmp-vd-re-cainit90-cn0d125-Pe5k},
the drop velocity (scaled by the characteristic velocity $U_{c, \textrm{inv}}$)
increases as the Reynolds number increases.
This can be understood because the viscosity decreases at larger $Re$,
resulting in smaller hydrodynamic resistance.
We would like to highlight that it is important to scale the velocity by
$U_{c, \textrm{inv}}$ because it provides a common base
for different Reynolds numbers;
otherwise, the comparisons are not meaningful.
Figure \ref{fig:cmp-vdsteady-re-cainit90-cn0d125-Pe5k}
shows the variation of the drop velocity in steady state $V_{\textrm{drop}}$
with the Reynolds number $Re_{\sigma}$, 
and it suggests a linear relation between these two quantities, i.e.,
\begin{equation}
\label{eq:vdsteady-resigma-lin}
\frac{V_{\textrm{drop}}}{U_{c, \textrm{inv}}} = \alpha_{c} Re_{\sigma}  ,
\end{equation}
where the proportional coefficient $\alpha_{c}$ is found to be about $0.01067$.

\begin{figure}[htp]
  \centering
(a) \includegraphics[scale = 0.6]{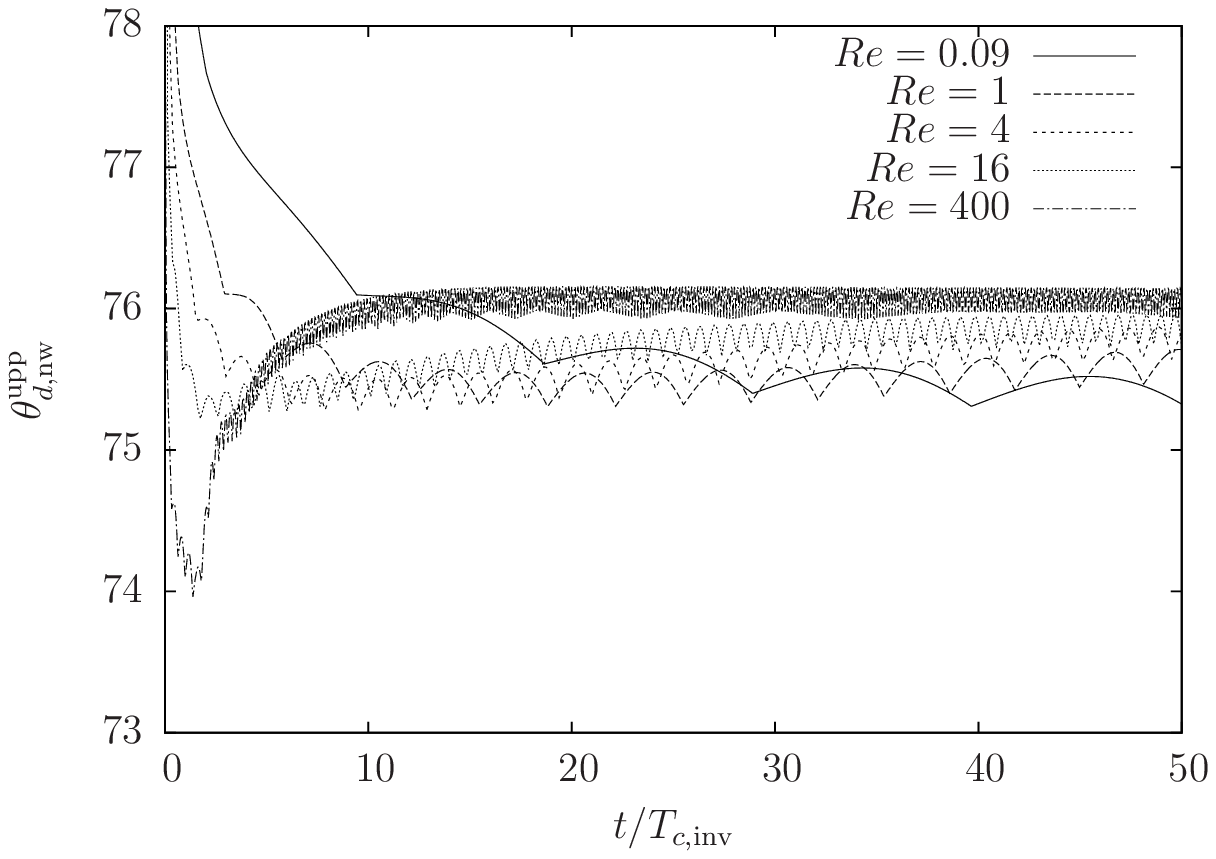}
(b) \includegraphics[scale = 0.6]{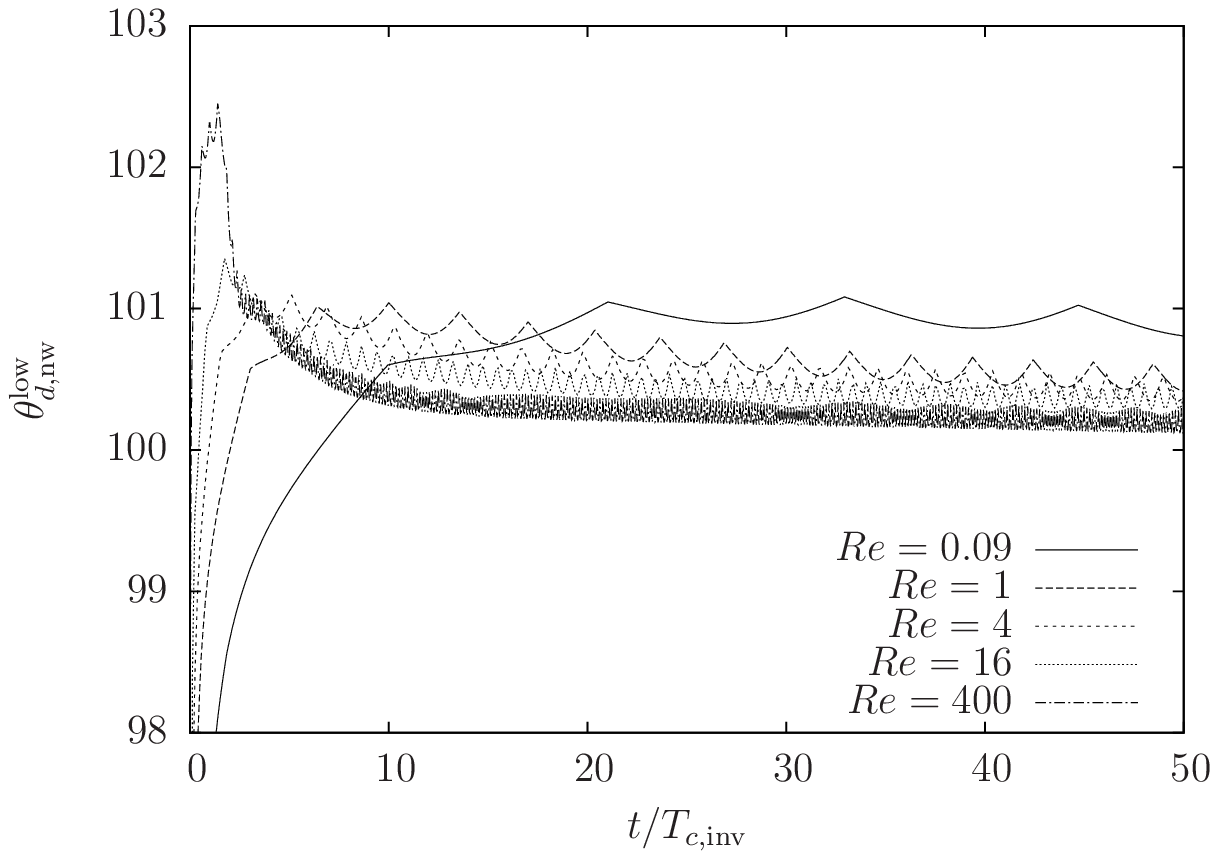}
  \caption{Evolutions of the dynamic contact angles
  near the upper and lower TPPs (measured at the next-to-outermost layer),
  $\theta^\textrm{upp}_{d, \textrm{nw}}$ and $\theta^\textrm{low}_{d, \textrm{nw}}$,
  of the drop subject to a stepwise WG at different Reynolds numbers:
  $Re = 0.09, \ 1, \ 4, \ 16$, and $400$.
The common parameters are $r_{\nu} = 1$, 
$\theta^{\textrm{upp}} = 75^{\circ}$, $\theta^{\textrm{low}} = 105^{\circ}$,
$\theta_{H}^{\textrm{upp}} = \theta_{H}^{\textrm{low}} = 0$,
$L_{x} = 4$, $L_{y} = 16$, 
$Cn=0.125$, $Pe = 5 \times 10^{3}$, 
$N_{L} = 32$, $N_{t} = 160$.}
  \label{fig:cmp-ca-upp-low-nw-re-cainit90-cn0d125-Pe5k}
\end{figure}

Figure \ref{fig:cmp-ca-upp-low-nw-re-cainit90-cn0d125-Pe5k}
compares the evolutions of the dynamic contact angles (DCAs)
near the upper and lower 
(Figs. \ref{fig:cmp-ca-upp-low-nw-re-cainit90-cn0d125-Pe5k}a and 
\ref{fig:cmp-ca-upp-low-nw-re-cainit90-cn0d125-Pe5k}b)
TPPs at these Reynolds numbers:
$Re = 0.09, \ 1, \ 4, \ 16$, and $400$.
An obvious difference is seen for both the upper and lower DCAs 
between those at high and low Reynolds numbers from
Fig. \ref{fig:cmp-ca-upp-low-nw-re-cainit90-cn0d125-Pe5k}.
At high $Re$, the DCA shows an overshoot initially
before it gradually reaches a (nearly) constant value.
As $Re$ decreases, the amplitude of the overshoot decreases,
and it even disappears when $Re$ is low enough (e.g., at $Re=0.09$).
This could be attributed to the inertial effects,
which become more significant at high $Re$.
Another observation in DCA is that it shows regular
periodic oscillations after the initial adjustment stage:
as $Re$ increases, the frequency of the oscillation increases;
for all the Reynolds numbers considered,
the amplitudes of the oscillation are small (less than $1^{\circ}$).

\begin{figure}[htp]
  \centering
 \includegraphics[scale = 1.0]{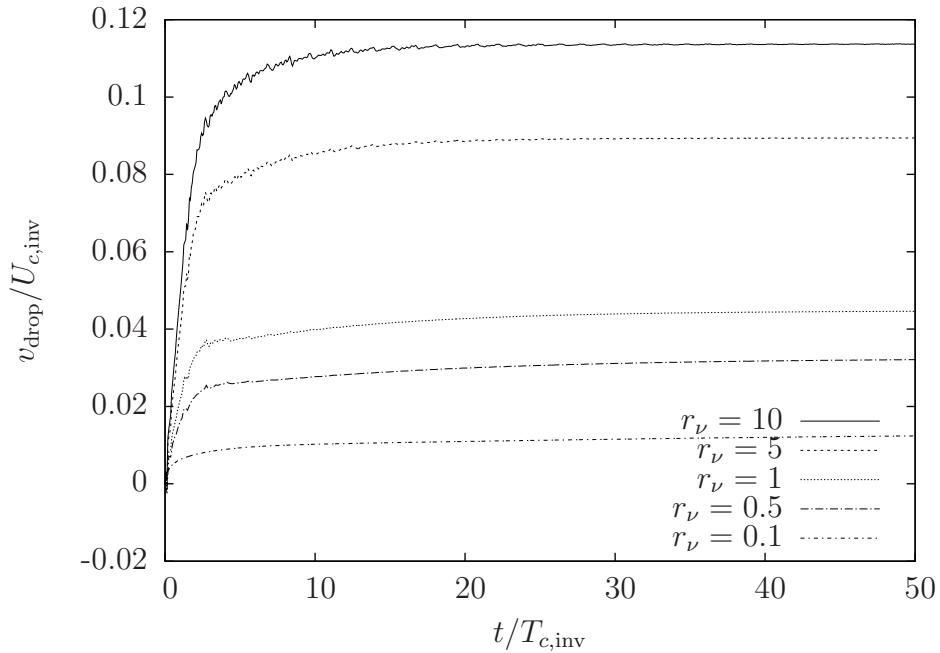}
  \caption{Evolutions of the centroid velocity of the drop $v_{\textrm{drop}}$
subject to a stepwise WG on a wall at different viscosity ratios
$r_{\nu} = 0.1, \ 0.5, \ 1, \ 5$, and $10$.
The common parameters are $Re = 16$, 
$\theta^{\textrm{upp}} = 75^{\circ}$, $\theta^{\textrm{low}} = 105^{\circ}$,
$\theta_{H}^{\textrm{upp}} = \theta_{H}^{\textrm{low}} = 0$,
$L_{x} = 4$, $L_{y} = 16$, 
$Cn=0.125$, $Pe = 5 \times 10^{3}$, 
$N_{L} = 32$.}
  \label{fig:cmp-vd-vr-Re16-cainit90-cn0d125-Pe5k}
\end{figure}

\begin{figure}[htp]
  \centering
 \includegraphics[scale = 1.0]{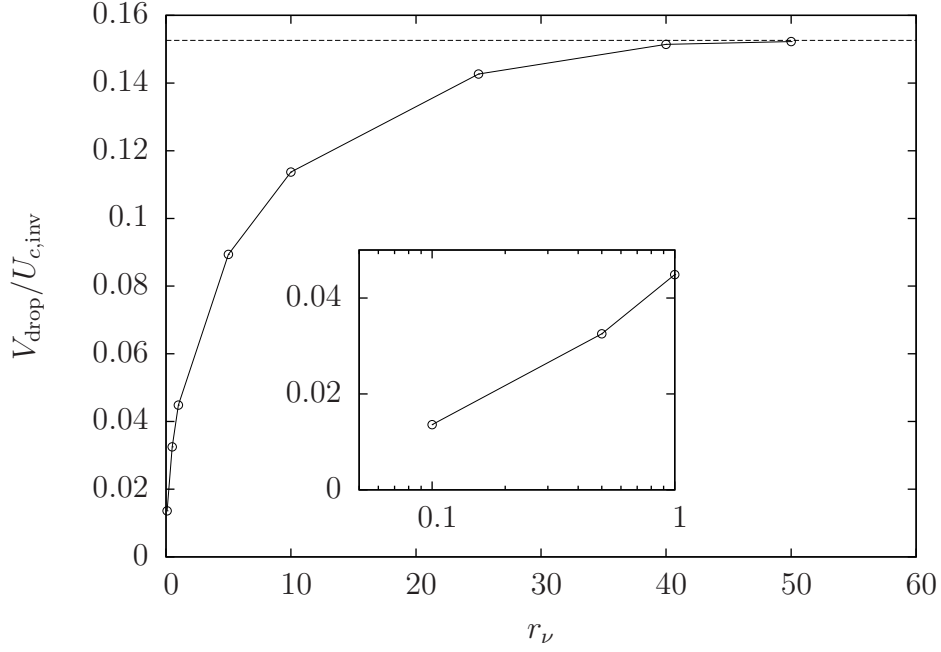}
  \caption{Variation of the centroid velocity of the drop in steady state $V_{\textrm{drop}}$
  subject to a stepwise WG with the viscosity ratio
$r_{\nu}$.
The common parameters are $Re = 16$, 
$\theta^{\textrm{upp}} = 75^{\circ}$, $\theta^{\textrm{low}} = 105^{\circ}$,
$\theta_{H}^{\textrm{upp}} = \theta_{H}^{\textrm{low}} = 0$,
$L_{x} = 4$, $L_{y} = 16$, 
$Cn=0.125$, $Pe = 5 \times 10^{3}$, 
$N_{L} = 32$.}
  \label{fig:cmp-vdsteady-vr-Re16-cainit90-cn0d125-Pe5k}
\end{figure}

Next, the effects of the viscosity ratio $r_{\nu}$ are studied
while the Reynolds number is fixed at $Re = 16$.
Several viscosity ratios, 
including $r_{\nu} = 0.1, \ 0.5, \ 1, \ 5, \ 10, \ 25, \ 40$, and $50$,  
were tested.
In order to keep both relaxation parameters 
$\tau_{f, A}$ and $\tau_{f, B}$ in a suitable range,
the temporal discretization parameter $N_{t}$ was varied for different $r_{\nu}$:
$N_{t} = 320$ for $r_{\nu} \geq 0.5$,
and $N_{t} = 1600$ for $r_{\nu} = 0.1$.
It is worth noting again that the viscosity ratio is defined as $r_{\nu} = \nu_{A} / \nu_{B}$
(i.e., the kinematic viscosity of the drop (fluid A) over that of the ambient fluid (fluid B)).
With a fixed Reynolds number $Re$,
a larger $r_{\nu}$ means a less viscous ambient fluid B.
Figure \ref{fig:cmp-vd-vr-Re16-cainit90-cn0d125-Pe5k}
shows the evolutions of the drop velocity $v_{\textrm{drop}}$
at five different viscosity ratios:
$r_{\nu} = 0.1, \ 0.5, \ 1, \ 5$, and $10$.
It is obvious that the viscosity of the ambient fluid has a significant effect
on the drop motion: to increase $r_{\nu}$ (in other words, to reduce $\nu_{B}$)
allows the drop to move faster.
Figure \ref{fig:cmp-vdsteady-vr-Re16-cainit90-cn0d125-Pe5k}
shows the variation of the steady drop velocity $V_{\textrm{drop}}$
(scaled by $U_{c, \textrm{inv}}$)
with the viscosity ratio $r_{\nu}$ based on the results obtained at the eight viscosity ratios tested, 
$r_{\nu} = 0.1, \ 0.5, \ 1, \ 5, \ 10, \ 25, \ 40$, and $50$.
From Fig. \ref{fig:cmp-vdsteady-vr-Re16-cainit90-cn0d125-Pe5k} we have the following observations:
when $r_{\nu} \leq 1$ the drop velocity $V_{\textrm{drop}}$ increases very fast
as $r_{\nu}$ becomes larger;
in contrast, when $r_{\nu}$ is much larger than unity (e.g., $r_{\nu} \geq 25$),
the rate of increase 
$\Delta V_{\textrm{drop}} / \Delta r_{\nu}$ 
gradually approaches zero
as $r_{\nu}$ increases, which indicates an upper limit for $V_{\textrm{drop}}$
(found to be approximately $0.1526 U_{c, \textrm{inv}}$ based on 
the exponential fit of the data at $r_{\nu} =40$ and $50$);
in between ($1 < r_{\nu} < 25$) is a transition region.

\begin{figure}[htp]
  \centering
(a) \includegraphics[scale = 0.6]{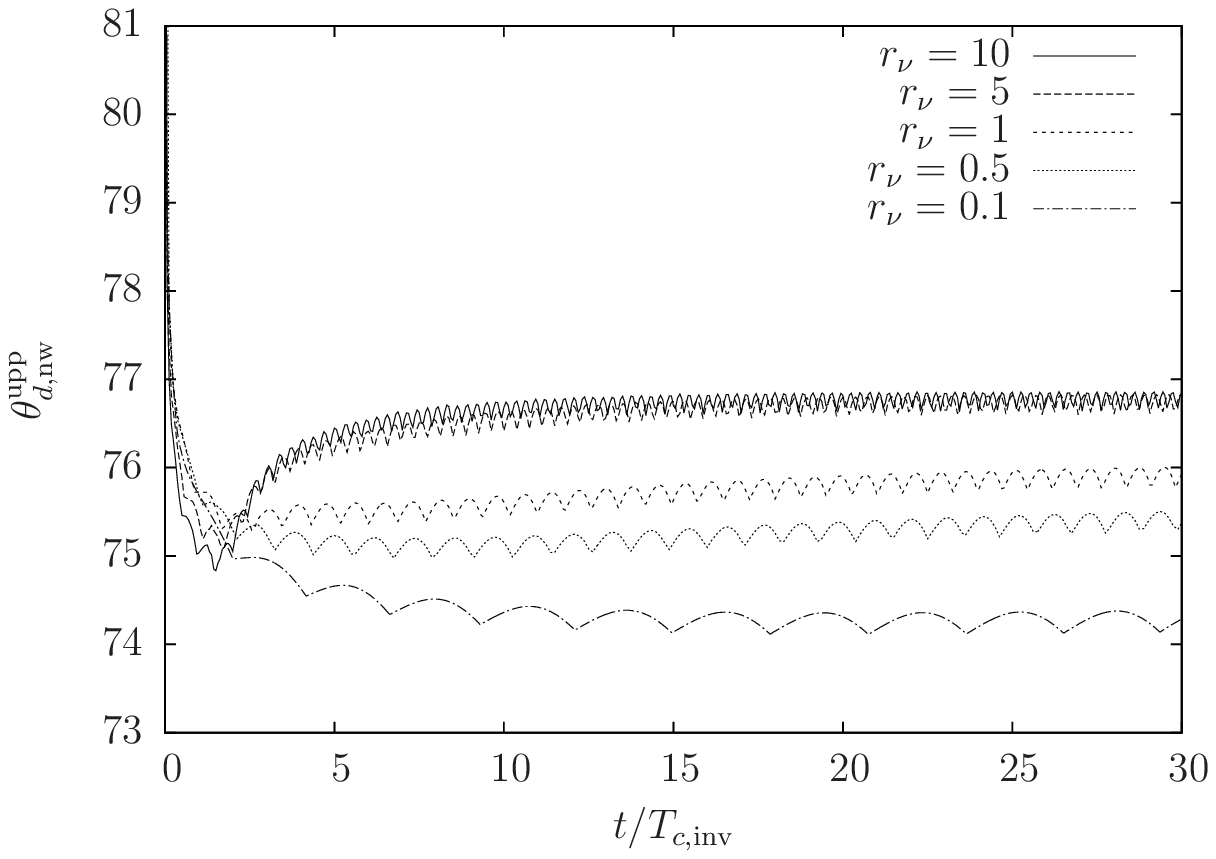}
(b) \includegraphics[scale = 0.6]{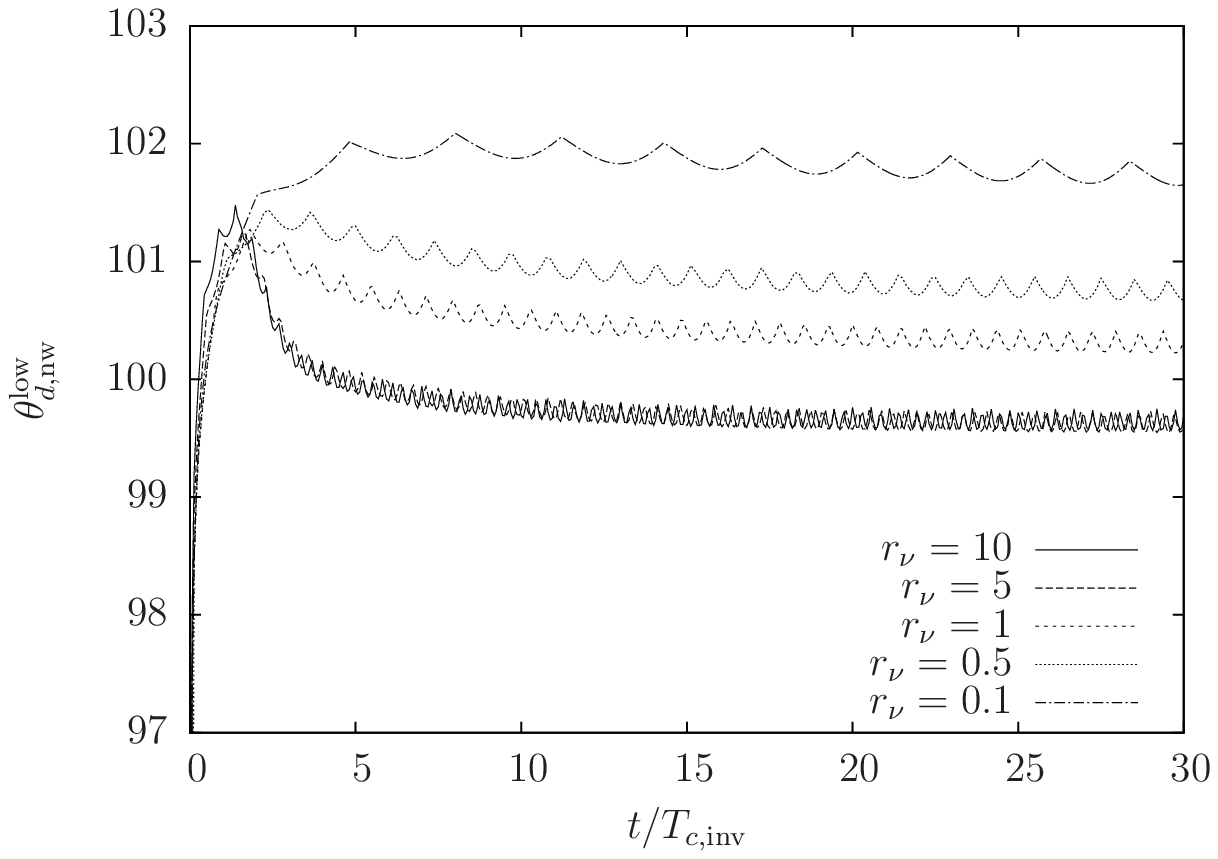}
  \caption{Evolutions of the dynamic contact angles
  near the upper and lower TPPs (measured at the next-to-outermost layer),
  $\theta^\textrm{upp}_{d, \textrm{nw}}$ and $\theta^\textrm{low}_{d, \textrm{nw}}$,
  of the drop subject to a stepwise WG at different viscosity ratios
$r_{\nu} = 0.1, \ 0.5, \ 1, \ 5$, and $10$.
The common parameters are $Re = 16$, 
$\theta^{\textrm{upp}} = 75^{\circ}$, $\theta^{\textrm{low}} = 105^{\circ}$,
$\theta_{H}^{\textrm{upp}} = \theta_{H}^{\textrm{low}} = 0$,
$L_{x} = 4$, $L_{y} = 16$, 
$Cn=0.125$, $Pe = 5 \times 10^{3}$, 
$N_{L} = 32$.}
  \label{fig:cmp-ca-upp-low-nw-vr-Re16-cainit90-cn0d125-Pe5k}
\end{figure}

As before, the DCAs are also examined.
Figure \ref{fig:cmp-ca-upp-low-nw-vr-Re16-cainit90-cn0d125-Pe5k}
compares the evolutions of the DCAs
near the upper and lower 
(Figs. \ref{fig:cmp-ca-upp-low-nw-vr-Re16-cainit90-cn0d125-Pe5k}a and 
\ref{fig:cmp-ca-upp-low-nw-vr-Re16-cainit90-cn0d125-Pe5k}b)
TPPs at the five viscosity ratios as in Fig. \ref{fig:cmp-vd-vr-Re16-cainit90-cn0d125-Pe5k}
($r_{\nu} = 0.1, \ 0.5, \ 1, \ 5$, and $10$).
A difference is observed for both the upper and lower DCAs 
between those at high and low viscosity ratios from
Fig. \ref{fig:cmp-ca-upp-low-nw-vr-Re16-cainit90-cn0d125-Pe5k}.
At high $r_{\nu}$ (i.e., with less viscous ambient fluid), 
the DCA shows an overshoot initially
before it gradually becomes (almost) constant.
As $r_{\nu}$ decreases (i.e., the ambient fluid becomes more viscous), 
the amplitude of the overshoot decreases,
and the overshoot disappears when the ambient fluid is sufficiently viscous
(e.g., at $r_{\nu} = 0.1$).
This is likely due to the high viscous damping at small $r_{\nu}$.
As above, the DCAs show regular
oscillations after the initial stage,
and a less viscous ambient fluid (corresponding to a larger $r_{\nu}$)
makes the oscillation frequency higher.

\begin{figure}[htp]
  \centering
 \includegraphics[scale = 1.0]{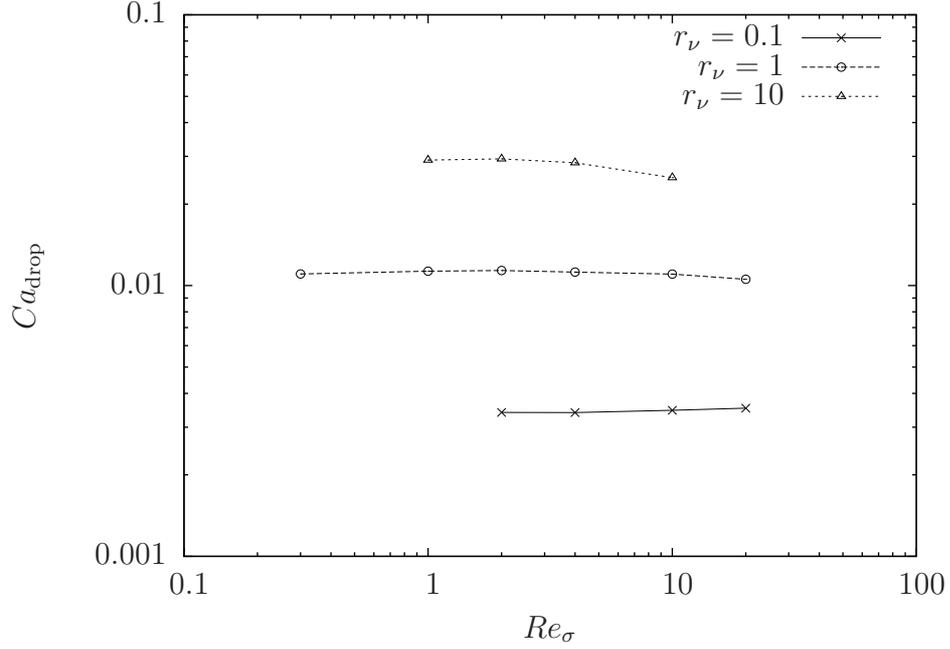}
  \caption{Variation of the capillary number of the drop 
  in steady state $Ca_{\textrm{drop}}$
  with the (input) Reynolds number $Re_{\sigma}$ 
  at three viscosity ratios, $r_{\nu} = 0.1, \ 1$, and $10$.
The common parameters are 
$\theta^{\textrm{upp}} = 75^{\circ}$, $\theta^{\textrm{low}} = 105^{\circ}$, 
$\theta_{H}^{\textrm{upp}} = \theta_{H}^{\textrm{low}} = 0$,
$L_{x} = 4$, $L_{y} = 16$, 
$Cn=0.125$, $Pe = 5 \times 10^{3}$, 
$N_{L} = 32$.}
  \label{fig:cmp-cadrop-resigma-cainit90-cn0d125-Pe5k}
\end{figure}

Now we study the effects of both the (input) Reynolds number
and the viscosity ratio by examining the two dimensionless numbers:
$Ca_{\textrm{drop}}$ and $Re_{\textrm{drop}}$.
Figure \ref{fig:cmp-cadrop-resigma-cainit90-cn0d125-Pe5k}
plots the variation of the capillary number of the drop in steady state 
$Ca_{\textrm{drop}}$ 
with the Reynolds number $Re_{\sigma}$
at three viscosity ratios, $r_{\nu} = 0.1, \ 1$, and $10$.
Note that $Ca_{\textrm{drop}}$ actually would correspond to the proportional coefficient $\alpha_{c}$, 
as seen from Eqs. (\ref{eq:ca-drop-relation}) and (\ref{eq:vdsteady-resigma-lin}).
As would be deduced from previous observations (c.f. Fig. \ref{fig:cmp-vdsteady-re-cainit90-cn0d125-Pe5k}),
it is found that the capillary number $Ca_{\textrm{drop}}$
remains \textit{almost constant} for each viscosity ratio.
This suggests that Eq. (\ref{eq:ca-drop-func}) may be reduced to,
\begin{equation}
\label{eq:ca-drop-func-reduced}
Ca_{\textrm{drop}} = f (r_{\nu}, 
\theta^{\textrm{upp}}, \theta_{H}^{\textrm{upp}},
\theta^{\textrm{low}}, \theta_{H}^{\textrm{low}})  .
\end{equation}
The present results indicate that the capillary number of the drop
is independent of the drop size,
which seems to differ from that by~\cite{l04-drop-grad-surf-ratchet},
where it was predicted that (c.f. Eq. (1) in that article),
\begin{equation}
\label{eq:ca-drop-ref-continuousWG}
Ca = \alpha R \frac{d (\cos \theta)}{dx}    ,
\end{equation}
with $x$ being the coordinate in the WG direction and $\alpha$ being a constant.
This is because~\cite{l04-drop-grad-surf-ratchet} 
considered a continuously varying WG with the contact angle $\theta$
having a distribution that satisfies $d (\cos \theta) / dx = const$
whereas the present work considers a stepwise WG that is independent of the drop radius $R$.
On the other hand, from Eq. (\ref{eq:ca-drop-ref-continuousWG})
one may deduce that $Ca$ is proportional to the change in $\cos \theta$
across the footprint of the drop.
For convenience, we denote this quantity as $\Delta \cos \theta$.
In the case of a stepwise WG as in the present work,
it may be expressed as 
$\Delta \cos \theta = \cos \theta^{\textrm{upp}} - \cos \theta^{\textrm{low}}$
and its effects will be studied in the next section.

\begin{figure}[htp]
  \centering
 \includegraphics[scale = 1.0]{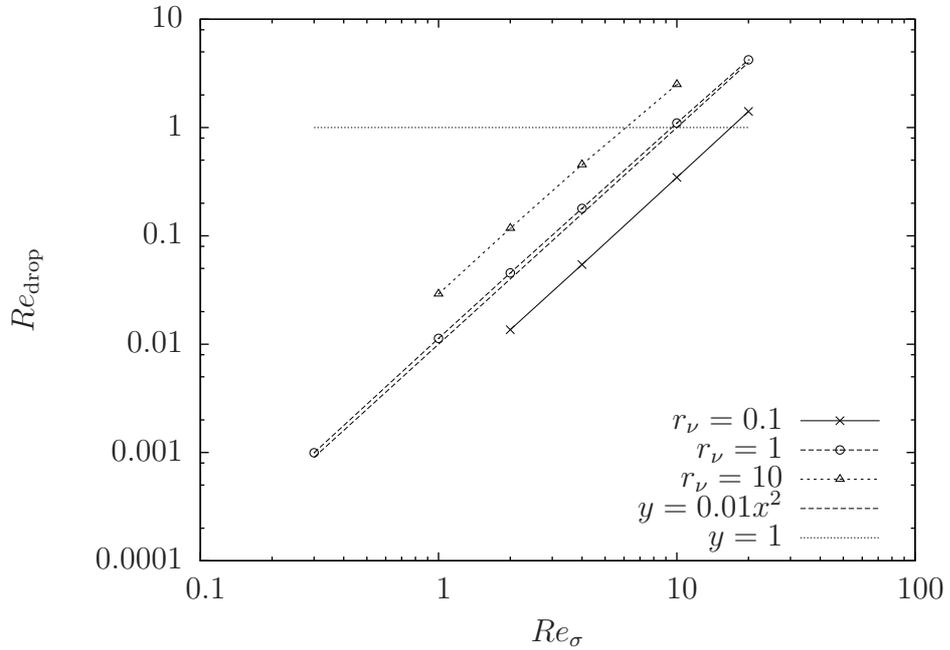}
  \caption{Variation of the Reynolds number of the drop 
  in steady state $Re_{\textrm{drop}}$
  with the (input) Reynolds number $Re_{\sigma}$
  at three viscosity ratios, $r_{\nu} = 0.1, \ 1$, and $10$.
The common parameters are 
$\theta^{\textrm{upp}} = 75^{\circ}$, $\theta^{\textrm{low}} = 105^{\circ}$, 
$\theta_{H}^{\textrm{upp}} = \theta_{H}^{\textrm{low}} = 0$,
$L_{x} = 4$, $L_{y} = 16$, 
$Cn=0.125$, $Pe = 5 \times 10^{3}$, 
$N_{L} = 32$.}
  \label{fig:cmp-redrop-resigma-cainit90-cn0d125-Pe5k}
\end{figure}

With Eq. (\ref{eq:ca-drop-func-reduced}) and the previous relations,
$Re_{\textrm{drop}} = Re Ca_{\textrm{drop}}$ and $Re_{\sigma} = \sqrt{Re}$, 
one can express $Re_{\textrm{drop}}$ as,
\begin{equation}
\label{eq:re-drop-func-reduced}
Re_{\textrm{drop}} = Re_{\sigma}^{2} f (r_{\nu}, 
\theta^{\textrm{upp}}, \theta_{H}^{\textrm{upp}},
\theta^{\textrm{low}}, \theta_{H}^{\textrm{low}})  ,
\end{equation}
from which one has,
\begin{equation}
\label{eq:re-drop-func-reduced-log}
log (Re_{\textrm{drop}}) = 2 log (Re_{\sigma}) + log [ f (r_{\nu}, 
\theta^{\textrm{upp}}, \theta_{H}^{\textrm{upp}},
\theta^{\textrm{low}}, \theta_{H}^{\textrm{low}}) ] .
\end{equation}
Figure \ref{fig:cmp-redrop-resigma-cainit90-cn0d125-Pe5k}
gives the variation of the Reynolds number of the drop in steady state 
$Re_{\textrm{drop}}$ 
with the (input) Reynolds number $Re_{\sigma}$ (with both axes in logarithmic scale)
at the above three viscosity ratios, $r_{\nu} = 0.1, \ 1$, and $10$.
Also shown in Fig. \ref{fig:cmp-redrop-resigma-cainit90-cn0d125-Pe5k}
is the function $y = 0.01 x^{2}$ for comparison.
It is found from Fig. \ref{fig:cmp-redrop-resigma-cainit90-cn0d125-Pe5k}
that Eq. (\ref{eq:re-drop-func-reduced-log}) well captures the variation
of $log(Re_{\textrm{drop}})$ with $log (Re_{\sigma})$
at each of the viscosity ratios.
Besides, it is found that for all the cases tested,
the Reynolds number $Re_{\textrm{drop}}$ is less than $10$;
in most cases, it is actually less than $1$ (i.e., below the horizontal line in 
Fig. \ref{fig:cmp-redrop-resigma-cainit90-cn0d125-Pe5k}),
thus the inertial effects are not quite significant.

\subsubsection{Effects of the magnitude of WG}

In Eq. (\ref{eq:ca-drop-func-reduced}), in addition to the viscosity ratio $r_{\nu}$
there are still four parameters that may affect the capillary number $Ca_{\textrm{drop}}$, namely,
the contact angles of the upper and lower parts, 
$\theta^{\textrm{upp}}$ and $\theta^{\textrm{low}}$, 
and the magnitudes of the CAH of the two parts, 
$\theta_{H}^{\textrm{upp}}$ and $\theta_{H}^{\textrm{low}}$.
If there is no CAH (i.e., $\theta_{H}^{\textrm{upp}} = \theta_{H}^{\textrm{low}} = 0$),
two remaining parameters, $\theta^{\textrm{upp}}$ and $\theta^{\textrm{low}}$, still play some role.
Their effects are studied in this section.
As mentioned above, the parameter 
$\Delta \cos \theta = \cos \theta^{\textrm{upp}} - \cos \theta^{\textrm{low}}$ 
is an important factor to determine the drop motion.
Thus, we will not only focus on the individual contact angle,
but also on this special parameter $\Delta \cos \theta$.
Two groups of the contact angle pair 
$(\theta^{\textrm{upp}}, \theta^{\textrm{low}})$,  each containing five pairs,
were investigated.
In both groups, the parameter 
$\Delta \cos \theta$ 
takes one of the following five values: 
$0.087, \ 0.259, \ 0.5, \ 0.707$, and $0.866$.
In one group, the lower contact angle $\theta^{\textrm{low}}$
was fixed to be $\theta^{\textrm{low}} = 90^{\circ}$
and the upper one $\theta^{\textrm{upp}}$ was varied: 
$\theta^{\textrm{upp}} = 85^{\circ}, \ 75^{\circ}, \ 60^{\circ}, \ 45^{\circ}$, and $30^{\circ}$.
In this group, the averages of the two contact angles,
$(\theta^{\textrm{upp}}+\theta^{\textrm{low}})/2$,
are less than $90^{\circ}$,
and it is called the hydrophilic group (GL).
In the other group, 
the upper contact angle $\theta^{\textrm{upp}}$ was fixed to be $\theta^{\textrm{upp}} = 90^{\circ}$,
and that on the lower part 
was varied:
$\theta^{\textrm{low}} = 95^{\circ}, \ 105^{\circ}, \ 120^{\circ}, \ 135^{\circ}$, and $150^{\circ}$.
In this group, the averages of $\theta^{\textrm{upp}}$ and $\theta^{\textrm{low}}$ 
are greater than $90^{\circ}$,
and it is called the hydrophobic group (GB).
The common parameters for all the cases in this part
are $Re = 16$, $r_{\nu} = 1$, 
$\theta_{H}^{\textrm{upp}}=\theta_{H}^{\textrm{low}} = 0$,
$L_{x} = 4$, $L_{y} = 16$, 
$Cn=0.125$, $Pe = 5 \times 10^{3}$, 
$N_{L} = 32$, $N_{t} = 320$.

\begin{figure}[htp]
  \centering
(a) \includegraphics[trim = 1mm 1mm 80mm 1mm, clip, scale = 0.35, angle = 90]{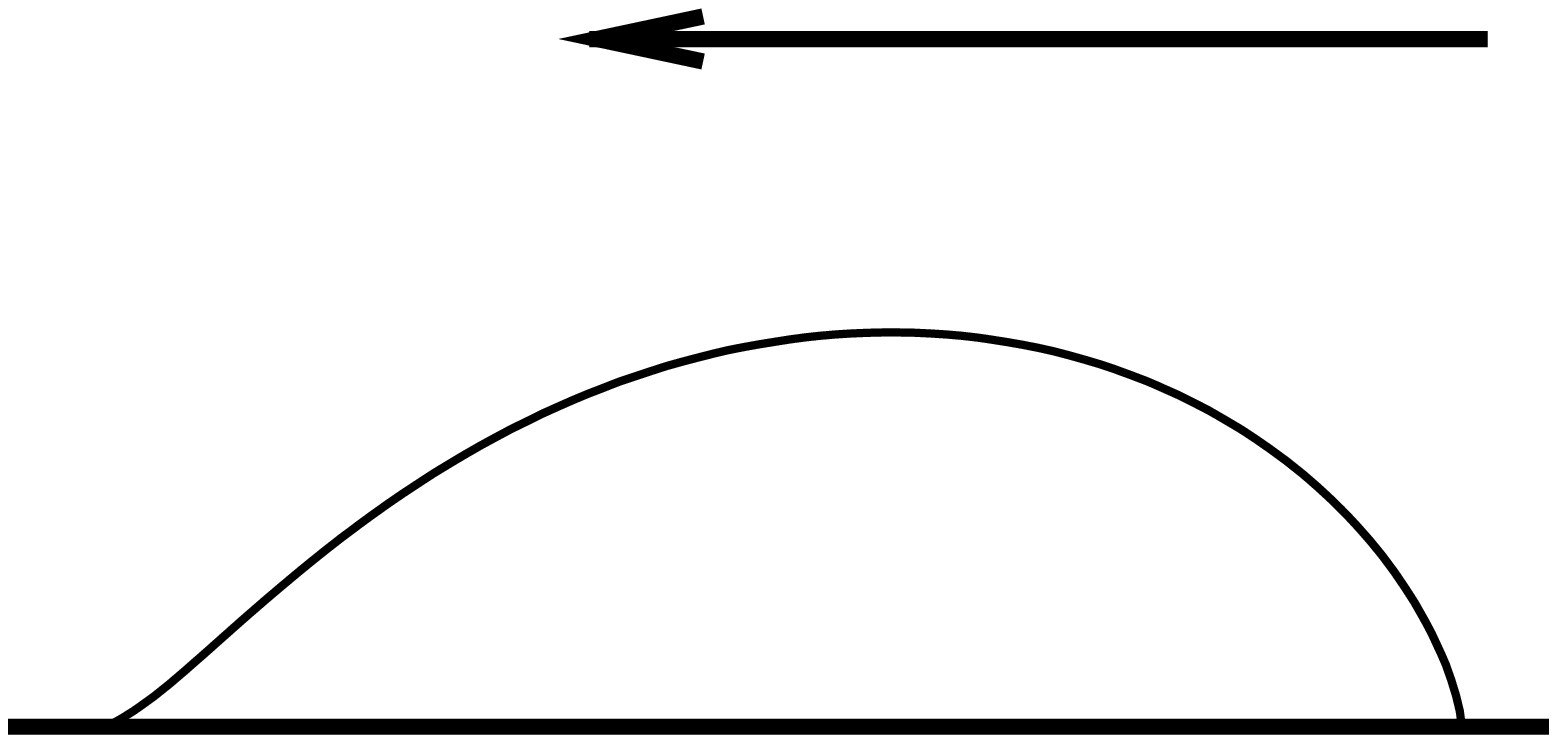}
(b) \includegraphics[trim = 1mm 1mm 80mm 1mm, clip, scale = 0.35, angle = 90]{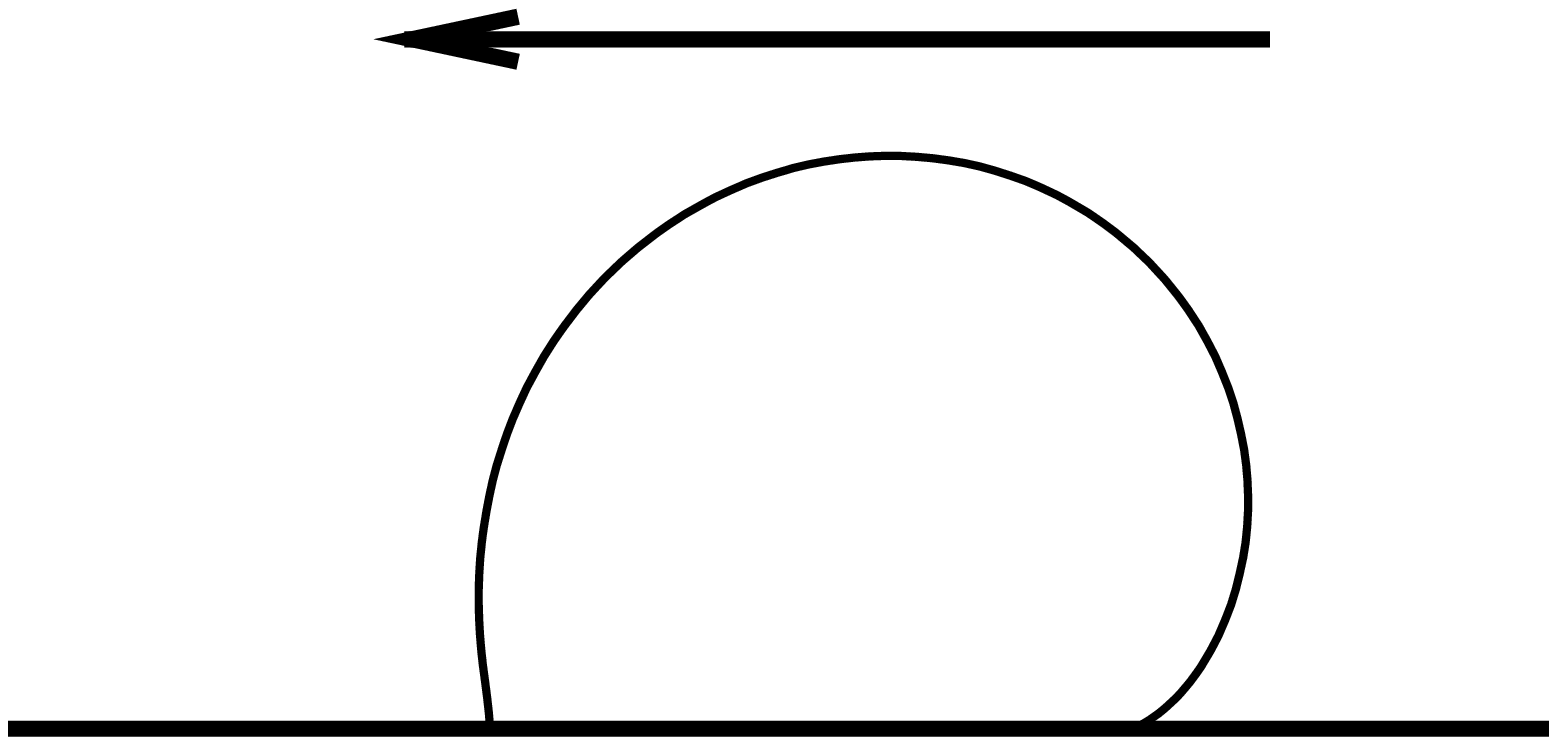}
  \caption{Drop shapes at $t = 50 T_{c, \textrm{inv}}$
  for two cases with the same WG ($\Delta \cos \theta = 0.866$):
  (a) $(\theta^{\textrm{upp}}, \theta^{\textrm{low}})= (30^{\circ}, 90^{\circ})$;
  (b) $(\theta^{\textrm{upp}}, \theta^{\textrm{low}}) = (90^{\circ}, 150^{\circ})$.
The common parameters are $Re = 16$, $r_{\nu} = 1$, 
$\theta_{H}^{\textrm{upp}}=\theta_{H}^{\textrm{low}} = 0$,
$L_{x} = 4$, $L_{y} = 16$, 
$Cn=0.125$, $Pe = 5 \times 10^{3}$, 
$N_{L} = 32$, $N_{t} = 320$.
The arrows denote the direction of motion.
The figures are rotated by $90^{\circ}$ in the anti-clockwise direction.}
  \label{fig:ff-Re16-gb-gl-cainit90-cn0d125-Pe5k}
\end{figure}

\begin{figure}[htp]
  \centering
 \includegraphics[scale = 1.0]{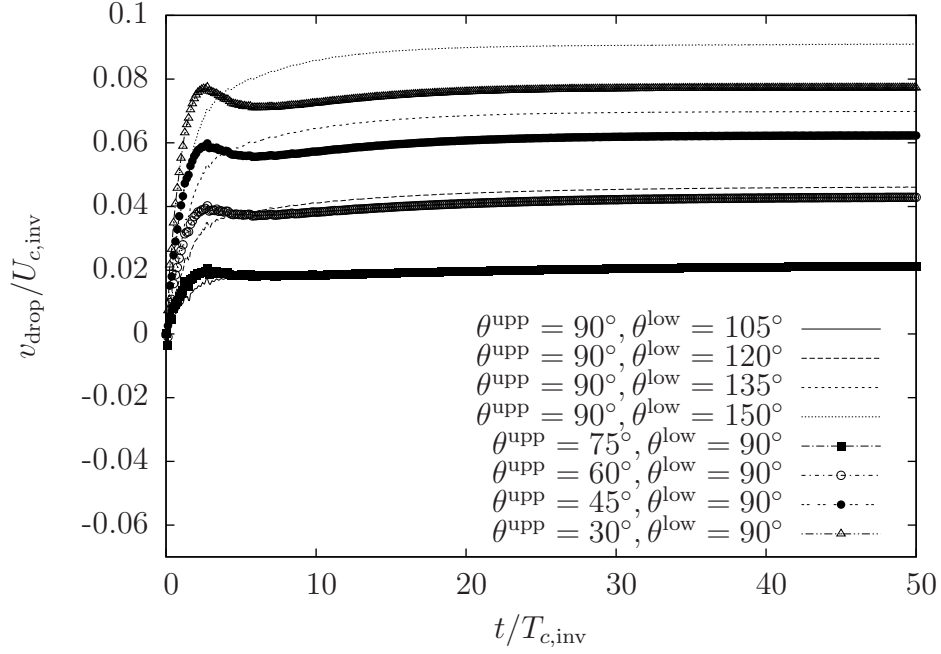}
  \caption{Evolutions of the centroid velocity of the drop $v_{\textrm{drop}}$
subject to different (stepwise) WGs on a wall.
The common parameters are $Re = 16$, $r_{\nu} = 1$, 
$\theta_{H}^{\textrm{upp}}=\theta_{H}^{\textrm{low}} = 0$,
$L_{x} = 4$, $L_{y} = 16$, 
$Cn=0.125$, $Pe = 5 \times 10^{3}$, 
$N_{L} = 32$, $N_{t} = 320$.}
  \label{fig:cmp-vd-dca-Re16-cainit90-cn0d125-Pe5k}
\end{figure}

Figure \ref{fig:cmp-vd-dca-Re16-cainit90-cn0d125-Pe5k} 
shows the evolutions of the drop velocity 
(scaled by $U_{c, \textrm{inv}}$, till $t = 50 T_{c, \textrm{inv}}$)
under the above two groups of different combinations of 
$\theta^{\textrm{upp}}$ and $\theta^{\textrm{low}}$
on the left wall. 
Note that the results for $\Delta \cos \theta = 0.087$ are not shown
in order to make the legends easy to recognize.
It is found from Fig. \ref{fig:cmp-vd-dca-Re16-cainit90-cn0d125-Pe5k} 
that, as expected, the velocity in steady state increases as $\Delta \cos \theta$ increases.
When $\Delta \cos \theta$ was small 
(e.g., $\Delta \cos \theta = \ 0.259$), 
the velocity seems to be less dependent on the
specific values of the upper and lower contact angles
(this also holds for another two cases with 
$\Delta \cos \theta = 0.087$ 
not shown here).
By contrast, at larger $\Delta \cos \theta$ (e.g., $\Delta \cos \theta = 0.5, \ 0.707, \ 0.866$), 
the drop velocity also depends on the specific values of $\theta^{\textrm{low}}$ and $\theta^{\textrm{upp}}$.
As seen from Fig. \ref{fig:cmp-vd-dca-Re16-cainit90-cn0d125-Pe5k},
after the initial acceleration stage the drop moves faster in the case with
$(\theta^{\textrm{upp}}, \theta^{\textrm{low}}) = (90^{\circ}, 150^{\circ})$ 
than in the case with 
$(\theta^{\textrm{upp}}, \theta^{\textrm{low}})= (30^{\circ}, 90^{\circ})$ 
though the value of $\Delta \cos \theta$ is the same ($\Delta \cos \theta = 0.866$)
for the two cases.
This is likely due to that in the case of GB 
the drop has less contact area with the wall,
thus having smaller viscous resistance.
Figure \ref{fig:ff-Re16-gb-gl-cainit90-cn0d125-Pe5k} shows the shapes of the drop
in steady motion for two cases with the same WG ($\Delta \cos \theta = 0.866$)
but with different upper and lower contact angles.
It is obvious that in the case of GL the drop spreads more on the wall.
Another observation from Fig. \ref{fig:cmp-vd-dca-Re16-cainit90-cn0d125-Pe5k} 
is that the initial acceleration stage seems to depend on the group (GL or GB),
especially at larger $\Delta \cos \theta$:
for the hydrophilic group, the drop experienced greater accelerations initially
and the drop velocity showed a \textit{bump} before it gradually approached the steady value;
by contrast, for the hydrophobic group, the drop was driven towards the steady state smoothly.
This could be attributed to the fact that the difference between the initial configuration
and the final steady shape of the drop is larger in the cases of GL 
(see Fig. \ref{fig:wgdrop-setup}
and Fig.  \ref{fig:ff-Re16-gb-gl-cainit90-cn0d125-Pe5k}), 
thus the drop was accelerated more
during the initial adjustment of configuration.

\begin{figure}[htp]
  \centering
 \includegraphics[scale = 1.0]{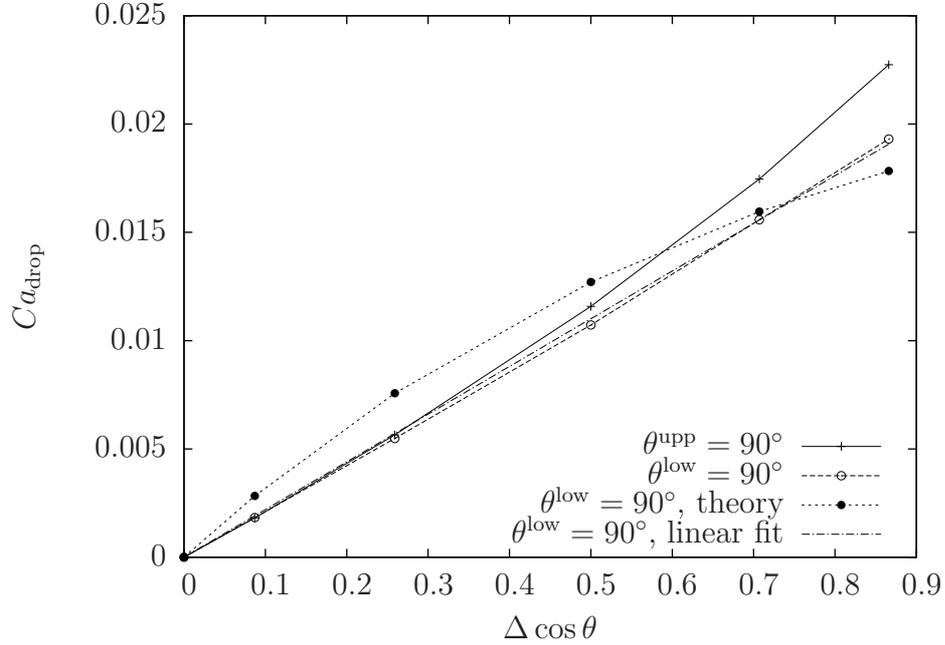}
  \caption{Variations of the capillary number $Ca_{\textrm{drop}}$ 
for a drop in steady state subject to different (stepwise) WGs
with the magnitude of the WG ($\Delta \cos \theta$).  
The common parameters are $Re = 16$, $r_{\nu} = 1$, 
$\theta_{H}^{\textrm{upp}}=\theta_{H}^{\textrm{low}} = 0$,
$L_{x} = 4$, $L_{y} = 16$, 
$Cn=0.125$, $Pe = 5 \times 10^{3}$, 
$N_{L} = 32$, $N_{t} = 320$.}
  \label{fig:cmp-cadrop-dcosca-Re16-cainit90-cn0d125-Pe5k}
\end{figure}

Figure \ref{fig:cmp-cadrop-dcosca-Re16-cainit90-cn0d125-Pe5k} 
shows the variations 
the capillary number $Ca_{\textrm{drop}}$ (when the drop is in steady state)
with the parameter $\Delta \cos \theta$
for the two groups of simulations.
Note that the steady velocity $V_{\textrm{drop}}$ used to calculate $Ca_{\textrm{drop}}$ was taken 
at different times for different cases to make sure that the criterion 
in Subsection \ref{sssec:obs-of-int} is satisfied. 
In Fig. \ref{fig:cmp-cadrop-dcosca-Re16-cainit90-cn0d125-Pe5k} 
we also show the linear fit for the hydrophilic group (GL with $\theta^{\textrm{low}} = 90^{\circ}$)
as well as the theoretical predictions based on the equations given 
by~\cite{l89-drop-gradient} for a drop subject to a continuous WG
and also used by~\cite{jphysii91-wg-drop}
for a drop subject to a stepwise WG .
With the current notations, one has the following prediction for 
$Ca_{\textrm{drop}}$
based on Eq. (A6) in~\citep{l89-drop-gradient}
(or alternatively, Eqs. (4) and (5) in~\citep{jphysii91-wg-drop}),
\begin{equation}
\label{eq:cadrop-steady-theory}
Ca_{\textrm{drop}} = \frac{1}{6 l}
 \sqrt{\frac{(\theta^{\textrm{low}})^{2} + (\theta^{\textrm{upp}})^{2}}{2}} 
 \bigg( \frac{(\theta^{\textrm{low}})^{2} - (\theta^{\textrm{upp}})^{2}}{2} \bigg)  ,
\end{equation}
where all angles should be in radian,
and $l$ is a constant prefactor usually taken to be of the order of $12$
~\citep{l89-drop-gradient, jphysii91-wg-drop},
which reflects the ratio of a macroscopic scale over a molecular scale in the problem of drop motion.
Here $l = 12$ was used to obtained the theoretical results in 
Fig. \ref{fig:cmp-cadrop-dcosca-Re16-cainit90-cn0d125-Pe5k}.
From this figure, it is seen that 
the capillary number $Ca_{\textrm{drop}}$ in steady state
almost increases linearly with the magnitude of the WG ($\Delta \cos \theta$)
for the hydrophilic group.
For the hydrophobic group, the linear relation still roughly holds when $\Delta \cos \theta$ is small;
but when the WG magnitude is large, a trend of nonlinear variation is observed
and the capillary number 
becomes slightly larger than that described by the linear variation.
The possible reason is already given above: the contact area 
between the drop and the wall becomes smaller and the viscous resistance is reduced in the hydrophobic group
at large $\Delta \cos \theta$.

The present findings seem to contradict those reported by~\cite{pre12-wg-drop}.
In their article, it was reported that the drop moved faster on the hydrophilic substrate
than on the hydrophobic one under "\textit{the same other conditions}".
This contradiction can be most likely attributed to the different definitions of \textit{the same other conditions}.
~\cite{pre12-wg-drop} also considered a 2-D drop, but it was driven by a continuous WG
with the contact angle $\theta$ having a distribution that satisfies $\frac{d (\cos \theta)}{d x} = const$
and $\frac{d (\cos \theta)}{d x} \ll 1$.
They mainly looked into the variation of the steady drop velocity
with the parameter $h_{0} \frac{\sigma}{\eta} \frac{d \cos \theta}{d x}$
where $h_{0}$ is the height of the drop.
Consider two drops, 
one on a hydrophilic surface with the average contact angle being $\theta_{1} < 90^{\circ}$ 
and the WG being $\frac{d (\cos \theta)}{d x}$
and the other on a hydrophobic surface with the average contact angle being $\theta_{2} > 90^{\circ}$
and the same WG.
When they have the same height $h_{0}$, they are regarded as under \textit{the same other conditions}
according to~\cite{pre12-wg-drop}.
However, the actual driving forces caused by the WG differ
because the distances between the two three-phase-points (TPPs)
(about twice of the contact radius $R_{c}$ in~\citep{pre12-wg-drop}) 
differ in these two cases.
Under a weak WG ($\frac{d (\cos \theta)}{d x} \ll 1$), 
the drop does not have significant deformations.
A straightforward calculation relates the average contact angle $\theta$,
the drop height $h_{0}$ and the contact radius $R_{c}$ as follows,
\begin{equation}
\label{eq:h0-Rc-theta}
R_{c} = h_{0}\frac{\sin \theta}{1 - \cos \theta}  .
\end{equation}
Since the WG satisfies $\frac{d (\cos \theta)}{d x} = const$, 
the driving force (per unit length, as we are considering 2-D problems)
$\vert \boldsymbol{F}_{d} \vert$ 
is found to be,
\begin{equation}
\label{eq:Fd-h0-Rc-theta}
\vert \boldsymbol{F}_{d} \vert  = 
\sigma \bigg[ (2 R_{c})  \frac{d (\cos \theta)}{d x} \bigg]
= \sigma \bigg[ \bigg( 2 h_{0}\frac{\sin \theta}{1 - \cos \theta}  \bigg)  \frac{d (\cos \theta)}{d x} \bigg],
\end{equation}
where the quantity in the square brackets is equivalent to $\Delta \cos \theta$ in the present work.
Then, it is easy to find that the ratio of the driving forces in the hydrophilic and hydrophobic cases is,
\begin{equation}
\label{eq:Fd-ratio-theta}
\frac{\vert \boldsymbol{F}_{d} \vert_{1}}{\vert \boldsymbol{F}_{d} \vert_{2}}  
= \frac{\sin \theta_{1} (1 - \cos \theta_{2}) }{(1 - \cos \theta_{1}) \sin \theta_{2}},
\end{equation}
Based on the lengths in Fig. 4 of~\citep{pre12-wg-drop},
the two cases their compared were estimated to have $\theta_{1} \approx 62.4^{\circ}$
and $\theta_{2} \approx 125.3^{\circ}$, which gives 
$\frac{\vert \boldsymbol{F}_{d} \vert_{1}}{\vert \boldsymbol{F}_{d} \vert_{2}}  \approx 2$.
Recall that in steady state, the driving forces are balanced by the viscous resistance forces
and the velocity is roughly proportional to the magnitude of the net driving force.
Then, it is not difficult to understand that this factor of $2$ found here
is quite close to the ratio of the coefficient $\alpha_{V}$
defined by~\cite{pre12-wg-drop} (as a measure to reflect how fast the drop moves under certain conditions)
for the hydrophilic case over that for the hydrophobic case,
which is about $2.06$ ($\alpha_{V} \approx 0.33$, and $0.16$ for the hydrophilic 
and hydrophobic cases respectively~\citep{pre12-wg-drop}).
In the present work, one of the requirements for two cases to be under \textit{the same other conditions}
is that the driving forces (or equivalently, the magnitudes of the stepwise WG) 
are equal (rather than any others based on the height of the drop).

In addition, 
from Fig. \ref{fig:cmp-cadrop-dcosca-Re16-cainit90-cn0d125-Pe5k}
it is found that the numerical results appear to be close to the theoretical predictions given by
Eq. (\ref{eq:cadrop-steady-theory}) 
with $l = 12$.
However, we would like to point out that this seemingly good agreement may be rather a coincidence.
In the derivation of Eq. (\ref{eq:cadrop-steady-theory}) 
a few assumptions were made
~\citep{l89-drop-gradient, jphysii91-wg-drop}.
For example, the pressure was assumed to reach equilibrium much faster than the drop's motion
and the profile of the drop remains an arc of circle.
What is more, the drop was assumed to be quite flat with the dynamic contact angles
being much smaller than unity.
In our simulations these conditions are not well satisfied.
Besides, in~\citep{l89-drop-gradient, jphysii91-wg-drop}
the resistance due to the ambient fluid was assumed to be negligible.
But the present work does not omit such effects.

\subsubsection{Effects of the CAH}

In the above, all the factors in Eq. (\ref{eq:ca-drop-func-reduced})
have been studied except the magnitude of the CAH,
$\theta_{H}^{\textrm{upp}}=\theta_{A}^{\textrm{upp}} - \theta_{R}^{\textrm{upp}}$ 
and $\theta_{H}^{\textrm{low}}=\theta_{A}^{\textrm{low}}-\theta_{R}^{\textrm{low}}$.
In this part, we consider the effects of the CAH.
For simplicity, we only study the cases with 
$\theta_{H}^{\textrm{upp}}=\theta_{H}^{\textrm{low}} = \theta_{H}$.
Two sets of upper and lower contact angles are considered:
(S1) $\theta^{\textrm{upp}} = 75^{\circ}$, $\theta^{\textrm{low}} = 105^{\circ}$;
(S2) $\theta^{\textrm{upp}} = 60^{\circ}$, $\theta^{\textrm{low}} = 120^{\circ}$.
The Reynolds number and the viscosity ratio are fixed at
$Re = 16$ and $r_{\nu} = 1$.
The remaining parameters are $L_{x} = 4$, $L_{y} = 16$, 
$Cn=0.125$, $Pe = 5 \times 10^{3}$, 
$N_{L} = 32$, $N_{t} = 320$.

\begin{figure}[htp]
  \centering
 \includegraphics[scale = 1.0]{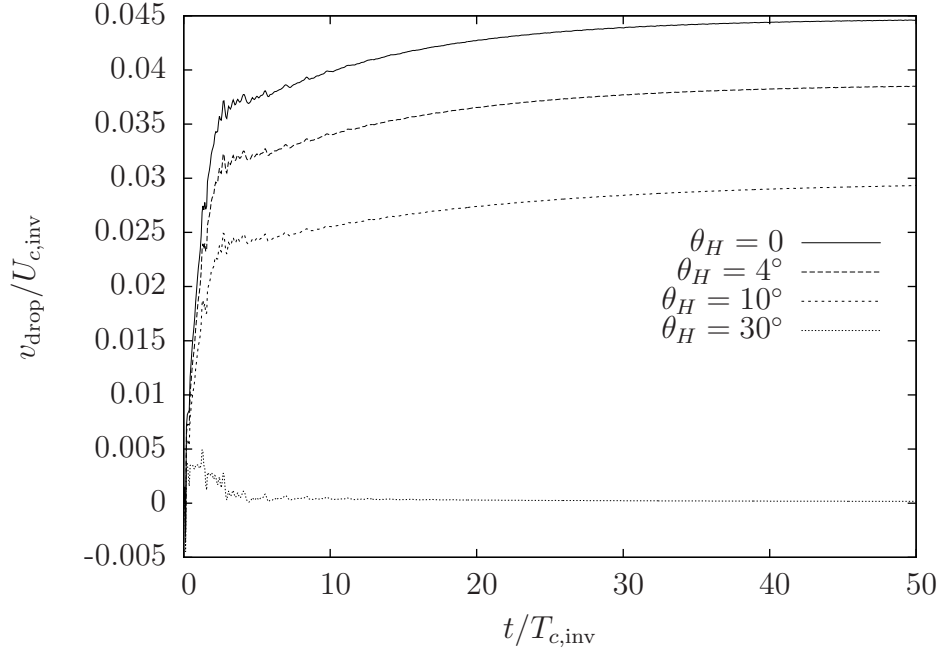}
  \caption{Evolutions of the centroid velocity of the drop $v_{\textrm{drop}}$
subject to a stepwise WG on a wall with CAH of 
(S1a) $\theta_{H} = 4^{\circ}$ 
($\theta_{A, R}^{\textrm{upp}} = 77^{\circ}, 73^{\circ}, \theta_{A, R}^{\textrm{low}} = 107^{\circ}, 103^{\circ}$);
(S1b) $\theta_{H} = 10^{\circ}$ 
($\theta_{A, R}^{\textrm{upp}} = 80^{\circ}, 70^{\circ}, \theta_{A, R}^{\textrm{low}} = 110^{\circ}, 100^{\circ}$);
(S1c) $\theta_{H} = 30^{\circ}$ 
($\theta_{A, R}^{\textrm{upp}} = 90^{\circ}, 60^{\circ}, \theta_{A, R}^{\textrm{low}} = 120^{\circ}, 90^{\circ}$).
Also shown is the case with no CAH ($\theta_{H} = 0^{\circ}$).
The common parameters are $Re = 16$, $r_{\nu} = 1$, 
$L_{x} = 4$, $L_{y} = 16$, 
$Cn=0.125$, $Pe = 5 \times 10^{3}$, 
$\theta^{\textrm{upp}} = 75^{\circ}$, $\theta^{\textrm{low}} = 105^{\circ}$,
$N_{L} = 32$, $N_{t} = 320$.}
  \label{fig:cmp-vd-ardca-Re16-cainit90-cn0d125-Pe5k}
\end{figure}

In each of the two sets, S1 and S2, three magnitudes of CAH 
were tried in addition to the cases with no CAH (i.e., $\theta_{H} = 0$).
In S1, $\theta_{H} = 4^{\circ}, \ 10^{\circ}$, and $30^{\circ}$,
and in S2, $\theta_{H} = 20^{\circ}, \ 40^{\circ}$, and $60^{\circ}$.
Thus, we have the following pairs of advancing and receding contact angles
for the upper and lower parts in S1,
\begin{itemize}
\item (S1a) $\theta_{A, R}^{\textrm{upp}} = 77^{\circ}, 73^{\circ}, \theta_{A, R}^{\textrm{low}} = 107^{\circ}, 103^{\circ}$,
\item (S1b) $\theta_{A, R}^{\textrm{upp}} = 80^{\circ}, 70^{\circ}, \theta_{A, R}^{\textrm{low}} = 110^{\circ}, 100^{\circ}$,
\item (S1c) $\theta_{A, R}^{\textrm{upp}} = 90^{\circ}, 60^{\circ}, \theta_{A, R}^{\textrm{low}} = 120^{\circ}, 90^{\circ}$,
\end{itemize}
and in S2 we have,
\begin{itemize}
\item (S2a) $\theta_{A, R}^{\textrm{upp}} = 70^{\circ}, 50^{\circ}, \theta_{A, R}^{\textrm{low}} = 130^{\circ}, 110^{\circ}$,
\item (S2b) $\theta_{A, R}^{\textrm{upp}} = 80^{\circ}, 40^{\circ}, \theta_{A, R}^{\textrm{low}} = 140^{\circ}, 100^{\circ}$,
\item (S2c) $\theta_{A, R}^{\textrm{upp}} = 90^{\circ}, 30^{\circ}, \theta_{A, R}^{\textrm{low}} = 150^{\circ}, 90^{\circ}$.
\end{itemize}
Figure \ref{fig:cmp-vd-ardca-Re16-cainit90-cn0d125-Pe5k} 
shows the evolutions of the drop velocity $v_{\textrm{drop}}$
(scaled by $U_{c, \textrm{inv}}$, till $t = 50 T_{c, \textrm{inv}}$)
in the four cases with different $\theta_{H}$ in S1.
It is found from Fig. \ref{fig:cmp-vd-ardca-Re16-cainit90-cn0d125-Pe5k} 
that when the CAH was not too large 
($\theta_{H} = 4^{\circ}$ for Case (S1a)), 
and $\theta_{H}= 10^{\circ}$ for Case (S1b))
the drop was accelerated initially and then gradually showed a trend to become steady.
This behavior is just like the reference case with no CAH
and the difference is that the velocity is reduced 
when CAH is present, as expected.
When the CAH was large enough 
($\theta_{H} = 30^{\circ}$ for Case (S1c)),
the drop almost remained static. 
This is because in Case (S1c) the initial drop shape is within the range of the equilibrium states
allowed by the given advancing and receding contact angles
($\theta^{i} = \theta_{A}^{\textrm{upp}}=\theta_{R}^{\textrm{low}} = 90^{\circ}$).
In the other set of simulations (S2) we have similar observations.

\begin{figure}[htp]
  \centering
 \includegraphics[scale = 1.0]{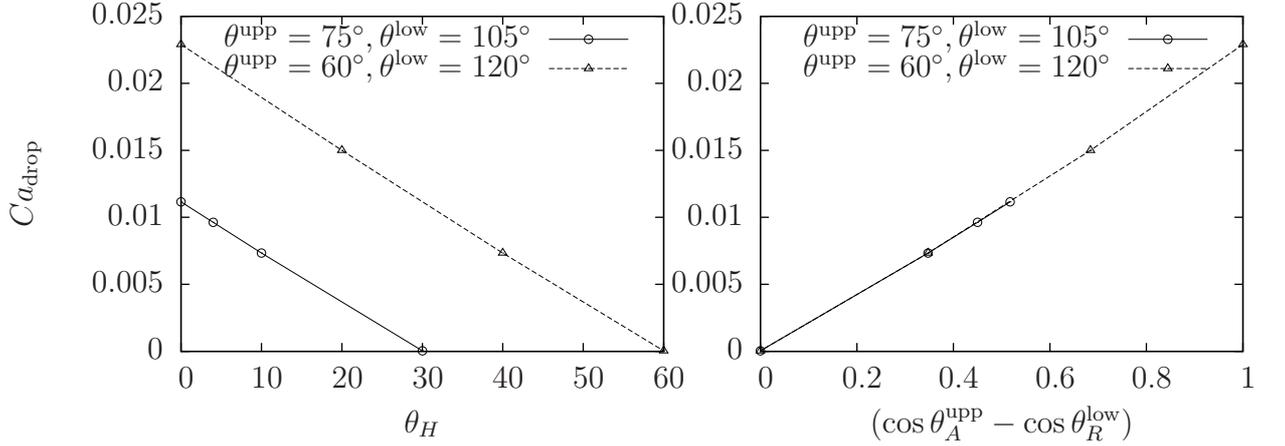}
  \caption{Variation of 
  the steady capillary number $Ca_{\textrm{drop}}$ 
  of a drop subject to a stepwise WG on a wall with CAH
 with the magnitude of the CAH ($\theta_{H}$) (left)
and the parameter 
$(\Delta \cos \theta)_{H} = \cos \theta^{\textrm{upp}}_{A} - \cos \theta^{\textrm{low}}_{R}$ (right)
for two sets with 
(S1) $\theta^{\textrm{upp}} = 75^{\circ}$, $\theta^{\textrm{low}} = 105^{\circ}$;
(S2) $\theta^{\textrm{upp}} = 60^{\circ}$, $\theta^{\textrm{low}} = 120^{\circ}$.
The common parameters are $Re = 16$, $r_{\nu} = 1$, 
$L_{x} = 4$, $L_{y} = 16$, 
$Cn=0.125$, $Pe = 5 \times 10^{3}$, 
$N_{L} = 32$, $N_{t} = 320$.}
  \label{fig:cmp-cadrop-dca-cah-Re16-cainit90-cn0d125-Pe5k}
\end{figure}

Figure \ref{fig:cmp-cadrop-dca-cah-Re16-cainit90-cn0d125-Pe5k}
plots the variations of the steady capillary number $Ca_{\textrm{drop}}$ 
with the magnitude of the CAH ($\theta_{H}$) (left panel)
and also with another quantity 
$(\Delta \cos \theta)_{H} = \cos \theta^{\textrm{upp}}_{A} - \cos \theta^{\textrm{low}}_{R}$  
(right panel)
for the two sets (S1 and S2).
It is found from the left panel of 
Fig. \ref{fig:cmp-cadrop-dca-cah-Re16-cainit90-cn0d125-Pe5k}
that the steady capillary number $Ca_{\textrm{drop}}$
decreases as the magnitude of the hysteresis ($\theta_{H}$) increases,
and the data points of the pair 
$(Ca_{\textrm{drop}}, \theta_{H})$
almost fall on a straight line for each set.
Besides, the two lines for S1 and S2 appear to be parallel.
From the right panel of 
Fig. \ref{fig:cmp-cadrop-dca-cah-Re16-cainit90-cn0d125-Pe5k}
it is seen that 
$Ca_{\textrm{drop}}$
increases roughly linearly with the quantity
$(\Delta \cos \theta)_{H}$ 
and the data points for both sets are almost on the same straight line.
These results suggest that for a (2-D) drop on a substrate with a stepwise WG and CAH,
the most important factors are the advancing contact angle of the more hydrophilic region
and the receding contact angle of the more hydrophobic region,
which somehow defines the equivalent magnitude of the WG in the presence of CAH.

\subsubsection{Analyses of the flow field and velocity profile}

Finally, we examine some details of the flow when the drop reaches steady state.
For this purpose, we select a few typical cases with $Re = 16$,  
$\theta^{\textrm{upp}} = 75^{\circ}$, $\theta^{\textrm{low}} = 105^{\circ}$,
$\theta_{H}^{\textrm{upp}}=\theta_{H}^{\textrm{low}} = 0$,
and the following viscosity ratios: 
$r_{\nu} = 0.1, \ 1, \ 5$, and $40$.
Other common parameters are $L_{x} = 4$, $L_{y} = 16$, 
$Cn=0.125$, $Pe = 5 \times 10^{3}$, 
$N_{L} = 32$, 
and the temporal discretization parameter $N_{t}$ varies for different cases (as already given before).

\begin{figure}[htp]
  \centering
(a) \includegraphics[scale = 0.24, angle = 90]{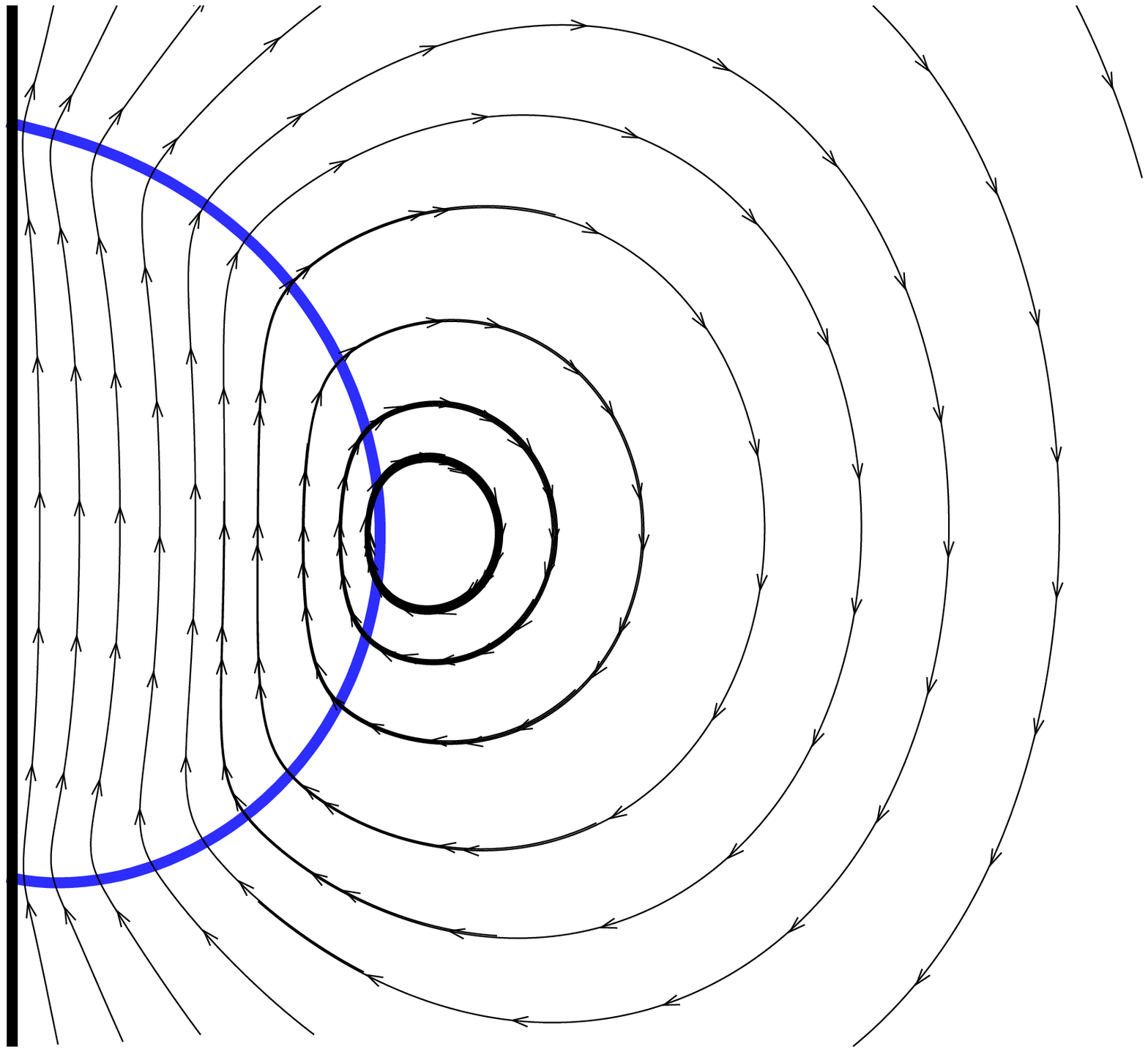}
(b) \includegraphics[scale = 0.24, angle = 90]{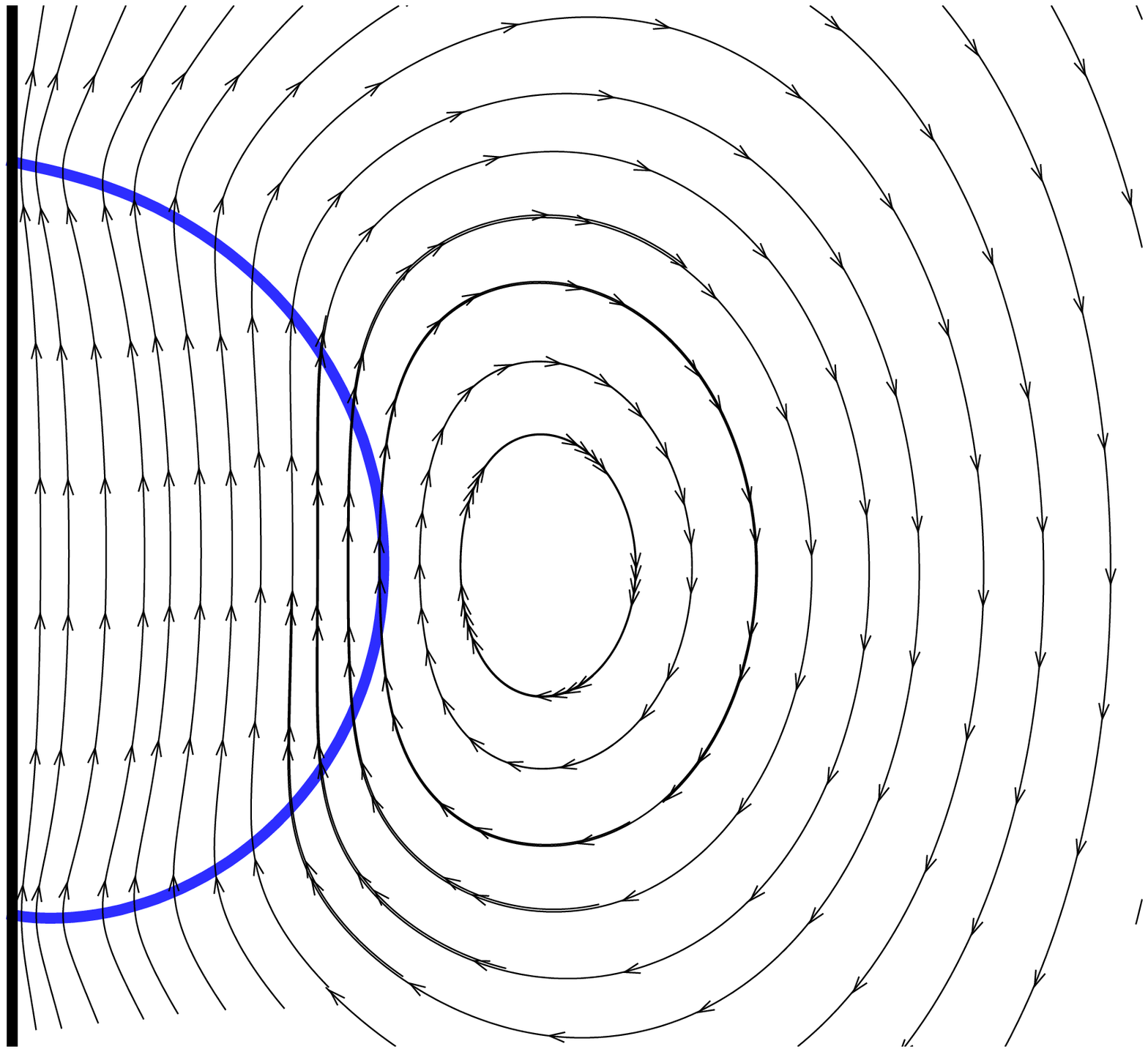}
(c) \includegraphics[scale = 0.24, angle = 90]{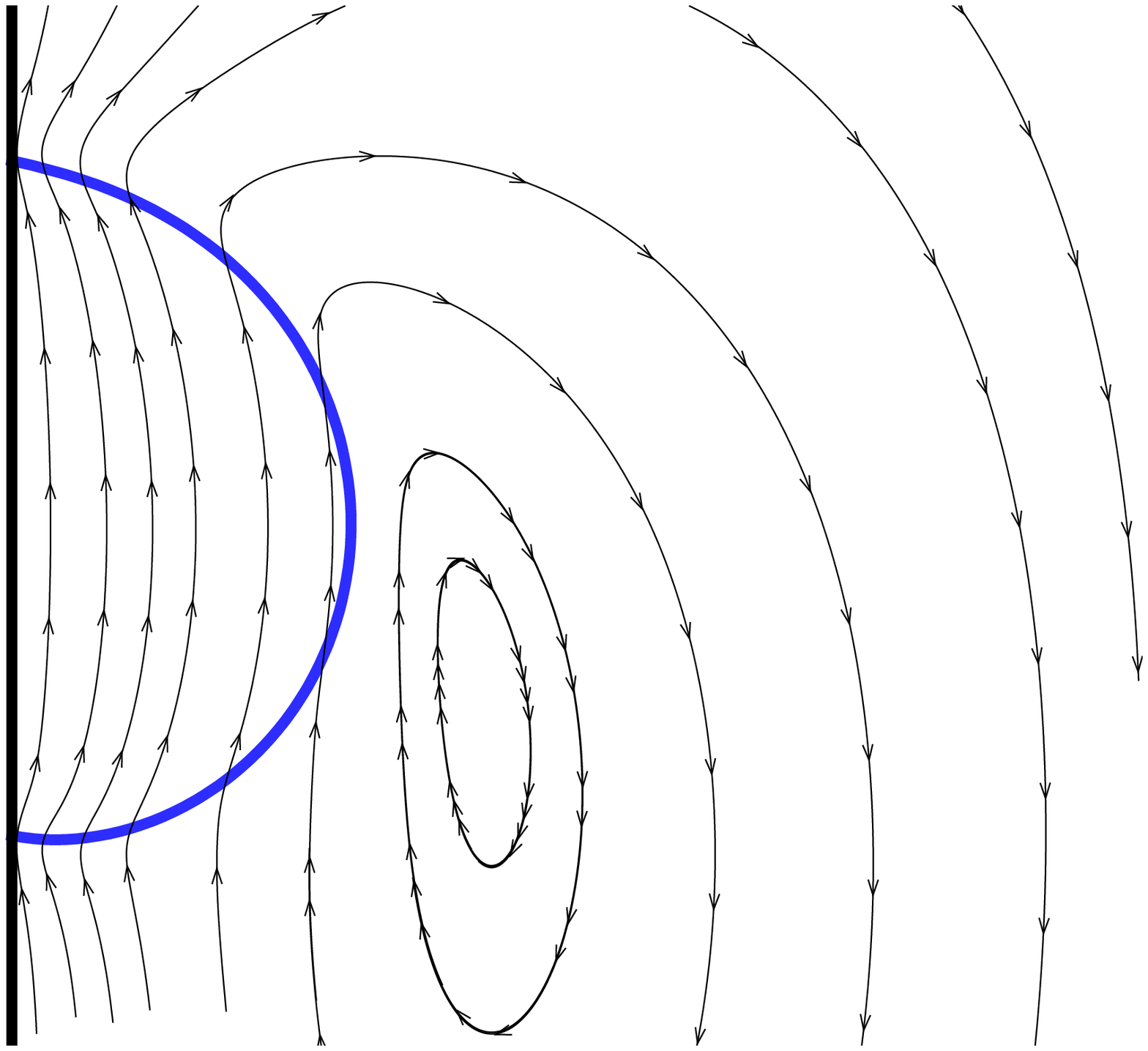}
(d) \includegraphics[scale = 0.24, angle = 90]{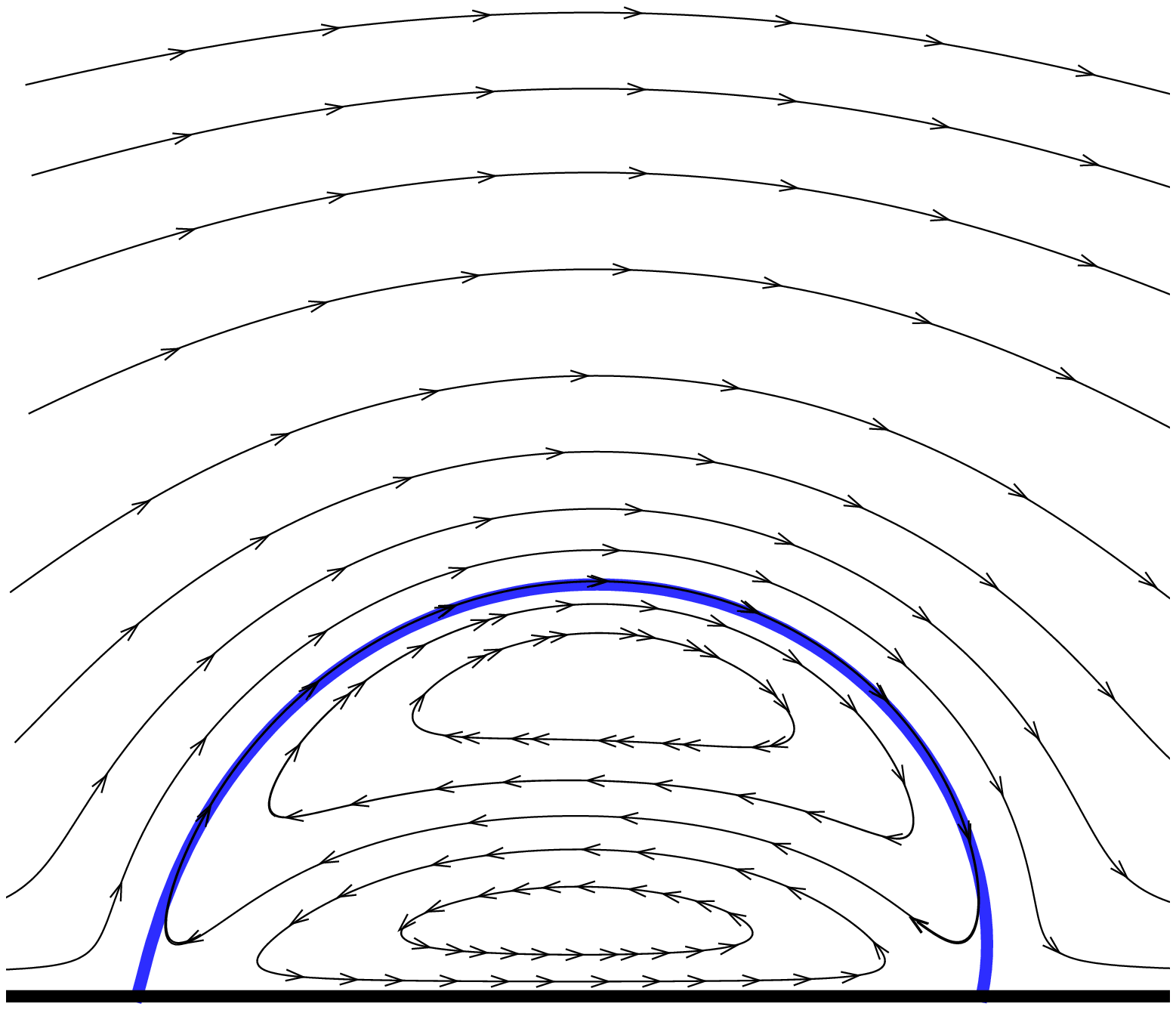}
(e) \includegraphics[scale = 0.24, angle = 90]{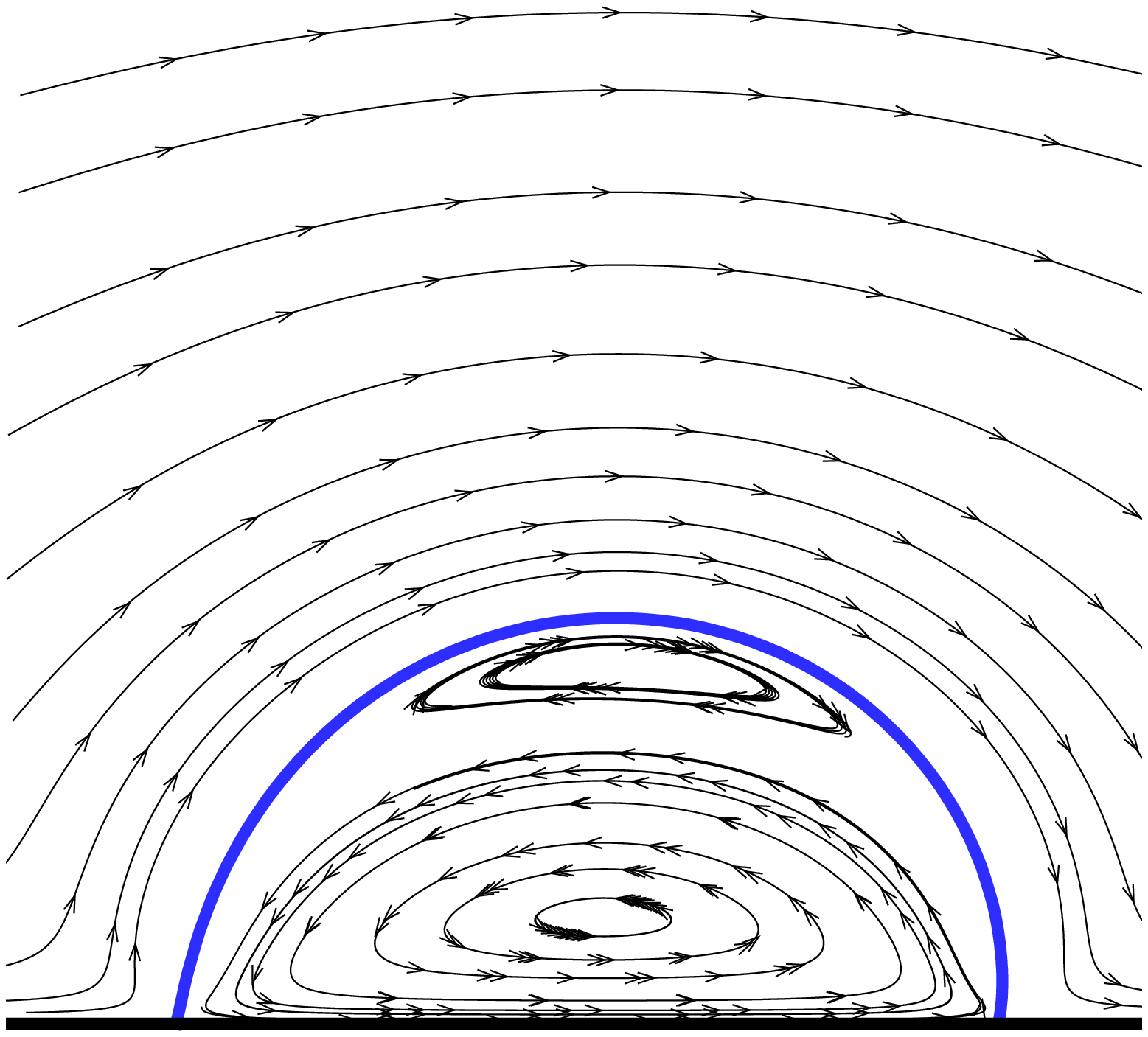}
(f) \includegraphics[scale = 0.24, angle = 90]{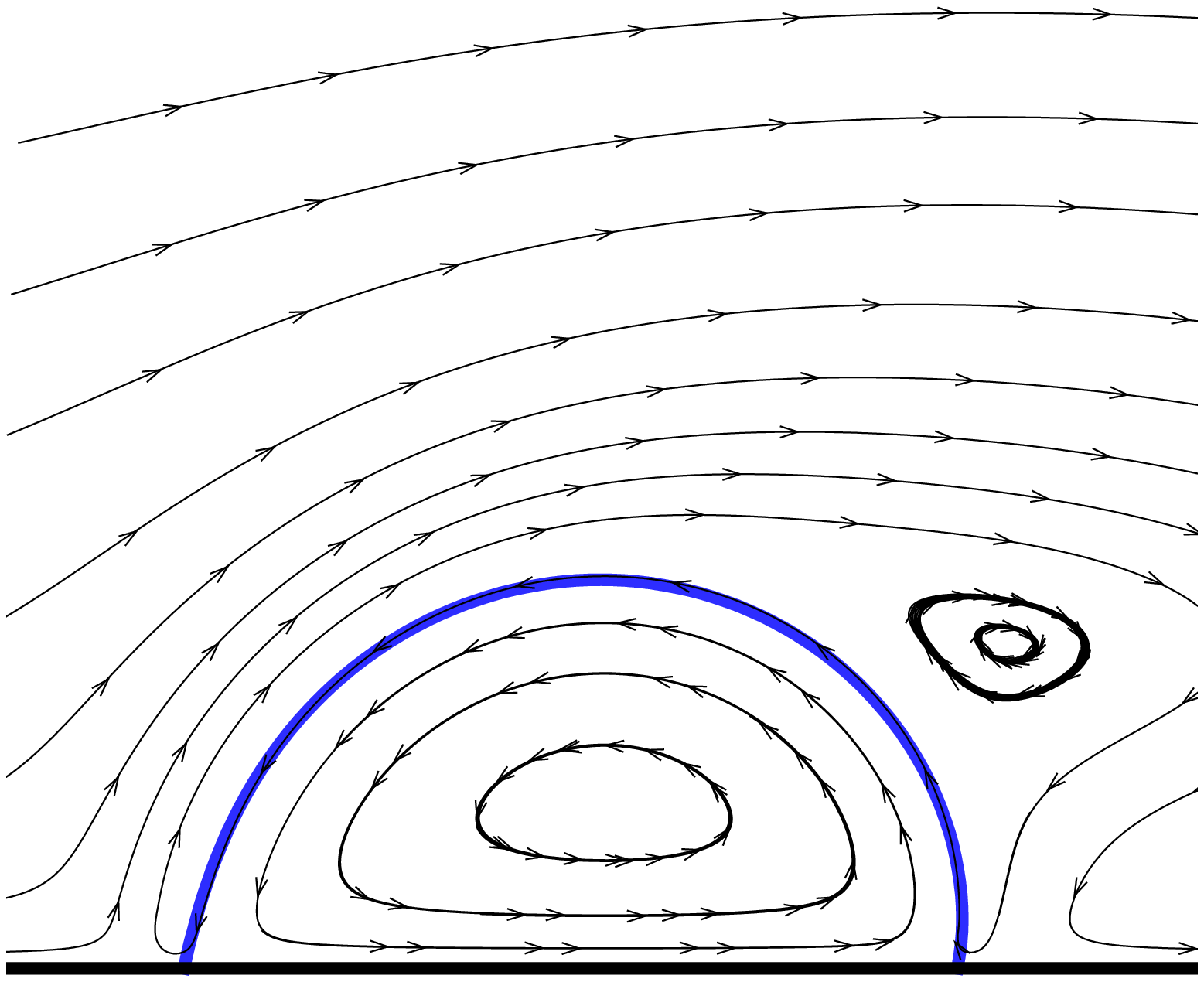}
  \caption{Drop shapes and streamlines around the drop subject to a stepwise WG
at $r_{\nu} = 0.1, \ 1,$ and $40$ (left, middle, and right columns, respectively).
The upper row shows streamlines as observed in the frame fixed on the wall
whereas the lower row shows those observed in the frame moving with the drop.
The common parameters are $Re = 16$, 
$\theta^{\textrm{upp}} = 75^{\circ}$, $\theta^{\textrm{low}} = 105^{\circ}$,
$\theta_{H}^{\textrm{upp}}=\theta_{H}^{\textrm{low}} = 0$,
$L_{x} = 4$, $L_{y} = 16$, 
$Cn=0.125$, $Pe = 5 \times 10^{3}$, 
$N_{L} = 32$. 
The figures are rotated by $90^{\circ}$ in the anti-clockwise direction.}
  \label{fig:ff-Re16-rnu-cainit90-cn0d125-Pe5k}
\end{figure}

Figure \ref{fig:ff-Re16-rnu-cainit90-cn0d125-Pe5k}
shows the drop shapes and the streamlines around the drop
(from left to right: 
(a, d) for $r_{\nu} = 0.1$ at $t = 150 T_{c, \textrm{inv}}$,
(b, e)  for $r_{\nu} = 1$ at $t = 100 T_{c, \textrm{inv}}$,
and (c, f) for $r_{\nu} = 40$ at $t = 50 T_{c, \textrm{inv}}$)
as observed in two different frames:
the frame fixed on the wall denoted as the \textit{absolute frame} shown in the upper row,
and the frame moving with the drop denoted as the \textit{relative frame} shown in the lower row.
From the upper row in 
Fig. \ref{fig:ff-Re16-rnu-cainit90-cn0d125-Pe5k},
when the observation is made in the \textit{absolute} frame,
a circulation is seen with its center being close to but above the top of drop.
As the viscosity ratio $r_{\nu}$ increases, 
the circulation center first moves upwards
and then moves downstream.
Besides, the streamlines pass through the drop and were slightly bent
when crossing the interfaces.
If the observation is made in the \textit{relative} frame,
the streamlines show distinctive pattens,
as found in the lower row of 
Fig. \ref{fig:ff-Re16-rnu-cainit90-cn0d125-Pe5k}.
The most noticeable feature is that two circulation regions form, with one above the other.
At low viscosity ratios (i.e., the ambient fluid is more viscous), both circulations are inside
the drop, but the upper one covers a larger area than the lower one at $r_{\nu} = 0.1$
whereas the opposite is true at $r_{\nu} = 1$.
At a high viscosity ratio ($r_{\nu} = 40$), the upper circulation forms outside the drop
and the lower one almost occupies the whole inner area of the drop.
It is suspected that the above change of streamline pattern is not only caused by 
the change of the viscosity ratio, but also
(probably more likely) caused by the change of the actual Reynolds number
$Re_{\textrm{drop}}$
($Re_{\textrm{drop}} = 0.054, \ 0.18$, and $0.61$ for $r_{\nu} = 0.1, \ 1$, and $40$, respectively).

\begin{figure}[htp]
  \centering
 \includegraphics[scale = 0.5]{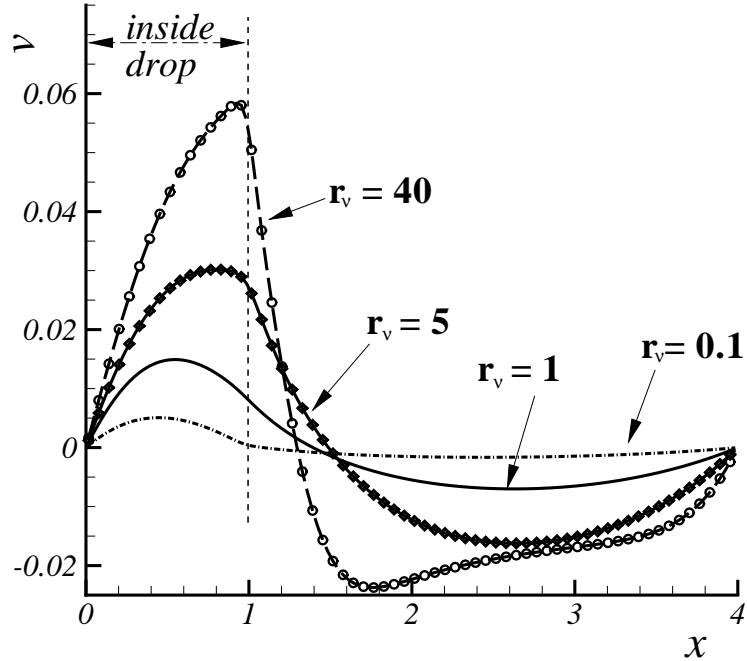}
  \caption{The profiles of the (absolute) velocity component $v (x)$ 
  along the horizontal line passing through the top of the drop
  in steady state for four cases with different viscosity ratios, 
  $r_{\nu} = 0.1, \ 1, \  5$, and $40$.
The common parameters are $Re = 16$, 
$\theta^{\textrm{upp}} = 75^{\circ}$, $\theta^{\textrm{low}} = 105^{\circ}$,
$\theta_{H}^{\textrm{upp}}=\theta_{H}^{\textrm{low}} = 0$,
$L_{x} = 4$, $L_{y} = 16$, 
$Cn=0.125$, $Pe = 5 \times 10^{3}$, 
$N_{L} = 32$, $N_{t} = 320$.
The dashed vertical line roughly denotes the heights of the drop
for the four cases
(which actually differ slightly, but are quite close).}
  \label{fig:v-x-Re16-rnu0d1-rnu1-rnu5-rnu40-cainit90-cn0d125-Pe5k}
\end{figure}

In addition, we examine the profile of the (absolute) velocity component $v$ in steady state
along the horizontal line passing through the top of the drop.
Four cases with different viscosity ratios, $r_{\nu} = 0.1, \ 1, \  5$, and $40$, 
are examined with other common parameters already given above.
Figure \ref{fig:v-x-Re16-rnu0d1-rnu1-rnu5-rnu40-cainit90-cn0d125-Pe5k}
compares the profiles of $v(x)$ for the four cases.
Note that the $x-$axis is put as in its normal position in 
Fig. \ref{fig:v-x-Re16-rnu0d1-rnu1-rnu5-rnu40-cainit90-cn0d125-Pe5k} 
without any rotation
(unlike in Fig. \ref{fig:ff-Re16-rnu-cainit90-cn0d125-Pe5k}),
and the velocity is measured in the characteristic velocity $U_{c}$.
It is seen that in all cases 
the profiles $v(x)$ along the selected lines
inside the drop (on the left of the dashed vertical line)
resemble that of a Poiseuille flow,
but the points of inflection, where the maximum velocities occur,
are below (i.e., on the left of) the top of the drop.
When the viscosity ratio is relatively low ($r_{\nu} = 0.1, \ 1$ and $5$), 
the point of inflection is relatively farther away from the top of the drop
(the lower $r_{\nu}$ is, the farther).
The profile at $r_{\nu} = 0.1$ (i.e., when the ambient fluid is ten times more viscous than the drop)
appears to be the closest to a full Poiseuille profile among all cases.
The observation on the inflection point at $r_{\nu} = 5$ 
seems to agree with that reported by~\cite{pre12-wg-drop},
in which the viscosity ratio was about $5$.
In contrast, the point of inflection 
becomes very close to the top of the drop at a large viscosity ratio ($r_{\nu} = 40$),
making the profile inside the drop look like half of the full Poiseuille profile.
This observation supports the previous assumption about the velocity profile
made by~\cite{l89-drop-gradient} 
for the derivation of theoretical results.

\subsubsection{Some discussions about the parameters}

Some remarks on the parameters in the present work may be useful.
First, we assume the two fluids have the same density, $\rho_{A} = \rho_{B} = \rho_{c}$,
thus the density ratio is unity. This differs significantly from the liquid-air systems
under usual conditions, in which the liquid/air density ratio can be as high as $10^{3}$.
At the same time, this setting is fairly close to some liquid-liquid systems
under usual conditions.
For instance, ~\cite{apl06ew-mixing} used droplets of water-glycerol-NaCl mixtures
with densities about $1000 kg / m^{3}$
in silicone oil (Wacker AK5 with a density about $920 kg / m^{3}$)
in their experiments.
And recently,~\cite{phf12-filmthickness} used two-fluid systems
composed of water/glucose syrup mixture 
with densities about $1170 kg / m^{3}$
and silicone oil
(of three different types, Wacker AK5, AK20 and AK50
with their densities being about $920 kg / m^{3}$, $945 kg / m^{3}$ and $960 kg / m^{3}$, respectively).
Besides, for some cases with low Reynolds numbers the inertial effects
may be negligible, making the density ratio not so important.
Of course, this may not hold for all of the cases studied here,
and for cases with intermediate (or even higher) $Re$ 
the effects of density ratio may be worth pursuing
(this is left for future work).
Second, we compare some parameters in the experiments by
~\cite{apl06ew-mixing} and~\cite{phf12-filmthickness} 
for some real liquid-liquid systems
with the present work
(only one case selected for each work).
The comparisons are given in Table \ref{tab:typical-parameters-exp}.
Note that the drop radius in~\citep{apl06ew-mixing} was estimated
from the drop volume given in that article.
The comparisons are made just to show the connections 
(in terms of the fluid properties and key dimensionless parameters) 
between the present numerical simulation
and the real world while we are aware that the three pieces of work study quite different drop problems.
In Table \ref{tab:typical-parameters-exp} the fluid properties 
(e.g., the density, viscosity, interfacial tension and drop radius) 
in the present work are not uniquely determined;
they are just one of the possible sets that would render a Reynolds number of
$Re = \sigma R/ (\rho_{c} \nu_{A}^{2}) = 400$.

\begin{table}[htp]
\caption{Parameters in the experiments by~\cite{apl06ew-mixing}, by~\cite{phf12-filmthickness}
and in the present work. Note that only one case in each work is selected for comparison.}
\begin{center}
\begin{tabular}{|c|c|c|c|}\hline
Parameter $\backslash$ Source      &      ~\citep{apl06ew-mixing}  & ~\citep{phf12-filmthickness}    & Present  \\\hline
Drop Density ($kg / m^{3}$)  &      1000    &      920 &      1000    \\\hline
Drop Viscosity (Dynamic)  ($mPa \  s$) &    5 
&    $5.5$ &    $5$      \\\hline
Ambient Fluid Density ($kg / m^{3}$) &   920 &   1170 &   1000           \\\hline
Ambient Fluid Viscosity (Dynamic)  ($mPa \  s$) &   $5$  & $5.8$ &  $5$        \\\hline
Interfacial Tension ($mN / m$) &    34   &    21&    25    \\\hline
Drop Radius ($mm$) & $0.7$  & $3.6$ & $0.4$       \\\hline
Density Ratio  (-) &     1.09   &     0.79 &     1.0       \\\hline
Viscosity Ratio (Dynamic)  (-) &     1.0 
&     0.95 &    1.0 
\\\hline
Characteristic Velocity  $U_{c}$ ($m / s$)&   6.8     &   3.82 &   5.0      \\
 $\quad \quad \quad \quad  \quad \quad  \quad  \quad  \quad$ 
$U_{c, \textrm{inv}}$ ($m / s$)&   0.22     &   0.08 &   0.25      \\\hline
Characteristic Time $T_{c}$ ($s$)&   $1.03 \times 10^{-4}$   &   $9.43 \times 10^{-4}$ &   $8 \times 10^{-5}$      \\
 $\quad \quad \quad \quad  \quad \quad  \quad  \quad  \quad$
$T_{c, \textrm{inv}}$ ($s$)&  $3.18 \times 10^{-3}$     &  $4.52 \times 10^{-2}$ &  $1.6 \times 10^{-3}$        \\\hline
Reynolds number $Re$ (-) &    952    &    2299 &    400 
\\
$\quad \quad \quad \quad  \quad \quad  \quad  \quad$
$Re_{\sigma}$ (-) &    30.9     &    48 &   20 
\\\hline
Ohnesorge number $Oh$ (-) & 0.032      & 0.021  & 0.05 
\\\hline
\end{tabular}
\end{center}
\label{tab:typical-parameters-exp}
\end{table}

\section{Concluding Remarks}
\label{sec:conclusion}

To summarize, we have investigated through numerical simulations
a 2-D drop on a wall with a stepwise wettability gradient (WG)
specified by two distinct contact angles 
under a broad range of conditions, covering different Reynolds numbers
and viscosity ratios, different magnitudes of WG and contact angle hysteresis (CAH).
Almost under all conditions (except when the CAH is sufficiently large), 
the drop was accelerated very quickly in the initial stage
and gradually reached a steady state.
The input Reynolds number (based on the physical properties of the fluids and the drop dimension)
was found to have little effect on the capillary number of the drop in steady state.
The steady capillary number increases with the viscosity ratio 
significantly
when the viscosity ratio is small, but its dependence on the viscosity ratio 
becomes weaker at large viscosity ratios. 
Besides, this capillary number shows linear dependence on the magnitude
of the WG under most situations.
In the presence of CAH,
the motion of the drop is largely determined
by the advancing contact angle of the more hydrophilic region
and the receding contact angle of the more hydrophobic region
if the CAH is not too large.
When the hysteresis is high enough, the drop remains static
because it is within the range of possible configurations
allowed by the advancing and receding contact angles of both regions.
In future, an important further step should be the extension of both the model and investigations to 3-D cases.
The 2-D problems with CAH are relatively simple and the WG in the presence of CAH
may be characterized by an equivalent parameter straightforwardly. 
However, it will not be as easy for 3-D problems; it remains to be explored
whether such an equivalent parameter exists, and if so, how it can expressed in terms of other known parameters.

\bf Acknowledgement \rm

This work is supported by the National Natural Science Foundation of China 
(NSFC, Grant No. 11202250) 
and the Fundamental Research Funds for the Central Universities
(Project No. CDJZR12110001, CDJZR12110072).
We thank Prof. Hang Ding from the University of Science and Technology of China
for helpful discussions.

\end{document}